\documentclass[a4paper,11pt]{article}
\pdfoutput=1
\usepackage{jheppub}
\usepackage{xcolor}
\colorlet{mdtRed}{red!50!black}
 \usepackage[]{hyperref}
\hypersetup{colorlinks=true,% Set to false to disable coloring links
linkcolor=gray,% The color of references to document elements (sections, figures, etc)
urlcolor=mdtRed
}
\usepackage[utf8]{inputenc}
\usepackage{tikz}
\usetikzlibrary{decorations.pathmorphing}
\usepackage{overpic}
\usepackage{subcaption}
\usepackage{fancybox}
\usepackage{float}
\usepackage{amsfonts}
\usepackage{amsmath}
\usepackage{amsbsy}
\usepackage{graphicx}
\usepackage{amssymb}
\usepackage{amsthm}
\usepackage{bm}
\usepackage{comment}
\usepackage{epsfig}
\usepackage{latexsym}
\usepackage{pdflscape}
\usepackage{color}
\usepackage{slashed}
\usepackage{tikz}
\usetikzlibrary{arrows.meta,decorations.markings}

\tikzset{
  mid arrow/.style={
    postaction={
      decorate,
      decoration={
        markings,
        mark=at position 0.5 with {\arrow{Stealth[length=6pt,width=6pt]}}
      }
    }
  }
}

\usepackage{cancel}
\usepackage{xcolor}
\usepackage{braket}
\usepackage{empheq}
\usepackage{graphicx} % Allows including images
\usepackage{braket}
\usepackage{tikz}
\usepackage{tikz-3dplot}
\tdplotsetmaincoords{20}{110}
\usetikzlibrary{decorations.pathreplacing,calc}
\usetikzlibrary{decorations.pathmorphing}

\newcommand{\inlinefig}[2][8]{
    \raisebox{-0.37\totalheight}{\includegraphics[height=#1\fontcharht\font`\B]{#2}}
}

\definecolor{blue5}{rgb}{0, 0, 0.66666}
\definecolor{red4}{rgb}{0.8, 0, 0}
\definecolor{navyblue}{rgb}{0, 0.2745, 0.67843}
\definecolor{cornflowerblue}{rgb}{0.388235, 0.6941176, 0.898039}
\definecolor{forestgreen}{rgb}{0.13, 0.55, 0.13}
\usetikzlibrary{decorations.text,decorations.pathmorphing}

\makeatletter
\def\@fpheader{\relax}
\makeatother

\usepackage{orcidlink}

\DeclareMathOperator{\vol}{vol}

\newcommand{\beq}{\begin{eqnarray}}
\newcommand{\eeq}{\end{eqnarray}}
\newcommand{\bea}{\begin{eqnarray}}
\newcommand{\eea}{\end{eqnarray}}
\newcommand{\be}{\begin{equation}}
\newcommand{\ee}{\end{equation}}
\newcommand{\bq}{\begin{equation}}
\newcommand{\eq}{\end{equation}}

\newcommand{\half}{\frac{1}{2}}
\newcommand{\nn}{\nonumber}

\def\k{\kappa}

\newcommand{\bto}{\bar{\tau}^{\textbf{I}}}
\newcommand{\btt}{\bar{\tau}^{\textbf{II}}}

\def\l{\lambda}
\def\lab{\label}

\def\t{\tau}

\def\half{\frac12}

\def\m{\mu}
\def\n{\nu}

\def\6{\partial}

\def\a{\alpha}
\def\b{\beta}
\def\lab{\label}

\def\lab{\label}

\def\6{\partial}

\def\t{\tau}

\def\k{\kappa}

\title{A Menagerie of Wormholes and Cosmologies in the Gravitational Path Integral}

\author[a]{Panos Betzios\orcidlink{0000-0002-5350-9404},}
\author[b]{Paul Ghiringhelli,}
\author[c]{Ioannis D.~Gialamas\orcidlink{0000-0002-2957-5276},}
\author[b]{Olga Papadoulaki\orcidlink{0000-0001-5302-2930}}
\affiliation[a]{\href{https://www.ugent.be/we/physics-astronomy/en}{Department of Physics and Astronomy},
Ghent University, \\ Krijgslaan, 281-S9, 9000 Gent, Belgium}
\affiliation[b]{\href{https://www.cpht.polytechnique.fr/?q=en}{CPHT, CNRS, École polytechnique, Institut Polytechnique de Paris},  91120 Palaiseau, France}
\affiliation[c]{\href{https://kbfi.ee/high-energy-and-computational-physics/?lang=en}{Laboratory of High Energy and Computational Physics}, National Institute of Chemical Physics and Biophysics, 
R{\"a}vala pst.~10, Tallinn, 10143, Estonia}
 
\emailAdd{panos.betzios@ugent.be}
\emailAdd{paul.ghiringhelli@polytechnique.edu}
\emailAdd{ioannis.gialamas@kbfi.ee}
\emailAdd{olga.papadoulaki@polytechnique.edu}

\abstract{We analyse a variety of Euclidean saddles in the gravitational path integral, with asymptotic AdS boundary conditions, in a class of Einstein-Scalar-Maxwell models. These include single boundary solutions, usual and wineglass wormholes, as well as more exotic (quasi)-oscillatory saddles. Our construction shows how an unbound number of oscillations gets tamed, when flat directions of the potential get lifted. We find several interesting phase transitions between these solutions. The Euclidean wormhole backgrounds can be analytically continued to Lorentzian FLRW universes. Some of them contain an early period of inflation. We delineate the conditions under which they can be the dominant saddles in the gravitational path integral and use them to estimate ratios of probabilities for different cosmological outcomes.}

\keywords{Euclidean Wormholes, Phase Transitions, Holography, Cosmology, Inflation}

\begin{document}
\maketitle
\flushbottom

%%%%%%%%%%%%%%%%%%%%%%%%%%%%%%%%%%%%%%%%%%%%%%%%%%%%%%%%%%
\section{Introduction}
\label{sec:in}

The gravitational path integral plays a central role in our understanding of gravitational physics, ranging from black hole thermodynamics to holography and cosmology, yet its precise definition remains an open problem. In contrast to ordinary quantum field theories, gravity is background independent whilst very sensitive to the asymptotic structure of spacetime, perturbatively nonrenormalizable, and in Euclidean signature afflicted by an unbounded conformal mode. As a consequence, the path integral definition of the integration over geometries is quite subtle and naturally depends sensitively on the choice of contour in the space of metrics. Despite these difficulties, the gravitational path integral continues to provide a useful organizing framework, particularly within the semiclassical regime.

In the semiclassical approximation in particular, the gravitational path integral is typically organized as a sum over saddle points of the Euclidean action. In many settings, multiple saddle geometries coexist and compete, with the dominant contribution depending on external parameters such as boundary conditions or the couplings of the specific gravitational theory. Transitions in dominance between distinct saddles give rise to a rich phase structure in the physics of the gravitational systems~\cite{Hawking:1982dh}, analogous but conceptually even richer to thermodynamic phase transitions of non-gravitating local field theory systems. Understanding this competition between saddles is essential for interpreting the gravitational path integral and for determining the different physical regimes of the system at hand.

Among the relevant saddle points that we study in this work are Euclidean wormhole geometries, which connect multiple (in our examples two) asymptotic regions or boundaries and represent intrinsically nonperturbative configurations/solutions of the semiclassical equations of motion (EOMs). In holography and anti-de Sitter/conformal field theory correspondence (AdS/CFT), such saddles lead to the definition of multi-boundary observables and raise sharp questions concerning the factorization of the path integral and the interpretation of bulk gravitational correlations in terms of a dual quantum field theory system~\cite{Maldacena:2004rf}. Their role in the gravitational path integral is therefore closely tied to foundational issues in holography and to the broader problem of defining the gravitational sum over geometries.

A particularly interesting aspect of Euclidean wormhole saddles is their relation, via analytic continuation, to Lorentzian cosmological spacetimes. Different Euclidean saddles may admit continuations leading to qualitatively distinct time-dependent geometries. In particular, while some Euclidean wormhole saddles continue to cosmologies that undergo a big-bang/big-crunch evolution, see~\cite{Maldacena:2004rf,Betzios:2019rds,Betzios:2021fnm,VanRaamsdonk:2021qgv,Antonini:2022blk,Antonini:2022ptt} for some early works in the context of holography that emphasized this perspective\footnote{Similar wormhole saddles can also be constructed by considering correlated pairs of operators in two CFT copies~\cite{Betzios:2023obs,Maloney:2025tnn,VanRaamsdonk:2026tnv} or by appropriate pinching limits of higher genus Riemann surfaces in $3d$ gravity~\cite{Belin:2025ako}.}, there exist others that can remarkably give rise to inflationary cosmological spacetimes~\cite{Betzios:2024oli,Betzios:2024zhf}\footnote{We should emphasize here that this construction is different from quantum tunnelling/Coleman de-Lucia processes~\cite{DeAlwis:2019rxg} or the ``bag-of-gold'' spacetimes behind black hole horizons~\cite{Fu:2019oyc}, where a portion of Lorentzian de Sitter could appear inside Lorentzian anti-de Sitter. These later cases seem to suffer from a version of the information paradox~\cite{Freivogel:2005qh}, whereby a pure state has to evolve into a mixed state.}  - see also~\cite{Lavrelashvili:1988un} for an early incarnation of this idea in asymptotically flat space. The competition between Euclidean saddles in the path integral thus translates, after continuation, into a competition between distinct cosmological outcomes. Clarifying how this semiclassical phase structure manifests itself in Lorentzian signature provides a new perspective on the emergence of cosmological spacetimes from holography.

In this work, in section~\ref{sec:wavefunction}, we start with a general discussion on the relation between different classes of Euclidean saddle points of the gravitational path integral and how they can be used in a cosmological setting to define normalised probabilities of different outcomes of the analytically continued cosmologies at the semi-classical level. Due to various issues inherent to the no-boundary (NB) proposal of Hartle and Hawking~\cite{Hartle:1983ai,Lehners:2023yrj,Maldacena:2024uhs}, we are driven to consider the possibility that well defined non-trivial cosmological states/wavefunctions and probabilities can only exist in models that admit asymptotically AdS gravitational saddles~\cite{Betzios:2024oli,Betzios:2024zhf}, whereby different dominant saddles correspond to distinct cosmological outcomes, including crunching and expanding universes.

In order to make our general discussion precise, in section~\ref{sec:ESRmodel} we study a three parameter family of Einstein-Scalar-Maxwell models. The three parameters determine the shape of the scalar potential. Starting from an analytic Friedmann–Lema\^itre–Robertson–Walker (FLRW) ansatz for the metric and potential that depend on a single Euclidean time coordinate $\tau$, we solve the saddle point EOM in terms of  the said three parameters and three additional parameters (we use a six parameter ansatz) that are fixed by the asymptotic value of the metric/scalar and gauge fields on the Euclidean AdS (EAdS) boundaries.%\footnote{The solutions we consider are $\mathbb{Z}_2$ symmetric and these values are the same on both boundaries.}.

In our models, since we find that the conformal dimension of the dual scalar operator $\Delta_{\mathcal{O}_s} < 3/2$, both scalar asymptotic modes are normalizable, and we have two choices leading to different quantizations of the scalar field. We analyze both choices (with and without a scalar source). The same also happens for the gauge field(s) that provides the appropriate flux supporting the wormhole geometries, leading to four possibilities (Dirichlet vs. Neumann\footnote{More general mixed BCs. also exist but we do not analyze them in this work.}) in total and we analyze all of them in section~\ref{sec:MenagerieSaddles}. In particular (and dependent also on the boundary conditions (BCs) for the various fields) we find a plethora of saddles, with a non-trivial gauge field flux and with or without a running scalar - this menagerie of possibilities is classified and described in detail at the beginning of section~\ref{sec:MenagerieSaddles}.

In section~\ref{sec:comparison}, we compare the various saddles (both factorised and non-factorised) by computing their on-shell action differences with the method of background subtraction. We perform this analysis explicitly for a four (in total) parameter sub-family of models, for which most of the calculations can be performed analytically. The results are summarised in figs.~\ref{fig:saddle_dom} and~\ref{fig:saddle_dom_2} for all the various choices of boundary conditions.

We finally conclude with a discussion on possible future directions in section~\ref{sec:Discussions}.

\section{Cosmological wavefunctions, transition amplitudes and probabilities}
\label{sec:wavefunction}

In this section, we discuss the cosmological interpretation of our results\footnote{More details that include perturbations of fields will be presented in future work~\cite{Betzios2025toappear}.}, noticing that Euclidean wormhole backgrounds can be analytically continued to Lorentzian FLRW cosmologies~\cite{Maldacena:2004rf,Betzios:2017krj,Betzios:2019rds,Betzios:2024oli,Betzios:2024zhf,VanRaamsdonk:2021qgv,Antonini:2022blk,Antonini:2022ptt}.

We first start with the Hartle-Hawking wavefunction~\cite{Hartle:1983ai}, that can be computed using the Euclidean path integral
\be\label{Noboundarywvf}
\langle N.B. | h_{i j}, \phi_0 \rangle = \Psi_{N.B.}[h_{i j}, \phi_0] = \int^{g_{i j}(0) = h_{i j}} \mathcal{D}g \int^{\phi(0) = \phi_0} \mathcal{D} \phi e^{- S_E[g_{\m \n}, \phi]} \, .
\ee
This path integral is over compact Euclidean geometries in the Euclidean past (``no-boundary''), so that there is only a single boundary on which we define a bulk gravity state/wavefunction (here we take it to be at $\tau = 0$---at the ``birth'' of the Lorentzian Universe). We can then evolve this wavefunction to a real Lorentzian time $t$. The resulting complete gravitational path integral is described at the semi-classical level by a complex geometry that interpolates between Euclidean and Lorentzian de-Sitter (assuming $V(\phi) > 0$). While being a quite elegant and beautiful construction, the no-boundary proposal~\cite{Hartle:1983ai} is plagued by notorious issues, reviewed in~\cite{Quantum:Cosmology,Lehners:2023yrj,Maldacena:2024uhs,Betzios:2024oli}, the most important being that it is at clash with inflationary theory, predicting the smallest possible number of inflationary e-folds. Additionally, it cannot be derived or justified within any known consistent model of quantum gravity such as holography or string theory. In a recent work~\cite{Abdalla:2026mxn}, even the original interpretation of the gravitational path integral in eq.~\eqref{Noboundarywvf} as a wavefunction by Hartle and Hawking came at question, and upon using a standard normalization of gravitational transition amplitudes (i.e. Born's rule)
\be\label{bulktransitionnorm}
\langle h_{i j}^{(1)}, \phi^{(1)} | h_{i j}^{(2)}, \phi^{(2)} \rangle_{\text{norm.}} = \frac{\langle h_{i j}^{(1)}, \phi^{(1)} | h_{i j}^{(2)}, \phi^{(2)} \rangle}{\sqrt{\langle h_{i j}^{(1)}, \phi^{(1)} | h_{i j}^{(1)}, \phi^{(1)} \rangle} \sqrt{\langle h_{i j}^{(2)}, \phi^{(2)} | h_{i j}^{(2)}, \phi^{(2)} \rangle}} \, ,
\ee
the authors were led to the remarkable conclusion that classically distinct data $| h_{i j}, \phi \rangle$, actually define nearly parallel states of the gravitational Hilbert space up to non-perturbatively small differences in the Newton's constant $G_N$. This would imply that the (normalised) transition probability from one closed universe state to any other is also approximately one, which seems to disagree with even the most basic quantum mechanical rules and experiments. Given this lack of predictability, the authors of~~\cite{Abdalla:2026mxn} also considered the possibility of projecting out the no-boundary state (for example by redefining the rules of the gravitational path-integral). One is thus naturally led to consider alternative proposals for the wavefunction of the universe that incorporate the presence of asymptotic boundaries---one such possibility being that EAdS wormholes such as those in~\cite{Maldacena:2004rf,Betzios:2019rds,Betzios:2021fnm,VanRaamsdonk:2021qgv,Antonini:2022blk,Antonini:2022ptt,Betzios:2024oli,Betzios:2024zhf}, are the semiclassical saddles that should be used to define non-trivial bulk gravitational states, see~\cite{Betzios:2024oli} for more details on this proposal in the context of inflationary cosmology.  

In this proposal, one considers models that admit a collection of EAdS wormhole saddles (so $V(\phi)$ has to be negative for some field range of $\phi$, or more generally the steepest descend contour on which the gravitational path integral is defined needs to pass through such geometries), that contain (at least) two asymptotic boundaries. Given a Euclidean gravitational saddle with a $\mathbb{Z}_2$ (or CPT) symmetry,
one can then slice it in half to define a consistent wavefunction of the (bulk) gravitational path integral on the slice. While still being a subject of ongoing debate regarding their precise microscopic interpretation, the presence of the AdS boundaries for the wormholes conforms with holographic (AdS/CFT) expectations and allows the potential embedding of the resulting cosmologies in a well defined framework for quantum gravity%\footnote{This is true regardless on whether we wish to adopt an averaged or non averaged holographic perspective for such two boundary wormhole configurations.}
, giving rise to a non-trivial bulk Hilbert space, descending precisely from the non-trivial Hilbert space of the dual Holographic QFT. In this case, the state/wavefunction associated with the time reflection symmetric point ($\tau = 0$) is defined at the semi-classical level by a sliced (half) wormhole manifold. In such an asymptotically EAdS (half) wormhole space, one finds the presence of an additional conformal boundary at the infinite (Euclidean) past, in contrast with the Hartle-Hawking prescription. This means that the ``wavefunction'' at $\tau = t = 0$
is now given by the following Euclidean path integral corresponding to an actual transition amplitude
\be
\langle g^{\partial}_{i j}, J^{\partial}_\phi | h_{i j}, \phi_0 \rangle \, \equiv \, \Psi[ g^{\partial}_{i j}, J^{\partial}_\phi \, ; \, h_{i j}, \phi_0] = \int^{g_{i j}(0) = h_{i j}}_{g_{i j}(\partial \mathcal{M}) \rightarrow g_{i j}^{\partial}} \mathcal{D}g \int^{\phi(0) = \phi_0}_{\phi(\partial \mathcal{M}) \rightarrow J^{\partial}_\phi} \mathcal{D} \phi e^{- S_E[g_{\m \n}, \phi]} \, .
\ee
The result for the wavefunction/state at $\tau = 0$ is now dependent on the asymptotic EAdS boundary conditions of the various fields and in particular on the sources for the metric and any other matter fields at the EAdS boundary, as well as the choice of state. Of course from a holographic perspective there is always a preferred canonical choice of state---that of the holographic QFT vacuum (assuming it exists)---and the sources then play the role of ``knobs" that we can introduce in order to differentiate the functional and compute correlation functions of operators on top of the vacuum and other excited states. In our setup, after evolving the wavefunction into Lorentzian signature, these become cosmological correlators for the resulting cosmology. It is evident now that in contrast with the NB proposal, this construction is endowed with a non-trivial and potentially rich Hilbert space of states. Of course this is to be expected since it parallels our quite explicit examples of the holographic correspondence in the case of single boundary saddles\footnote{Whether one views the wormhole as arising from some microscopic construction in string theory, or from averaging a microscopic (single) boundary model, is immaterial for the presence of the non-trivial Hilbert space. What is different in these options is the (in-principle) accessibility to the complete microscopic state or only to a coarse grained version thereof, when using the bulk gravitational theory.}.

At this point we should emphasize that due to different asymptotic conditions in the Euclidean past, one cannot really compare the no-boundary with the wormhole saddles in the gravitational path integral---they actually seem to descend from different theories of quantum gravity\footnote{Unless there is a way to make sense of the sum over the class of all boundary conditions including the cases of not even having a boundary as well as having sources or no sources. This would correspond to a sum over ``states'' of some sort of ``metatheory'' and transitions between such ``states'' do not seem possible to happen for any physical system that we know so far.}. An alternative way to phrase this is that the space of all asymptotic possibilities/configurations can be split into a ``superselection sector sum'' $\mathcal{H} = \cup_{\rm asym.} \mathcal{H}_{\rm asym.}$, where with ``asym." we label classes with inherently different asymptotics i.e AdS vs. flat boundaries vs. no-boundaries and different sources, for which no finite energy operation can induce cross-transitions between them. This leads to an equivalent conclusion: different asymptotics define different theories of quantum gravity when there is no physical operation that can induce transitions between them (see~\cite{Banks:2025nfe,Sen:2025bmj} for similar ideas). From those possibilities, and within our current knowledge of microscopic models of quantum gravity, the asymptotic AdS wormhole proposal is the only one admitting a well defined UV completion with a well defined underlying Hilbert space of states encoded in the Holographic dual QFT.

This also means that one needs to modify slightly and enrich the notion of probabilities in this setting compared to the recent analyses of the no-boundary proposal, for which the Hilbert space of closed Universes is found to be one-dimensional. As emphasized in~\cite{Betzios:2022oef} one has on the one hand the (UV) asymptotic AdS boundary ($AB$) Hilbert space of states $\mathcal{H}_{AB}$ (descending from the Hilbert space of the dual QFT and leading to different vacuum expectation values (VEVs) for bulk fields in a region where the geometry and fields become non fluctuating) and on the other hand the interior (bulk) boundary Hilbert space of states $\mathcal{H}_{IB}$ ($IB$) (which is defined upon slicing a geometry on a locus where the metric and fields are dynamical and fluctuate and where the gravitational constraints are most important). Any mapping between these two Hilbert spaces of states is bound to be highly non-trivial and rests upon the precise definition and operational details of holography. The meaningful probabilities that one can define in this setting and which are also relevant for cosmology, are computed using transition amplitudes from an $AB$-state to some $IB$-state
\be
P(AB \rightarrow IB) = \frac{|\langle AB | IB \rangle|^2}{\langle AB | AB \rangle \langle IB | IB \rangle} \, .
\ee
In the examples that we study in this work, the numerator and the first term in the denominator admit a semi-classical saddle point approximation. Two typical backgrounds that we discuss in section~\ref{sec:MenagerieSaddles} are wormholes with a single minimum of the scale factor, and ``wineglass'' wormholes with two local minima and one local maximum of the scale factor (factorised single-boundary saddles are also always present). As we elaborate upon in section~\ref{sec:MenagerieSaddles}, the resulting cosmologies correspond to those of a crunching-Universe or an inflating Universe respectively. The resulting probability for these two possibilities at the semi-classical level is determined from the ratios (see figure~\ref{fig:wormholes} for the different types of EAdS saddles analyzed in this work)
\begin{align}
    P(AB \rightarrow IB_{1,2}) = \frac{\left|\, \raisebox{0.2ex}{\tiny $AB$\,}
 \inlinefig[3.5]{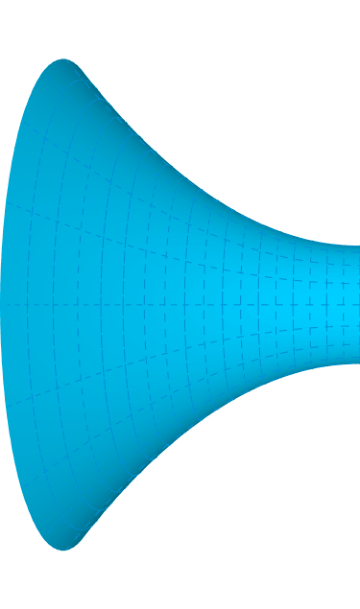} \raisebox{0.2ex}{\tiny \,$IB_1$}  \quad\text{or}\quad \raisebox{0.2ex}{\tiny $AB$\,}\inlinefig[3.2]{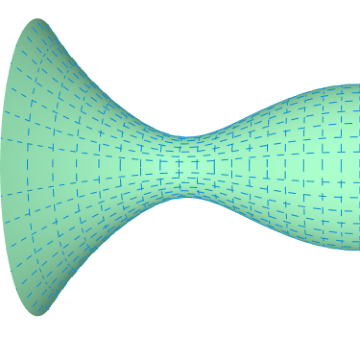}\raisebox{0.2ex}{\tiny \,$IB_2$} \quad\text{or} \, \,  \cdots \, \, \right|^2}{\left(\raisebox{0.2ex}{\tiny $AB$\,}\inlinefig[2.8]{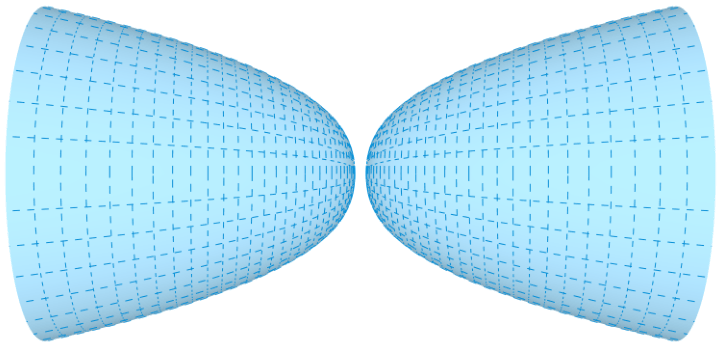}\raisebox{0.2ex}{\tiny \,$AB$}+\raisebox{0.2ex}{\tiny $AB$\,}\inlinefig[3.5]{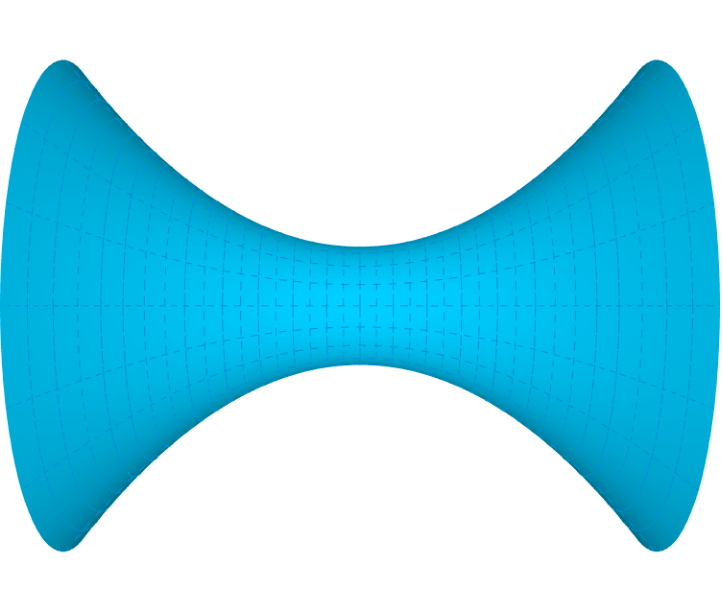}\raisebox{0.2ex}{\tiny \,$AB$}+\raisebox{0.2ex}{\tiny $AB$\,}\inlinefig[3.2]{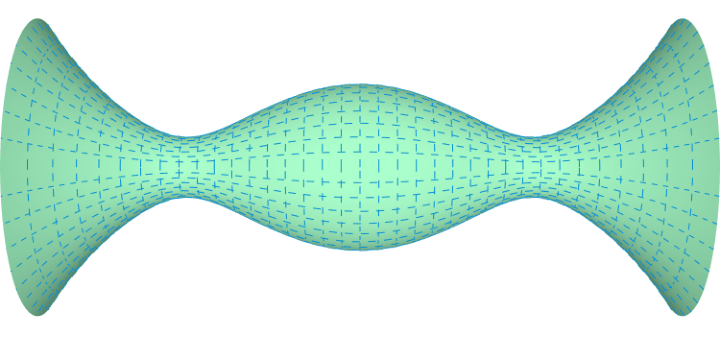}\raisebox{0.2ex}{\tiny \,$AB$} + \cdots \, \right)\, \langle IB_{1,2} | IB_{1,2} \rangle},
\end{align}
where with the ellipsis $(...)$ we denote wineglass wormholes with a further number of oscillations in their throat (should they exist)
and where $\langle IB_{1,2} | IB_{1,2} \rangle$ is to be computed using~\eqref{bulktransitionnorm}, depending also on whether we admit the presence of a bulk no-boundary state (i.e. $\langle N.B. | IB \rangle$) and the resulting factorised contribution or not\footnote{One could imagine that $\langle IB | IB \rangle$ should be computable by inserting a complete sum over all microscopic bulk states (if there is such a complete set), or over all the microscopic QFT states. In practise though this computation is difficult to perform.}. We immediately observe that the role of single boundary geometries and their products, such as the ones related to either the Hartle-Hawking no-boundary state or asymptotically AdS geometries with a single boundary, affect only the normalization of the states making all the probabilities less than one, but they never contribute to the relevant cosmological transition amplitude in the numerator. Hence in the AdS wormhole proposal to cosmology, physically meaningful ratios of probabilities have common $(AB)$-boundary conditions and read
\be
r_{12} = \frac{P(AB \rightarrow IB_1)}{P(AB \rightarrow IB_2)} \, .
\ee
In practice, they can be computed at the semiclassical level using ratios of the on-shell actions of connected (half) wormhole geometries with wavefunctions at $\tau =0 $ corresponding to the various $IB_i$ states. More generally, if one wishes to compute probabilities that involve later real times $t$, one has to consider complex geometries, part of which corresponds to the (half)-wormhole and a part of which depends on the specific inflationary model/region of the scalar potential (we leave this more complete analysis for the future~\cite{Betzios2025toappear}). 

In the model we analyze in this work, we find that for certain choices of UV BCs. ($AB$) a wineglass (half) wormhole manifold dominates, while for others a simple connected (half) wormhole dominates. This simply means that in the first case, it is much more probable that the Universe gets nucleated at $\t =0$ and then starts to inflate, while in the second case it is more probable that the Universe will undergo a big-Crunch collapse. From the $AB$-cases for which inflation is the most probable outcome, the dominant initial condition ($IB_i$) for the state of the Universe is determined from the classical solution to the bulk EOMs\footnote{As mentioned previously there are always off-shell or complex contributions for transitions to various $IB_i$ states, but these are all suppressed at the semi-classical level, due to the existence of a real classical saddle.} - in the wineglass case, this is found to be high up in the inflationary plateau of the effective potential $V(\phi)$, giving further credence to the proposal of~\cite{Betzios:2024oli,Betzios:2024zhf} as providing a good initial state for the beginning of an inflationary Universe.
The basic question remaining is to determine the conditions under which ``wineglass'' wormholes (or quasi-oscillatory wormholes of the type that we describe in section~\ref{sec:MenagerieSaddles}) can be the dominant connected saddles of the gravitational path integral, rendering inflation as the most probable outcome for the early Universe cosmology.

%%%%%%%%%%%%%%%%%%%%%%%%%%%%%%%%%%%%%%%%%%%%%%%%%%%%%%%%%%

\section{The Einstein-scalar-radiation model}\label{sec:ESRmodel}

Our starting point is the Einstein–Hilbert action with a minimally coupled scalar field $\phi$ (the inflaton), in the presence of gauge/radiation fields
%\footnote{These could also be axionic fields~\cite{Betzios:2024oli}.} 
which play an important role in shaping the solutions of interest.
This is a quite general model that forms a subsector of the Standard Model in the case that the scalar inflaton is identified with the Higgs boson~\cite{Bezrukov:2007ep,Barvinsky:2008ia,DeSimone:2008ei,Barbon:2009ya,Barvinsky:2009fy,Barvinsky:2009ii,Bezrukov:2010jz,Bezrukov:2014ipa,Betzios:2024zhf,Gialamas:2025kef}. We shall work in four dimensions motivated from phenomenological purposes, but similar solutions can be found in an arbitrary number of dimensions.

The four dimensional Euclidean action of the model is given by
\begin{equation}\label{background}
\mathcal{S}_E = \int {\rm d}^4x \sqrt{g_E} \left(-\frac{1}{2\kappa}\mathcal{R} + \frac12 (\partial_\mu{\phi})^2 +V(\phi) + \mathcal{L}_{\rm rad}^E \right)+\mathcal{S}_{\rm GHY}\,,
\end{equation}
where $\mathcal{R}$ denotes the scalar curvature, $V(\phi)$ is the scalar field potential, $\kappa=1/M_{\rm Pl}^2$ and $\mathcal{S}_{\rm GHY}$ is the Gibbons-Hawking-York boundary term which has been added to render the variational problem well defined. Higher–order curvature corrections and possible non-minimal couplings between gravity and the scalar field are also expected to be generated at the quantum level. Such terms could, in principle, have a significant impact on the inflationary dynamics of the system, for example by flattening the scalar potential and modifying the inflationary trajectory leading to different inflationary parameters (see~\cite{Bezrukov:2010jz,Bezrukov:2014ipa,Gialamas:2025kef} and~\cite{Betzios:2024zhf} for an application in the present context). Nevertheless, in what follows we shall keep the discussion and analysis general within our minimal setup, since it can be applied to any particular inflationary model.

In a Euclidean FLRW (EFLRW) type of ansatz (homogeneous and isotropic)
\be
\label{eq:ansatz}
{\rm d} s^2_{\rm EFLRW} =  {\rm d} \tau^2 +a^2(\tau)  {\rm d} \Sigma_3^ 2\, , \quad \phi =\phi(\tau) \, , \quad \rho^E_{\rm rad}=\rho^E_{\rm rad}(\tau) \, ,
\ee
the inflaton and gravitational Euclidean EOMs take the form
\begin{align}
\tilde{\phi}''+3\frac{a'\tilde{\phi}'}{a}-\frac{{\rm d}\tilde{V}(\phi)}{{\rm d} \tilde{\phi}} &= 0\,, \nonumber
\\
\frac{2a''}{a}+\frac{a'^2}{a^2}-\frac{k}{a^2}  + 3 \left(\tilde{V}(\phi)+\frac{\tilde{\phi}'^2}{2} \right) +  \frac{\tilde{\rho}^E_{\rm rad}}{a^4} & =0  \,, \nonumber
\\ 
\frac{a'^2}{a^2} -\frac{k}{a^2}  + \left(\tilde{V}(\phi)-\frac{\tilde{\phi}'^2}{2} \right) - \frac{\tilde{\rho}^E_{\rm rad}}{a^4} & =0 \,.
\label{eq:eoms1}
\end{align}
Here the prime denotes a derivative with respect to the Euclidean time $\tau=i t$ , and the parameter $k = 1, 0, -1$ corresponds to closed (spherical), flat, and open (hyperbolic) spatial slices, respectively. For convenience, we defined rescaled quantities $\tilde{V}(\phi) = \k V(\phi)/3$, $\tilde{\phi} = \sqrt{\k/3}\phi$ and $ \kappa  a^4 \rho_{\rm rad}^E/3 = \tilde{\rho}^E_{\rm rad} = \rm const. $. 
Only the first and last equations in~\eqref{eq:eoms1} are independent, so introducing a logarithmic variable for the scale factor, $a(\tau) = e^{A(\tau)}$ , the independent equations can be rewritten in a more compact form as
\bea
\tilde{\phi}''+3 A' \tilde{\phi}' -\frac{{\rm d}\tilde{V}(\tilde\phi)}{{\rm d} \tilde{\phi}} &= 0\,, \nonumber
\\
A'^2 - k e^{- 2 A}  + \left(\tilde{V}(\tilde\phi)-\frac{\tilde{\phi}'^2}{2} \right) -  \tilde{\rho}^E_{\rm rad} e^{- 4 A} & =0 \label{eq:equAp} \, .
\eea
Taking a derivative of the last equation and using the first we find
\be
\label{eq:diffA}
2 A'' + 2 k e^{- 2 A} + 4 \tilde{\rho}^E_{\rm rad} e^{- 4 A} + 3 \tilde{\phi}'^2 = 0 \, ,
\ee
from which we observe that once $\tilde{\phi}(\tau)$ is given, all the rest of the functions can be determined, if we can solve for the scale factor, and vice versa, if the scale factor is known. Written in terms of the original scale factor the velocity reads
\begin{equation}\label{eq:velocity}
\tilde{\phi}'^2(\tau) = -\frac{2}{3a(\t)^4}\left[a(\t)^2(k-a'(\t)^2+a''(\t) a(\t) )+2\tilde{\rho}^E_{\rm rad}\right]\,. 
\end{equation}
At the AdS UV boundary, where the scale factor goes to infinity $a \sim a_{\rm UV} e^{H_{\rm UV} |\tau|}\rightarrow\infty$, the velocity goes to zero for any type of slices.

The dominance of $SO(4)$-symmetric Euclidean configurations in the path integral~\cite{Coleman:1985rnk}, which reflects the universe's observed isotropy and homogeneity, motivates the use of the closed (spherical) case, $k=1$ , which we shall adopt in the rest of this work. An important aspect to consider is the sign of the electromagnetic energy density $\tilde{\rho}^E_{\rm rad}$. To obtain the Euclidean wormhole solutions for the cosmological purposes of interest~\cite{Betzios:2024oli}, it is necessary that $\tilde{\rho}^E_{\rm rad} < 0\,$.  This requirement becomes clear from the third equation in~\eqref{eq:eoms1}, where the curvature term $-1/a^2$ tends to reduce the scale factor, whereas the term $-\tilde{\rho}^E_{\rm rad}/a^4$, with $\tilde{\rho}^E_{\rm rad} < 0$ , contributes to its expansion. The competition between these terms gives rise to the wormhole neck and supports the subsequent accelerated expansion of the universe. The condition $\tilde{\rho}^E_{\rm rad} < 0\,$ can be naturally realized if the magnetic field contribution dominates over the electric one in the early universe~\cite{Betzios:2024zhf}\footnote{A similar mechanism can also work in braneworld models such as the ones studied in~\cite{Antonini:2024bbm} and it would be interesting to extend our analysis in this direction.}.
In particular, the Euclidean Lagrangian of the electromagnetic field in the $U(1)$ case is $\mathcal{L}_{\rm rad}^E = F_{\m\n}^E F^{E\,\m\nu}/4=(E^2+B^2)/2\,$, which, through the corresponding electromagnetic energy-momentum tensor, ${T^E_{\rm rad}}^\m_{\,\,\n}$ , leads to a Euclidean radiation energy density of the form
\begin{equation}\label{generaleuclideanradiation}
\rho^E_{\rm rad} = {T^E_{\rm rad}}^0_{\,\,0}=\frac12(E^2-B^2)\,.  
\end{equation}
Unlike in the Lorentzian case, where the radiation energy density is always non–negative, in the Euclidean case it can become negative whenever the magnetic contribution dominates. 

A similar mechanism can be demonstrated for axions (or two form) fields instead of gauge (or one form) fields, see for example~\cite{Betzios:2024oli}. This again stems from the fact that the axionic contribution to the energy density $\rho_{\rm axion}$ arising from a three form flux $H_{i j k}$ is negative in the Euclidean signature, since it is always of the magnetic type (no $\tau$ component).
  A final type of mechanism giving rise to Euclidean wormhole saddles is to consider slices that have negative curvature i.e. $k = - 1\,$ (combined now with dominant electric radiation density). The slices should be then compactified (in analogy to higher genus Riemann surfaces in two dimensions), if the wormhole interpretation of the backgrounds is to be retained~\cite{Betzios:2019rds}.

\subsection{An exactly solvable \texorpdfstring{$U(1)^3$}{U(1)3} model for the radiation}

In this section we briefly introduce a specific model for the radiation part of the action, for which the complete set of equations (including the ones for the gauge fields) can be solved exactly. This is a model with a $U(1)^3$ symmetry (or in other words containing a triplet of gauge fields $A_\mu^{(I)}, \, I = 1,2,3$). It was first studied in a similar context in~\cite{Marolf:2021kjc} (for a fixed negative cosmological constant), but here we consider its extension, since we include the presence of the non-trivial inflaton field $\phi$, see also~\cite{Lan:2024gnv}.
We shall also improve the previous analyses of this model using differential forms for the gauge field, allowing us to define the solutions in a more invariant manner.

The Euclidean action for the radiation part is then
\be
\label{eq:radiationaction_0}
\mathcal{S}^E_{\rm rad} = \int {\rm d}^4 x \sqrt{g_E} \frac{1}{4} \sum_{I=1}^3 F^{(I) \, \m \n} F^{(I)}_{\m \n} \, ,
\ee
and the isotropic ansatz that we consider\footnote{This is invariant under the left action of $SU(2)$ and hence isotropic on $S^3$. A similar ansatz for an $SU(2)$ non-Abelian gauge field gives rise to the Meron wormhole, see~\cite{Maldacena:2004rf,Betzios:2019rds} and references within. See also~\cite{Rey:1989th} for a similar ansatz and manipulations, in the non-abelian case.} is
\be\label{fieldstrenght}
A^{(I)} = H(\tau) \omega^{I} \, , \qquad F^{(I)} = {H'(\tau)} {\rm d}\tau \wedge \omega^{I}  + \frac{H(\tau)}{2} \epsilon^I_{ J K} \omega^{J} \wedge \omega^K \, ,
\ee
with $\omega^{I}$ the Maurer-Cartan forms on the $S^3$ obeying
\be
{\rm d} \omega^I = \half \epsilon^I_{ J K} \omega^{J} \wedge \omega^K \, ,  \quad \sum_{I=1}^{3}\left(\omega^I\right)^2 = 4\rm{d}\Omega_3^2 \, .
\ee
The EFLRW metric can be written in the form
\be\label{EFLWR metric}
{\rm d}s^2 = \rm{d}\tau^2+\frac{1}{4}a^2(\tau)\omega^I \otimes \omega^I = {\rm d} \tau^2 + e^I \otimes e^I \, , \qquad e^I = \frac{a(\tau) \omega^I}{2} \, .
\ee
With this ansatz and in the local orthonormal frame $e^I$ we can define the electric and magnetic components of the total field strength
\be \label{eq:fieldstrenght}
F = E_I {\rm d} \tau \wedge e^I + \half B_I   \epsilon^I_{ J K}  e^J \wedge e^K \, , \quad E_I = \frac{2H'}{a} \, , \quad B_I =  \frac{4H}{a^2} \, .
\ee
The Hodge dual (Electric/Magnetic duality) is simply given by
\be
\star F = B_I {\rm d} \tau \wedge e^I + \half E_I   \epsilon^I_{ J K}  e^J \wedge e^K \, .
\ee
Using these results, the Euclidean energy density is found to be
\be\label{EuclEnergy}
 \rho^E_{\rm rad} = \frac{3}{2} \left[ \left(\frac{2H'}{a} \right)^2 - \left(\frac{4H}{a^2}\right)^2 \right] = 6\left[ \left(\frac{H'}{a} \right)^2 - \frac{4H^2}{a^4} \right] \, .
\ee
and the EOM for each of the gauge fields takes a very simple form
\be\label{EOMpotential}
 {\rm d} \star F = 0 \quad \Rightarrow H'' + \frac{a' H'}{a} -  \frac{{4}H}{a^2} = 0 \, .
\ee
The Electric/Magnetic (E/M) EOM can also be derived using the reduced $U(1)^3$ E/M action ($\text{Vol}(S^3) = {\frac{1}{8}} \int  \omega^1 \wedge  \omega^2 \wedge  \omega^3$)
\be\label{radiationaction}
\mathcal{S}^E_{\rm rad} =  \frac{1}{2} \int  F \wedge \star F  = 6 \text{Vol}(S^3) \int {\rm d} \tau \left( a H'^2 + \frac{4H^2}{a} \right) \, .
\ee
Performing one integral of the EOM, we find
\be\label{euclideanenerrgy}
 \rho^E_{\rm rad} = 6 \left[ \left(\frac{H'}{a} \right)^2 - \frac{4H^2}{a^4} \right] = \frac{\rm const.}{a^4} = \frac{3\tilde{\rho}^E_{\rm rad}}{\k a^4}\, ,
\ee
that exhibits the classical scale invariance of a $U(1)$ gauge theory in four dimensions.
As we mentioned previously, in order for $\tilde{\rho}_{\rm rad}^{E}<0$ , the magnetic contribution to the stress-energy tensor should be non-zero and dominant for our wormhole solutions of interest.

We can explicitly solve the gauge field EOM by passing to the conformal gauge $d \tau = a\, d u$, to find the simple saddle point equation and solution
\be\label{eq:equationH}
H_{uu} - 4H = 0 \, , \qquad H(u) =c_1 e^{2u} + c_2 e^{-2u} \, , \qquad u = \int_0^{\tau} \frac{{\rm d} \tau'}{a(\tau')} \, .
\ee
So given the scale factor $a(\tau)$ we can completely determine the gauge field. From this equation we also observe that the radiation density is related to the integration constant of the gauge field EOM i.e.
\be
\rho^E_{\rm rad} =   6 \left[\frac{H_u^2}{a^4} - \frac{4H^2}{a^4}\right] = - 96 \frac{c_1 c_2}{a^4} \, , 
\ee
so that it is non zero when both $c_{1,2}$ are non zero, else the background does not carry stress-energy.
\\
We can easily rewrite the electromagnetic on-shell action in the conformal gauge
\begin{equation}
    \begin{aligned}
        \mathcal{S}^E_{\rm rad} = 6\text{Vol}(S^3) \int_{u(\tau_1)}^{u(\tau_2)} {\rm d} u \left( H_u^2(u)+4H^2(u)\right)=12\text{Vol}(S^3)\Bigg(c_1^2\Big[e^{4u}\Big]^{u(\tau_2)}_{u(\tau_1)}-c_2^2\Big[e^{-4u}\Big]^{u(\tau_2)}_{u(\tau_1)}\Bigg)\,,
    \end{aligned}
\end{equation}
where, once $u(\tau)$ and the integration bounds are known, the radiative part of the on-shell action can be computed exactly and takes the form of a boundary term. Finally, if $u(\tau)$ is an odd function and the bounds are opposite $\tau_1=-\tau_2$ , as in the case of wormhole solutions one finds
\begin{equation}\label{eq:analyticradwormhole}
\begin{aligned}
    H(u)&=H_0\cosh(2u)\,, \qquad \pi_A(u) \equiv \frac{H_u}{2} = H_0 \sinh (2u) \, , \qquad  H_0 = 2 c_1 = 2 c_2 \, ,  \\
    \mathcal{S}^E_{\rm rad}(u) &= 24\, \text{Vol}(S^3)\,|H(u) \pi_A(u) | = 24\, \text{Vol}(S^3)\,|H(u)| \sqrt{H(u)^2-H_0^2}\, , 
\end{aligned}
\end{equation}
where we introduced an absolute value, to emphasize that the action is positive, being an integral over a positive definite Lagrangian density.

At this point we would like to understand the holographic and physical interpretation of these results for the gauge field background. We note that upon using the AdS asymptotics of the scale factor $a(\tau) \sim e^{H_{\rm AdS} |\tau|}$ , we can solve the equation of motion for the gauge fields perturbatively i.e. we set
\begin{equation}
H(\tau) = H^{(0)} e^{-\l H_{\rm AdS}|\tau|}+H^{(1)}e^{-(\l+1)H_{\rm AdS}|\tau|}+\mathcal{O}\left(e^{-(\l+2) H_{\rm AdS}|\tau|}\right) \, ,
\end{equation}
to find at order $e^{-\l H_{\rm AdS} |\tau|}$
\begin{equation}
    \l^2-\l=0~\quad \l_\pm=1,0\,.
\end{equation}
This was expected since the near boundary behavior of a vector field is 
\begin{equation}\label{eq:asymptoticexpansionvectorfield}
    A_{\mu}(\tau)=A^{-}_\mu\, e^{-(\Delta_--1)H_{\rm AdS}|\tau|}+...+A^{+}_{\mu}\,e^{-(\Delta_+-1)H_{\rm AdS}|\tau|}+...\,,
\end{equation}
where $A^{-}_\mu,~A^{+}_\mu$ correspond to the two possible modes at the boundary. For a massless gauge field the conformal dimensions are given by $\Delta_\pm=\frac{d}{2}\pm\left(\frac{d}{2}-1\right)\Rightarrow\Delta_+=2\,,\, \Delta_-=1$ where we specialize to $d=3$. This is in agreement with the values of $\l_{\pm}$ found just above since $\l_{\pm}=\Delta_{\pm}-1$.
This simple check clarifies that the leading asymptotic behavior of the electromagnetic potential is a radially independent constant corresponding to a boundary source using the standard Dirichlet BCs. With these conditions the $U(1)^3$ symmetry at the boundary is a global symmetry and the magnetic field approaches a source at the boundary (to leading order) since $B_I \sim H^{(0)}$, while the electric field only contains a VEV part (since $E_I \sim H'(\tau) \sim H^{(1)} $). Notice that very importantly, in three boundary dimensions both leading and subleading modes are normalizable, since $\Delta_- = 1 < d/2 = 3/2$ (and $d>2)$. This means that we can also interpret the leading behaviour as a VEV and the subleading as a source (alternate quantization/Neumann BCs). In this later case the $U(1)^3$ is a local gauge symmetry on the boundary. The transformation between these two options is affected by a boundary Legendre transform. This transformation is also one part/generator of the $SL(2,Z)$ group action that performs Electric/Magnetic (E/M) duality in the bulk. In other words E/M duality contains both the Legendre transform together with a Hodge duality operation that exchanges the E/M fields~\cite{Hawking:1995ap,Witten:2003ya}\footnote{In fact one is also free to add a Cherns-Simons action $S_{C.S.} \sim k \int A \wedge d A$ at the boundary and the T-operation of E/M duality ($SL(2,Z)$) shifts the C.S. level $k$ by one unit.}.
In order to perform the change from Dirichlet to Neumann BCs (alternate quantization), one needs to add the following boundary action (for a single boundary component\footnote{In the case of wormholes (two boundary components), provided the wormhole is $\mathbb{Z}_2$-symmetric at $\tau=0$ ($a$ even) and $H(\tau)$ is even, the contribution is simply doubled.})
\bea
S_{\partial \mathcal{M}}^{\rm U(1)^3} &=& - \int_{\partial \mathcal{M}} A_I \wedge \star F_I  = - \sum_{I=1}^3 \int_{\partial \mathcal{M}} {\rm d}^3 x \sqrt{h} n_\mu F^{\m \n}_I A_\n^I \nn \\ 
&=& - 12 \vol(S^3)  a(\tau) H'(\tau) H(\tau)\vert_{\partial \mathcal{M}_{\rm}}  
= - 24 \vol(S^3) \pi_A(u) H(u)\vert_{\partial \mathcal{M}_{\rm}} \, ,
\label{eq:bt_U1}
\eea
so that while the Dirichlet action obeys
\be
\delta S_D^{U(1)^3} = \sum_{I=1}^3 \int_{\partial \mathcal{M}} {\rm d}^3 x \sqrt{h} n_\mu F^{\m \n}_I \delta A_\n^I  = 24 \vol(S^3)  \pi_A(u) \delta H(u) \vert_{\partial \mathcal{M}}  \, ,
\ee
the Neumann action $S_N = S_D + S_{\partial \mathcal{M}}$ obeys
\be
\delta S_N^{U(1)^3} = - \sum_{I=1}^3 \int_{\partial \mathcal{M}} {\rm d}^3 x \sqrt{h} n_\mu \delta( F^{\m \n}_I )  A_\n^I = - 24 \vol(S^3)  H(u) \delta \pi_A(u) \vert_{\partial \mathcal{M}}  \, .
\ee
The complete on-shell action on the (two components) background solution then reads
\be 
\mathcal{S}^{E, N}_{\rm rad}(u) = - 24\, \text{Vol}(S^3)\, |\pi_A(u) H(u)| = - 24\, \text{Vol}(S^3)\, |\pi_A(u) \sqrt{\pi_A^2(u) + H_0^2}|  = - \mathcal{S}^{E, D}_{\rm rad}(u) \, ,
\ee
where the superscripts ``D" and ``N" stand for Dirichlet and Neumann BCs respectively.
Let us finally mention that, for wormholes, the electromagnetic energy density can be naturally expressed in terms of the electromagnetic field at the center of the wormhole $H_{0}$
\begin{equation}\label{eq:energydensitywormhole}
    \tilde{\rho}^E_{\rm rad} = -8\kappa H_0^2\,.
\end{equation}
This implies that holographically the electromagnetic energy density is an IR quantity and not a UV parameter of the model, so that backgrounds with different E/M density can be compared, as long as the gauge field has the same UV (source) asymptotics. These UV sources correspond to $H_{\rm UV} = H(u)\vert_{\partial \mathcal{M}}$ for Dirichlet BCs and to $- \pi_A \vert_{\partial \mathcal{M}}$ for Neumann BCs.

\subsection{Background subtraction/renormalization procedure in EFLRW metric}\label{sub:procedure_background}

For asymptotically AdS spacetimes the on-shell action generically diverges due to the infinite volume near the conformal boundary, with the precise divergence structure depending on which boundary modes (sources) are turned on. To compare gravitational saddles one must impose identical boundary conditions, i.e.\ the same set of sources in the dual CFT. Indeed, bulk solutions with different sources correspond to different deformations of the CFT and should not be compared within the same ensemble of solutions.

There are two equivalent ways to obtain finite saddle-to-saddle action differences:
\begin{itemize}
    \item \textit{Holographic renormalization}~\cite{Skenderis:2002wp}: add local covariant counterterms on a cut-off surface to render each on-shell action finite. The resulting renormalized action is scheme dependent, with the minimal subtraction scheme adding only the minimal terms required to cancel divergences.
    \item \textit{Background subtraction}: introduce a cut-off regulating each manifold and choose the regulator so that the induced boundary data (in particular the boundary metric, and any other sources held fixed in the ensemble) coincide. After removing the cut-off, the difference of regulated on-shell actions remains finite provided the boundary sources match.
\end{itemize}
Although the renormalized on-shell action in holographic renormalization depends on the choice of finite counterterms i.e. choice of ``scheme'', differences between saddles evaluated with the same boundary conditions are well defined and scheme independent. In~\cite{Marolf:2021kjc} the authors used holographic renormalization; here we instead implement background subtraction (but we still analyze the appropriate counterterms and renormalized action for the scalar field since they help clarify the physics and the phase space of solutions).

In the EFLRW ansatz \eqref{eq:ansatz}, background subtraction proceeds as follows. For each geometry $\Sigma_i$ we introduce a cut-off at Euclidean time $\tau_i$ and fix $\tau_i=\tau_i(R)$ by requiring that the induced metric on the cut-off surface $\partial\Sigma_i$ is the same, i.e.
\begin{equation}\label{eq:prescription}
    a(\tau_i)=R\,.
\end{equation}
We stress that $\tau_i(R)$ can differ between saddles. We then evaluate the regulated on-shell actions, expand them at large $R$, and isolate the divergent terms. These divergences are universal functionals of the induced boundary data and cancel in the difference of actions when the boundary sources are identical. The finite saddle-to-saddle on-shell action difference is obtained by taking $R\to\infty$.

We now summarize general features of the solutions. The model consists of Einstein gravity coupled to $U(1)^3$ radiation and a scalar field. The saddles studied fall into two classes: (i) solutions with constant scalar $\phi=0$, and (ii) solutions with a running scalar profile $\phi(\tau)$. Both arise as Euclidean solutions for the same scalar potential $V(\phi)$, which near $\phi=0$ takes the form
\begin{equation}\label{eq:potential}
    V(\phi)  \underset{{\phi\to0}}{\sim}   V_{\rm AdS}+\frac{1}{2}m^2\phi^2\,,
\end{equation}
with $V_{\rm AdS}<0$ and $m^2<0$. Thus $\phi=0$ is an extremum of the potential and always solves the scalar equation of motion (cf.\ the first line of \eqref{eq:eoms1}). The $\phi=0$ saddles may equivalently be viewed as solutions of Einstein gravity with a negative cosmological constant $\Lambda=V_{\rm AdS}$; these are precisely the configurations considered in~\cite{Marolf:2021kjc}.

Finally, the scalar mass corresponds to a relevant deformation with $\Delta_-<d/2=3/2$ (see Sec.~\ref{piecewisewineglass}). In this window one may choose alternative quantizations. Passing from Dirichlet to Neumann (or mixed) BCs is implemented by adding the appropriate boundary terms/counterterms (equivalently, by a Legendre transform of the generating functional). We discuss their explicit form in \eqref{piecewisewineglass}. These additional contributions vanish identically for the constant scalar saddles, as the scalar profile contains no non-trivial normalizable or non-normalizable mode; consequently, the choice between Dirichlet and Neumann BCs is immaterial in this case.

\section{A menagerie of (Euclidean) gravitational saddles}\label{sec:MenagerieSaddles}

\begin{figure}[t!]
    \centering
    \begin{overpic}[width=0.9\linewidth]{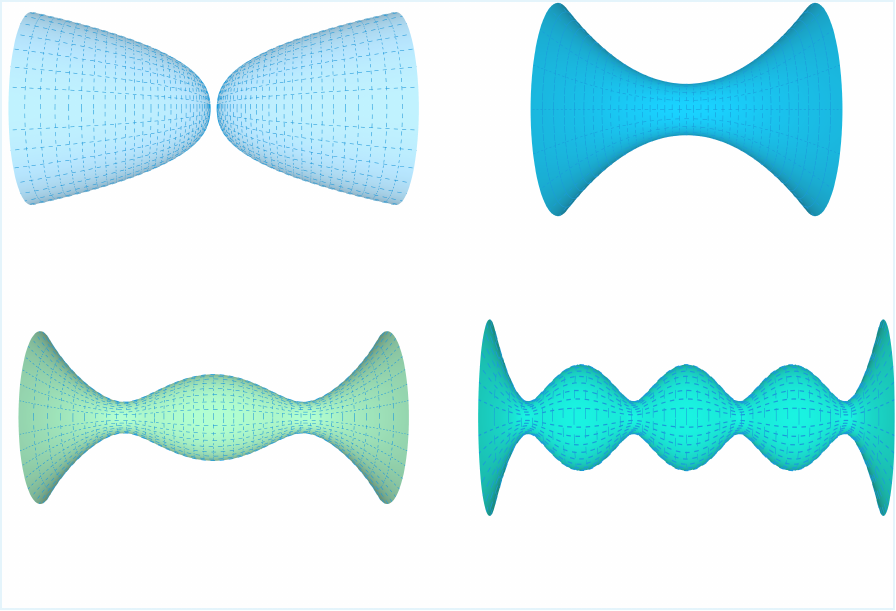}
\put(16,40){\textcolor{black}{\textbf{Disconnected}}}
\put(70,40){\textcolor{black}{\textbf{Connected}}}
\put(17,8){\textcolor{black}{\textbf{Wineglass}}}
\put(63,8){\textcolor{black}{\textbf{Oscillatory wineglass}}}
    \end{overpic}
    \caption{The different types of Euclidean AdS gravitational saddles that we analyze and compare in this work. The first two types lead to crunching cosmologies, while the last two to inflating/expanding cosmologies upon analytic continuation ($\tau = i t$).}
    \label{fig:wormholes}
\end{figure}
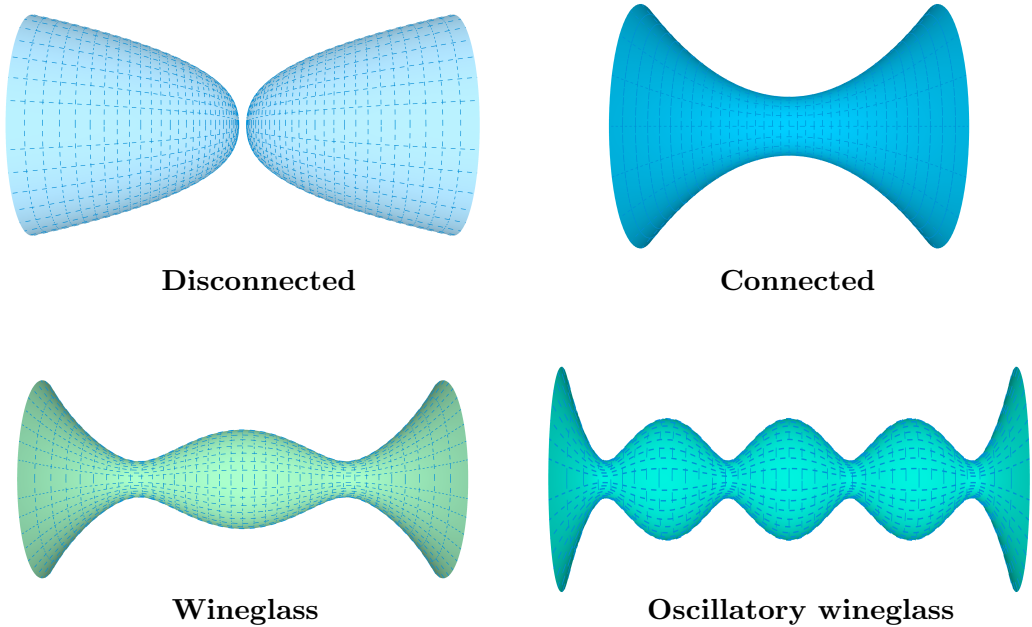

In what follows, we will compare four different types of geometries corresponding to solutions of the equations of motion with different physical properties, see fig.~\ref{fig:wormholes} for a depiction of them.
\begin{itemize}
    \item Disconnected (products of single boundary) geometries with a self dual radiation field, for which the stress-energy tensor is zero and with a constant inflaton. These are AdS background solutions at the negative valued local minima/maxima of the scalar/inflaton potential.
    
    \item Wormhole geometries supported by a negative Euclidean energy density for the radiation field $\rho^E_{\rm rad}$ , keeping the scalar/inflaton field constant as in the previous case.
    
    \item Wormhole geometries with radiation and a non trivial, running scalar/inflaton field. These backgrounds explore different regions of the scalar potential, and can lead to wineglass geometries if it acquires a positive value near the wormhole throat.

    \item More general oscillatory wormholes (see~\cite{Jonas:2023ipa} for some analogous asymptotically flat geometries), for which the scale factor oscillates in the throat region multiple times. Such geometries can be found in our class of models iff the scalar potential is also an oscillating function in the region where it is positive (we term this a quasi-oscillatory behaviour)\footnote{There is also a fine-tuned case, whereby the potential is a fixed positive cosmological constant. This case admits exactly oscillatory solutions with a potentially troublesome infinite number of oscillations, such as in the axion de-Sitter wormholes studied in~\cite{Halliwell:1989pu,Aguilar-Gutierrez:2023ril}. Lifting the flat direction of the potential precludes this possibility.}.

\end{itemize}

The first type of geometries does not contribute to cosmological transition amplitudes (only affects the overall normalization of probabilities, see our discussion in section~\ref{sec:wavefunction}) and concerns the probability that the system will remain in an AdS (non-cosmological) background. The second type of geometries leads to crunching cosmological solutions upon rotating to Lorentzian signature, see~\cite{Maldacena:2004rf,Betzios:2019rds,VanRaamsdonk:2021qgv,Betzios:2024oli} for discussions. The third type of backgrounds leads to expanding/inflating cosmologies~\cite{Betzios:2024oli,Betzios:2024zhf} and is the one relevant for phenomenological purposes. Expanding cosmologies can also be produced by the oscillatory wormholes, if the number of oscillations is odd, but these require (quasi)-periodic potentials with further non-trivial features as we shall prove.

Finally in the literature there are also appearances of single boundary ``centaur'' geometries~\cite{Anninos:2017hhn}, which can be thought of as a finely tuned case of the wineglass wormholes, when the second asymptotic region decouples and the manifold smoothly caps off. These configurations are not $\mathbb{Z}_2$ symmetric in the radial direction and in our setting they are expected to exist when the running scalar reaches a local minimum/maximum of the scalar potential with a positive value and stops there. We find that in our class of models, the presence of the flux that supports the minimum size of the throat, is in clash with a subsequent smooth closure of the Euclidean manifold. In other words such solutions can only be found if the stress-energy density becomes a time dependent function that while being non-trivial initially, it diminishes as the scalar field reaches the local maximum/minimum of the potential, leading to a pure dS solution in the deep interior of the geometry.

\subsection{Disconnected solutions}\label{sub:disconnected}

The simplest solutions to the complete set of EOMs, are found assuming that we are in a local stationary point of the inflaton potential $\partial_\phi V(\phi) = 0$ (without loss of generality we place this at $\phi = 0$, so we denote the potential there by $V_{\rm AdS}(\phi=0) < 0$) , and then $\phi = 0$ is a simple solution of its equation of motion. Then the scale factor is determined by the following simple equation using the variable $y(\tau) = a^2(\tau)$
\be\label{eq:scale_constpotential}
y'^2(\tau) - 4 y(\tau) + 4 \tilde{V}_{\rm AdS} y^2(\tau) - 4 \tilde{\rho}^E_{\rm rad} = 0 \, .
\ee
This can be solved via a direct integration of
\be
2 {\rm d} \tau = \pm \frac{{\rm d} y}{\sqrt{- \tilde{V}_{\rm AdS} y^2 + y + \tilde{\rho}^E_{\rm rad}}} \, .
\ee
Assuming our range of interest $\tilde{V}_{\rm AdS}\,,\, \tilde{\rho}^E_{\rm rad} < 0\,, \, y \geq 0$ , we observe that the scale factor should never vanish for this expression to be well defined (so we are in general describing a wormhole), unless $\tilde{\rho}^E_{\rm rad} = 0$. In this special case the stress-energy of the electric field cancels precisely the one from the magnetic field and the background carries a self-dual flux $F = \star F$. The single boundary (disconnected) geometry corresponds to an $AdS_4$ background
\be
a_D(\tau) = \frac{1}{\sqrt{- \tilde{V}_{\rm AdS}}} \sinh \sqrt{-\tilde{V}_{\rm AdS}} \tau \, , \qquad H_D(\tau)=H_{\rm UV}\,\tanh^2\left(\frac{\sqrt{-\tilde{V}_{\rm AdS}}\tau}{2}\right) \, .
\label{eq:aH_sol1}
\ee
For the solution of the gauge field, we fixed one integration constant, by demanding that it is regular at $\tau=0$ (leading to a finite E/M action). This amounts to taking $c_2=0\,,c_1=H_{\rm UV}$, so that $H_D(0) = 0$ and where the boundary source is identified with $H_{D}(\infty) = H_{\rm UV}$. The asymptotic expansion yields
\be
H_D(\tau) = H_{\rm UV} - 4 H_{\rm UV} e^{-\sqrt{-\tilde{V}_{\rm AdS}} \tau} \, + \, \mathcal{O} (e^{-2 \sqrt{-\tilde{V}_{\rm AdS}} \tau} ) \, , 
\ee
so that the source and VEV $\left(\pi_{A} = \frac{a(\tau) H_D'(\tau)|_{\tau = \infty}}{2}= H_{\rm UV}\right)$ are equal  (for a self dual configuration $E_I=B_I$ leads to $a(\tau) H_D'(\tau) = 2H_D(\tau)\,\, \forall \t$). If we perform alternate quantization, their roles simply get exchanged.

Evaluating the Euclidean action~\eqref{background} for the disconnected solution on-shell, one finds (see Appendix~\ref{sec:eucl_action in U(1)^3})
\begin{equation}
\begin{aligned}
    \mathcal{S}^{D}_E = -\frac{\text{Vol}(S^3)}{\kappa}\left[3\int_{0}^{\tau_{D}} {\rm d}\tau \,a_D^3\,\tilde{V}_{\rm AdS}\,+\partial_{\tau}a_D^3\Big|^{\tau_D}\right]\,+12\,\text{Vol}(S^3)\,H_{\rm UV}^2\,x_D^4\,,
\label{eq:act_disc_with_reg}
\end{aligned}
\end{equation}
with $x({\tau})=\tanh(\sqrt{- \tilde{V}_{\rm AdS}}{\t}/2)$. In this expression we have introduced the regulator $\tau_{D}$ to render the action finite and be able to perform subtractions between the on-shell actions of the different geometries. In particular, while the exact value of the on-shell action after performing Holographic Renormalization is scheme dependent, the differences of the on-shell actions between different backgrounds with the same asymptotic BCs are well defined and scheme independent. Note that the radiative part of the action remains finite so we can already remove the regulator and write
\begin{equation}
\begin{aligned}
    \mathcal{S}^{D}_E = -\frac{\text{Vol}(S^3)}{\kappa}\left[3\int_{0}^{\tau_{D}} {\rm d}\tau \,a_D^3\,\tilde{V}_{\rm AdS}\,+\partial_{\tau}a_D^3\Big|^{\tau_D}\right]\,+12\,\text{Vol}(S^3)\,H_{\rm UV}^2\,.
\label{eq:act_disc}
\end{aligned}
\end{equation}
Finally, if we adopt the alternative quantization with Neumann BCs, for the gauge field, the boundary term in~\eqref{eq:bt_U1} must be included in~\eqref{eq:act_disc}. For the disconnected solution this becomes
\begin{equation}
    S_{\partial \mathcal{M}}^{U(1)^3,\, {D}}  = - 24\,\text{Vol}(S^3) H_{\rm UV}^2\,.
\label{eq:neu_Disc}
\end{equation}

\subsection{Simple wormhole solutions with constant scalar}\label{sub:simpleconnected}

\begin{figure}[t!]
    \centering
    \includegraphics[width=0.5\linewidth]{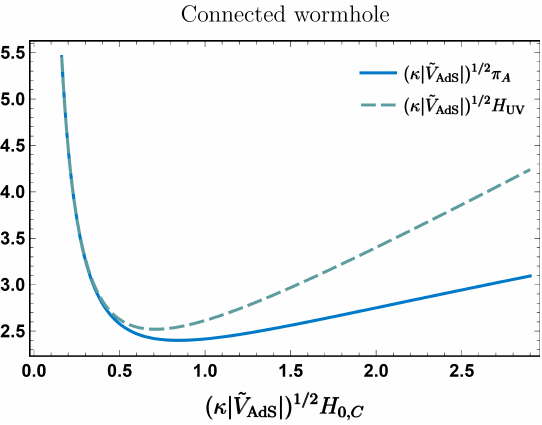}
    \caption{$H_{\rm UV}$ and $\pi_A$ as functions of $H_{0,C}$ for the connected wormhole, as given by Eq.~\eqref{eq:Huv_connected}. There is a minimum value of $H_{\rm UV}\, (\text{or}\,\pi_A)$, namely $H_{\rm UV}^{\rm min} \simeq 2.52/(\k|\tilde{V}_{\rm AdS}|)^{1/2}$ ($\pi_A^{\rm min} \simeq 2.40/(\k|\tilde{V}_{\rm AdS}|)^{1/2}$), at which the two wormholes merge into a single one. Below this value, no wormhole exists. This plot can also be extended to negative $H_{\rm UV} \,(\text{or}\,\pi_A)\,,\,H_{0,C}$ being an odd function with a discontinuity at zero.}
    \label{fig:placeholder}
\end{figure}

Let us now pass to the simple connected wormhole with a constant scalar but with $\tilde{\rho}^E_{\rm rad} < 0$, such that the flux carries non-trivial stress-energy. The solution to equation~\eqref{eq:scale_constpotential} can be written as
\be\label{eq:scale_connected}
y_C(\tau) = a_C^2(\tau) = \frac{1}{2 \tilde{V}_{\rm AdS}} - \frac{\sqrt{1 + 4 \tilde{\rho}^E_{\rm rad} \tilde{V}_{\rm AdS}}}{2 \tilde{V}_{\rm AdS}} \cosh \left(2 \sqrt{- \tilde{V}_{\rm AdS}} {\t} \right) \, .
\ee
This solution describes a simple two boundary wormhole, with a single local minimum in the throat located at $\tau= 0$. One can compute the conformal time $u_C(\tau)$ defined in \eqref{eq:equationH}. It yields (see Appendix~\ref{computationdetails} for further details)
\begin{equation}
    u_C(\tau)=\left(\frac{2}{\sqrt{1+4\tilde{\rho}^E_{\rm rad}\tilde{V}_{\rm AdS}}-1}\right)^{1/2}\left.\mathcal{F}\left[\arcsin\left(\tanh{\tilde{\tau}}\right)\right| m\right]\,,
\end{equation}
where $\mathcal{F}$ denotes the incomplete elliptic integral of the first kind (see Appendix~\ref{definitionElliptic}) and $\tilde{\tau} = \sqrt{- \tilde{V}_{\rm AdS}} \tau$.
From the conformal time, one can easily derive $H_C$ using \eqref{eq:equationH}. Selecting the $\mathbb{Z}_2$ even solution, $H_{C}(\tau)=H_{0,C}\cosh[2u_C(\tau)]$ one obtains
\begin{equation}\label{connectedgaugesln}
H_C(\tau)=H_{0,C}\, \cosh\left[2\left(\frac{2}{\sqrt{1+4\tilde{\rho}^E_{\rm rad}\tilde{V}_{\rm AdS}}-1}\right)^{1/2}\left.\mathcal{F}\left[\arcsin\left(\tanh{\tilde{\tau}}\right)\right| m\right]\right]\,,
\end{equation}
where $H_{0,C}$ denotes the value of the potential $H_C$ when the wormhole reaches its minimum size ($\tau = 0$). Note that $\tilde{\rho}^E_{\rm rad}$ , $m$ are implicit functions of $H_{0,C}$, with
\begin{equation}
    \tilde{\rho}^E_{\rm rad} = \frac{\kappa}{3} a^4 \rho^E_{\rm rad} = -8\kappa\,H_{0,C}^2 \, , \quad m=-\frac{\a+1}{\a-1}<0\, , \quad \a= \sqrt{1+4\tilde{\rho}^E_{\rm rad}\tilde{V}_{\rm AdS}}>1\,.
    \label{eq:rho_H0}
\end{equation}
Taking the Euclidean time to infinity, we obtain the following equalities relating the electromagnetic source $H_{\rm UV}$ and the momentum/BCs $\lim_{\tau\to\infty} a(\tau)\,H'(\tau)= 2\pi_A$ (with Dirichlet BCs) to $H_{0,C}$,
\begin{equation}
    H_{\rm UV}=H_{0,C}\, \cosh\left[2\left(\frac{2}{\sqrt{1-32\kappa H_{0,C}^2\tilde{V}_{\rm AdS}}-1}\right)^{1/2}\mathcal{K}[m(H_{0,C})]\right]\,,
\label{eq:Huv_connected}
\end{equation}
\begin{equation}
     \pi_A = H_{0,C}\, \sinh\left[2\left(\frac{2}{\sqrt{1-32\kappa H_{0,C}^2\tilde{V}_{\rm AdS}}-1}\right)^{1/2}\mathcal{K}[m(H_{0,C})]\right]\,,
\label{eq:VEV_connected}
\end{equation}
where $\mathcal{K}[m]$ denotes the complete elliptic integral of the first kind. This implicitly defines $H_{0,C}$ as a function of the source  $H_{\rm UV}$ (resp. $\pi_A$ in Neumann quantization picture). For a given value of the source, there are as many wormholes as the solutions of this implicit equation, see fig.~\ref{fig:placeholder}. In particular as was observed in~\cite{Marolf:2021kjc}, there always exists two branches of wormholes above a certain value of the source. The novelty here is that we discuss both Neumann or Dirichlet boundary conditions for the gauge field. As it is clear from~\eqref{eq:scale_connected}, small (resp large) wormholes are associated with small (resp large) values of $\tilde{\rho}^E_{\rm rad}$ i,e $H_{0,C}$. In both quantization schemes, we can derive the asymptotic relation of the source of the electromagnetic field ($H_{\rm UV}$ for Dirichlet BCs and $\pi_A$ for Neumann BCs) as a function of $H_{0,C}$. We find:
\begin{itemize}
    \item Dirichlet BCs
    \begin{subequations}
    \begin{equation}\label{eq:asymptoticconnected}
    \begin{aligned}
        H_{\rm UV}&\underset{H_{0,C}\to 0^+}{=} \frac{1}{\kappa |\tilde{V}_{\rm AdS}|H_{0,C}}+\mathcal{O}\left(H_{0,C}\log(H_{0,C})\right) \qquad \text{small branch}\,,\\
        H_{\rm UV}&\underset{H_{0,C}\to \infty}{=} H_{0,C}+\frac{\mathcal{K}(-1)^2}{\sqrt{2\kappa |\tilde{V}_{\rm AdS}|}}+\mathcal{O}\left(\frac{1}{H_{0,C}}\right) \qquad \,\,\,\,\, \text{large branch}\,,
    \end{aligned}
    \end{equation}
    \item Neumann BCs
    \begin{equation}\label{eq:asymptoticconnectedNeumann}
    \begin{aligned}
        \pi_A&\underset{H_{0,C}\to 0^+}{=} \frac{1}{\kappa |\tilde{V}_{\rm AdS}|H_{0,C}}+\mathcal{O}\left(H_{0,C}\log(H_{0,C})\right) \qquad \text{small branch}\,,\\
        \pi_{A}&\underset{H_{0,C}\to \infty}{=} \frac{2^{1/4}\mathcal{K}(-1)}{\left(\kappa |\tilde{V}_{\rm AdS}|\right)^{1/4}}\sqrt{H_{0,C}}+\mathcal{O}\left(H_{0,C}^{-1/2}\right) \qquad \,\,\,\, \text{large branch}\,,
    \end{aligned}
    \end{equation}
    \end{subequations}
\end{itemize}

Evaluating the on-shell Euclidean action in the Dirichlet quantization picture for the connected wormhole solution, one finds (see Appendix~\ref{sec:eucl_action in U(1)^3})
\begin{equation}
\label{eq:simplewormholeaction}
    \mathcal{S}^{C}_E= -\frac{\text{Vol}(S^3)}{\kappa}\left[3\int_{-\tau_{C}}^{\tau_{C}} {\rm d}\tau a_C^3\tilde{V}_{\rm AdS}\,+\partial_{\tau}a_C^3\Big|_{-\tau_{C}}^{\tau_{C}}\right]+24\text{Vol}(S^3)\,|H_{\rm UV}| \sqrt{H_{\rm UV}^2-H_{0,C}^2}\,,
\end{equation}
where we have again introduced a regulator $\tau_C$ to render the gravitational part of the action finite. The radiative part remains finite, thus we have already taken the cut-off to infinity using \eqref{eq:analyticradwormhole}. Notice that the action should in principle be written as a function of $H_{\rm UV}$ and not $H_{0,C}$. This can be performed implicitly, or numerically inverting the relation \eqref{eq:Huv_connected}. Finally, if we adopt the alternative quantization with Neumann BCs, for the gauge field, the boundary term in~\eqref{eq:bt_U1} must be included in~\eqref{eq:simplewormholeaction}. For the connected wormhole this becomes
\begin{equation}
    S_{\partial \mathcal{M}}^{U(1)^3,\, {C}}  = - 48 \text{Vol}(S^3)  |H_{\rm UV}| \sqrt{H_{\rm UV}^2 - H_{0,C}^2} 
     = -48 \text{Vol}(S^3)|\pi_A|\sqrt{\pi_A^2+H_{0,C}^2}\,.
\label{eq:neu_Conne}
\end{equation}
and the E/M part of the on-shell action simply switches sign between Dirichlet and Neumann BCs. Note however that in Neumann boundary conditions, the action should be viewed as a function of the source $\pi_A$.

\subsection{Wineglass wormhole solutions with running scalar}\label{wineglass}

In a more general setting, the scalar/inflaton potential being a non-trivial function, leads to (Holographic) RG flows. The scalar becomes then a non-trivial function of the radial coordinate $\phi(\tau)$ , reaching negatively valued local minima/maxima of the potential in the asymptotic AdS boundaries. In the present setting, we shall be interested in flows for which the background geometry resembles a ``wineglass wormhole''\footnote{This can also be thought of as a Euclidean version of a ``bag of gold'' space (see~\cite{Fu:2019oyc} and references within). While in the Lorentzian case the ``bag of gold'' is a codimension one surface, here the complete Euclidean spacetime has the form of a ``bag of gold'', see fig.~\ref{fig:wormholes}.}. On such backgrounds the scalar explores a positive region of the potential, near the middle of the wormhole throat, before returning back to an AdS negative local maximum. In general it is difficult to find a completely analytic solution of the Euclidean saddle point EOM. Here we shall resort to a smooth piecewise (continuous and thrice differentiable) ansatz, to overcome this obstacle and yield the solutions of interest.

\subsubsection{Piecewise wineglass wormhole solutions}\label{piecewisewineglass}

Since it is very hard to analytically solve the EOMs for a given potential that supports wineglass wormhole solutions, let alone to do this for a class of potentials with the wanted characteristics, we shall proceed in a reverse manner in finding the solutions of interest.

We first propose an analytic $\mathbb{Z}_2$ symmetric ansatz for the scale factor comprised out of three regions: (\textbf{I}) for $-\infty<\t<\tau_{\rm min}$ and (\textbf{II}) for $\tau_{\rm min} \leq \t \leq  - \tau_{\rm min}$ and $(\textbf{I}')$ for $-\tau_{\rm min} < \tau < \infty$ where the scale factor is similar to that in region (\textbf{I})
\begin{equation}
 a^2(\tau) = 
    \begin{cases}
   \displaystyle   \frac{1}{4 H_{\rm AdS}^2}\left(A_{\rm AdS}\cosh{(2 H_{\rm AdS}(\tau-\tau_{\rm min}))}+ \frac{B_{\rm AdS}}{\cosh{(2 H_{\rm AdS}(\tau-\tau_{\rm min}))}}+ c_{\rm AdS}\right)\,, &  (\textbf{I})
\\[0.3cm]
   \displaystyle     \frac{1}{ 4H_{\rm dS}^2} \left( A_{\rm dS} \cos{(2 H_{\rm dS}\tau)}+ c_{\rm dS}\right)\,, & (\textbf{II}) 
\\[0.3cm]
  \displaystyle   \frac{1}{4 H_{\rm AdS}^2}\left(A_{\rm AdS}\cosh{(2 H_{\rm AdS}(\tau+\tau_{\rm min}))}+ \frac{B_{\rm AdS}}{\cosh{(2 H_{\rm AdS}(\tau+\tau_{\rm min}))}}+ c_{\rm AdS}\right)\,, &  (\textbf{I}')
    \end{cases}
\label{eq:background_ans}
\end{equation}
with $A_{\rm (A)dS}, H_{\rm (A)dS}, c_{\rm dS}>0$. We also have $\t_{\rm min} = -\pi/2H_{\rm dS}$, when the scale factor reaches its minimum size. The parameters $ H_{\rm (A)dS}$ have dimensions of mass, while all other quantities are dimensionless.
Matching the geometries (up to $\mathcal{O}(a, a', a'',a''')$) we obtain the conditions
\begin{equation}
\frac{A_{\rm AdS}+B_{\rm AdS}+c_{\rm AdS}}{H_{\rm AdS}^2} = \frac{c_{\rm dS}-A_{\rm dS}}{H_{\rm dS}^2}\, \qquad \text{and} \qquad A_{\rm AdS}= B_{\rm AdS}+A_{\rm dS}\,.
\label{eq:cond_wine}
\end{equation}
Note also that $a_{\rm max}^2 = (A_{\rm dS}+c_{\rm dS})/4 H_{\rm dS}^2 >a_{\rm min}^2 = (c_{\rm dS}-A_{\rm dS})/4 H_{\rm dS}^2$, and $a_{\rm min}'' =A_{\rm dS}/(2a_{\rm min})$ and $a_{\rm max}'' = - A_{\rm dS}/(2a_{\rm max})$ , while the requirement $c_{\rm dS}>A_{\rm dS}$ ensures that $a^2_{\rm min}$ is positive. Assuming now that this background solves the eqs.~\eqref{eq:equAp} and~\eqref{eq:diffA}, we can extract the velocity of the field and the potential as functions of the Euclidean time $\tau$\footnote{We can also solve implicitly for the potential as a function of $\phi$ through $V(\phi(\tau))$, the resulting expression can only be studied numerically. Nevertheless it is possible to plot the potential $V(\phi)$ for any value of the parameters.}. While analytic expressions for the velocity and potential in both regions are given in Appendix~\ref{appendix:Useful formulas}, here we focus on extracting useful information about the parameter space by studying their behavior at $\t=0,\, \tau_{\rm min}$ and $-\infty$.
\begin{figure}[t!]
\begin{center}
\includegraphics[width=0.49\textwidth]{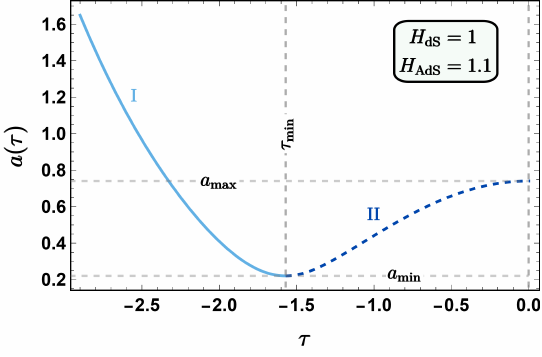}\\
\includegraphics[width=0.49\textwidth]{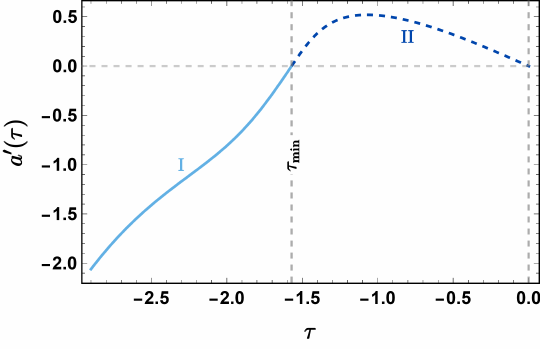}\hspace{0.15cm}
\includegraphics[width=0.49\textwidth]{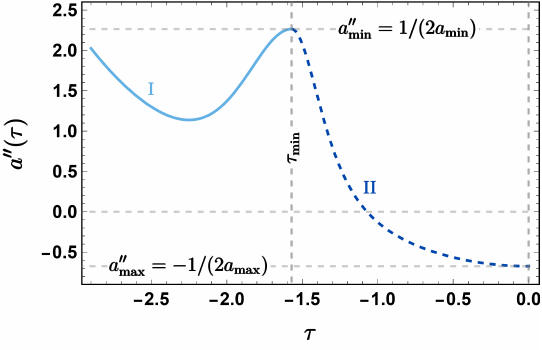} 
\end{center}
\caption{The scale factor (upper) and its first (lower left) and second (lower right) derivatives as functions of the Euclidean time $\t$ for the EAdS-like (blue), and EdS-like (blue dashed) regions described in eq.~\eqref{eq:background_ans}, all in $\k=1$ units. We have made the choice of parameters  $H_{\rm AdS} =1.1$, $H_{\rm dS} = 1$ and $A_{\rm dS}=1$.  The parameter $A_{\rm AdS}$ is in general freely tunable; however, we plot here the case that $c_{\rm AdS}=-2 \sqrt{A_{\rm AdS}(A_{\rm AdS}-A_{\rm dS})} =-2$, that corresponds to a configuration with a vanishing source for the scalar field. The resulting background inflates upon rotating $\tau = i t$ and evolving in Lorentzian signature after $t = 0$.}
\label{fig:scale_fac}
\end{figure}

We can set the velocity to zero ($\tilde{\phi}'=0$) at $\tau = 0$, by identifying
\begin{equation}
\tilde{\rho}^E_{\rm rad}  = - \frac{({2-A_{\rm dS}})(c_{\rm dS}+A_{\rm dS})}{16H_{\rm dS}^2} <0 \quad \Rightarrow \quad {A_{\rm dS}<2}\,,
\label{eq:rhorad_worm}
\end{equation}
eliminating one more parameter. The potential reads
\begin{equation}
\tilde{V}_0\equiv\tilde{V}(\tau=0) = \frac{(2+A_{\rm dS})H_{\rm dS}^2}{(c_{\rm dS}+A_{\rm dS})} > 0 \,,
\label{eq:Vzero}
\end{equation}
which is positive by construction. 
\begin{figure}[t!]
\begin{center}
\includegraphics[width=0.5\textwidth]{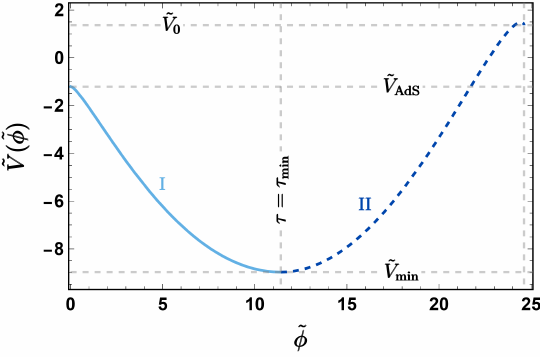}\\[0.2cm]
\includegraphics[width=\textwidth]{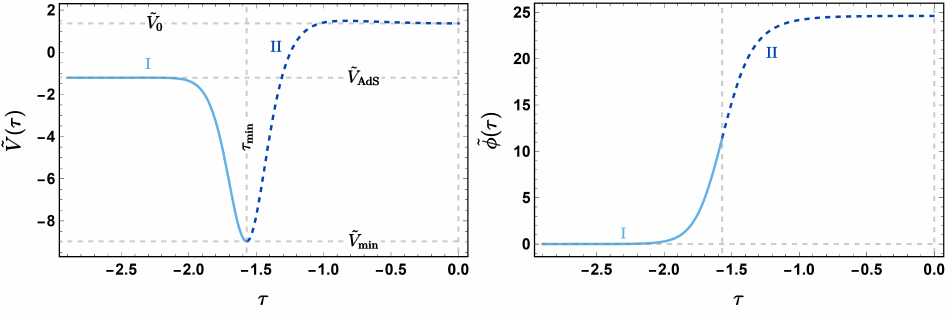}
\end{center}
\caption{$\tilde{V}$ vs. $\tilde{\phi}$ (upper), $\tilde{V}$ vs. $\tau$ (lower left) and $\tilde{\phi}$ vs. $\tau$ (lower right), all in $\k=1$ units. The parameters used are the same with those in fig.~\ref{fig:scale_fac}. The chosen field value at the EAdS boundary is $\tilde{\phi}(\tau\rightarrow-\infty)=0$.}
\label{fig:potential}
\end{figure}
At $\tau_{\rm min}$ we find that
\begin{align}
\tilde{\phi}'^2(\t = \tau_{\rm min}) &=  \frac{8A_{\rm dS}H_{\rm dS}^2}{3}\frac{2-c_{\rm dS}}{(c_{\rm dS}-A_{\rm dS})^2}\quad  \Rightarrow \quad
c_{\rm dS}<2\,,
\\
\tilde{V}_{\rm min}\equiv\tilde{V}(\t = \tau_{\rm min}) &= \frac{\left(3A_{\rm dS}^2+6c_{\rm dS}-A_{\rm dS}(c_{\rm dS}+10) \right)H_{\rm dS}^2}{3(c_{\rm dS}-A_{\rm dS})^2}\,,
\label{eq:Vmin_WG}
\end{align}
which indicates that the parameters $A_{\rm dS}, \, c_{\rm dS}$ control the sign of the potential at the minimum of the scale factor and at $\tau = 0$. 
Finally, the asymptotic behaviours of the scalar field and the potential at the UV boundary are 
\begin{align}
\label{eq:vel_inf}
\tilde{\phi}'^2(\t \rightarrow \infty)  \simeq &  -\frac{8H_{\rm AdS}^2(c_{\rm AdS}+2)}{3A_{\rm AdS}} e^{ -2 \tilde\tau} + \frac{16H_{\rm AdS}^2}{3A_{\rm AdS}^2} \big[A_{\rm AdS}(4A_{\rm dS}-5A_{\rm AdS}) \nonumber
\\
& +2c_{\rm AdS}+2c_{\rm AdS}^2-16H_{\rm AdS}^2 \tilde{\rho}^E_{\rm rad}\big] e^{-4 \tilde\tau} +\mathcal{O}(e^{-6 \tilde \tau})\,,
\\[0.2cm]
\tilde{V}(\t \rightarrow \infty) \simeq & \tilde{V}_{\rm AdS}+\frac{8H_{\rm AdS}^2(c_{\rm AdS}+2)}{3A_{\rm AdS}} e^{-2\tilde{\tau}} - \frac{4H_{\rm AdS}^2}{3A_{\rm AdS}^2}\big[ A_{\rm AdS}(4A_{\rm dS}-5A_{\rm AdS}) \nonumber
\\
&+5c_{\rm AdS}^2 + 8c_{\rm AdS} -16H_{\rm AdS}^2 \tilde{\rho}^E_{\rm rad} \big] e^{-4\tilde{\tau}} +\mathcal{O}\left(e^{-6\tilde{\tau}}\right)\,   ,
\\[0.2cm]
\tilde{V}_{\rm AdS}\equiv &   -H_{\rm AdS}^2 \, , \qquad \tilde{\tau} = H_{\rm AdS} \tau \,.
\label{eq:Vads_def}
\end{align}
We observe that $c_{\rm AdS} \leq -2$ , so that the velocity remains a real function\footnote{When the equality $c_{\rm AdS} = -2$ holds, an additional condition on the parameters must be imposed to guarantee that the next-to-leading-order term in~\eqref{eq:vel_inf} is positive. In the special case $c_{\rm AdS} = -2\sqrt{A_{\rm AdS}(A_{\rm AdS}-A_{\rm dS})}=-2$ , this condition becomes $H_{\rm AdS}^2/H_{\rm dS}^2 > (10+A_{\rm dS}^2 -4\sqrt{4+A_{\rm dS}^2}+A_{\rm dS}(3\sqrt{4+A_{\rm dS}^2}-4))/(4A_{\rm dS}(2-A_{\rm dS})) $.}. This also means that at the AdS boundary/ies, the potential reaches a negative local maximum.

Demanding a matching of the piecewise solutions and that the velocity is a real function, the total set of conditions reduces to
\begin{align}\label{setofconditions}
\frac{2A_{\rm AdS} + c_{\rm AdS} - A_{\rm dS}}{H_{\rm AdS}^2} &= \frac{c_{\rm dS} - A_{\rm dS}}{H_{\rm dS}^2}\,, \qquad B_{\rm AdS} =A_{\rm AdS}-A_{\rm dS}\,, \\[0.2cm]
0< A_{\rm dS}< c_{\rm dS} &< 2\,, \qquad  \qquad c_{\rm AdS} \leq -2\,, \qquad A_{\rm (A)dS}, B_{\rm AdS}, H_{\rm (A)dS}>0\,.
\end{align}
In figs.~\ref{fig:scale_fac} we depict the scale factor (top panel) and its first (bottom left) and second derivative (bottom right) and in figs.~\ref{fig:potential} the potential as a function of $\phi$ (top panel) and $\t$ (bottom left) and the scalar field value as a function of $\tau$ (bottom right), for $\t\leq 0$. The behaviour for $\t>0$ is identical by symmetry, since all functions involved are even in $\t$. We observe that both the potential and the running of the scalar are consistent with the criteria laid out in~\cite{Betzios:2024oli,Betzios:2024zhf} for the existence of Euclidean wineglass wormhole solutions and that all the analytic functions are smoothly joined in the interpolating region, up to third order in derivatives. The parameter choice used in figs.~\ref{fig:scale_fac} and~\ref{fig:potential} is $H_{\rm dS}=1$, $H_{\rm AdS}=1.1$, $A_{\rm dS}=1$ and $c_{\rm AdS} = -2\sqrt{A_{\rm AdS}(A_{\rm AdS}-A_{\rm dS})}=-2$ , which corresponds to the values of the physical parameters $\k \tilde{V}_0 \simeq 1.37$, $\k \tilde{V}_{\rm AdS} = -1.21$, and the derived parameter $\k \tilde{V}_{min} \simeq -8.97$.

Let us now proceed in discussing the holographic properties of these wineglass wormhole backgrounds and the physical role of the various parameters. In particular, assuming the matching conditions, we are left with $5$ independent parameters out of the $7$ parameters of the ansatz~\eqref{eq:background_ans}. Changing the value of $A_{\rm AdS}$  corresponds to rescaling the asymptotic value of the metric at infinity (i.e. metric source) and $H_{\rm AdS}$ is related to $\tilde{V}_{\rm AdS}$ from eq.~\eqref{eq:Vads_def}, therefore corresponding to a UV parameter of the model/Lagrangian. We are then left with three parameters. To proceed further, it is  important to understand the properties of the running scalar from a holographic perspective. In order to determine the conformal dimension $\Delta$ of the dual operator $\mathcal{O}_\Delta$ that drives the running/RG flow, we expand the potential to second order (we fix the integration constant of the integral of $\tilde \phi'$ setting
the value of $\tilde{\phi}(- \infty) = 0$)
\begin{equation}
\tilde{V}(\tilde{\phi})_{\tau \rightarrow - \infty} \simeq \tilde{V}_{\rm AdS} + \frac{\tilde{V}_{\tilde \phi \tilde \phi}(\tau\rightarrow-\infty)}{2} \tilde{\phi}^2 \, + \, ... \, ,
\end{equation}
where $m^2=\tilde{V}_{\tilde \phi \tilde \phi} = (\tilde{V}'' - \tilde{V}'\tilde{\phi}''/\tilde{\phi}')/\tilde{\phi}'^2 \overset{\t\rightarrow\infty}{=} -2 H_{\rm AdS}^2$. The conformal dimensions are then given by (in our conventions $L^2_{\rm AdS} = -1/\tilde{V}_{\rm AdS}$)
\be
\Delta_\pm = \frac{3}{2} \pm \sqrt{\frac{9}{4} + m^2 L_{AdS}^2 } \, , \qquad m^2 L_{AdS}^2 = \frac{\tilde{V}_{\tilde \phi \tilde \phi}(\t\rightarrow-\infty)}{-\tilde{V}_{\rm AdS}} = -2\,.
\ee
We therefore find $\Delta_+ = 2$ and $\Delta_- = 1$. In particular our deformation is an allowed relevant deformation, since
\be
- \frac{9}{4} \leq  m^2 L_{\rm AdS}^2 < 0 \, ,
\ee
where the lower bound is the BF-bound~\cite{Breitenlohner:1982bm,Breitenlohner:1982jf}. Moreover, since $\Delta_- = 1 < d/2 = 3/2$ , we are in a situation where we can choose multiple different quantizations for the scalar field, since both modes are normalizable. In particular if we expand it around infinity we find that
\be
\tilde{\phi}(\tau) \sim  \tilde{\phi}_+^{(0)} e^{ -\Delta_+ H_{\rm AdS}|\tau|} + \tilde{\phi}_+^{(1)} e^{ -(\Delta_+ + 1) H_{\rm AdS}|\tau|} + ... + \tilde{\phi}_-^{(0)} e^{- \Delta_- H_{\rm AdS} |\tau|} + \tilde{\phi}_-^{(1)} e^{ -(\Delta_- + 1) H_{\rm AdS}|\tau|} +...\,.  
\ee
Two typical choices correspond to setting the coefficient of the $\Delta_-$ term to be the source and that of $\Delta_+$ to be the VEV (Dirichlet BCs), and the other is the case in which $\tilde \phi_-$ corresponds to the VEV and $\tilde \phi_+$ to the source (Neumann BCs)\footnote{In general one can consider mixed BCs leading to generic multi-trace deformations of the boundary CFT.}. In addition, we observe that we are currently in a resonant setup, in the sense that we also have $\Delta_+ - \Delta_- =1$ , so that the first subleading term in the expansion of $\tilde \phi_-$ scales the same as the leading term in the expansion of $\tilde \phi_+$ and so forth\footnote{Nevertheless, no logarithmic term (associated with a conformal anomaly) appears in the expansion, since the conformal dimension satisfies $\Delta_+ \notin \frac{d}{2}+\mathbb{N}$ (see the criterion in \cite{Petkou_1999}).}.

In order to properly define quantization, we have to describe the symplectic structure of the solution space at infinity and perform appropriate holographic renormalization. In particular we define the (bare) momentum conjugate to 
$\tilde \phi$ and its asymptotic expansion

\be\label{bare}
\tilde \Pi_{\rm bare} = \sqrt{h} n^\mu \partial_\mu \tilde \phi = \sqrt{h} \tilde \phi' =  \sqrt{h} \left( \tilde \pi_{\Delta_-} e^{- \Delta_- H_{\rm AdS} |\tau|} + ... + \tilde \pi_{\Delta_+} e^{- \Delta_+ H_{\rm AdS} |\tau|} + ... \right)  \, .
\ee
We then write down the scalar counterterm action (for a single boundary component)\footnote{This counterterm action is the same for all different choices of quantization. In general it is derived integrating the relation $\delta S_{c.t.}(\tilde \phi) =  \int_{\partial \mathcal{M}} d^3 x \sqrt{h} \sum_{n < \Delta_+} \tilde \pi_{n} \delta \tilde \phi $ , where $\tilde \pi_n$ are the various terms in the expansion of the momentum.}, see~\cite{Papadimitriou:2007sj} and references within
\be
S_{ct.} =   \frac{3}{2 \kappa} \int_{\partial \mathcal{M}} {\rm d}^3 x \sqrt{h} \tilde \phi \left(  \Delta_- H_{\rm AdS} \tilde \phi + \frac{\nabla^2_h \tilde \phi } {H_{\rm AdS}(1 - 2 \Delta_-)} + \cdots \right) \, .
\label{eq:scalar_ct}
\ee
In our specific case where $\Delta_+ = 2 ,\, \Delta_- = 1 < 3/2$, only the first term in this action is needed for renormalization~\cite{Papadimitriou:2007sj}. We then compute the counterterm contribution to the momentum from \eqref{eq:scalar_ct} to be $\tilde \Pi_{c.t.} = \sqrt{h} \Delta_-  H_{\rm AdS} \tilde \phi $. Subsequently, the renormalized momentum is given by the sum of the bare momentum \eqref{bare} and the counterterm $\tilde \Pi_{c.t.}$ i.e.
\be
\tilde \Pi_{\rm ren.} = \sqrt{h} (n^\mu \partial_\mu \tilde \phi +  \Delta_-  H_{\rm AdS} \tilde \phi) = \sqrt{h}  (\tilde  \phi' + H_{\rm AdS} \tilde  \phi)  \, .
\ee
This renormalised momentum has a leading term in the asymptotic expansion $\tilde \Pi_{ren. } \sim \sqrt{h} \tilde \pi_{\Delta_+}^{\rm ren.} e^{- \Delta_+ |\tau|}$. One can then show that the variation of the renormalized on-shell action produces the term 
\be\label{eq:renormalizedvariationalproblem}
\delta S_{\rm ren.}^{\rm on-shell} = \delta (S_{\rm bare} + S_{c.t.}) \vert_{\rm on-shell} = \frac{3}{\kappa} \int_{\partial \mathcal{M}} {\rm d}^3 x  \,\tilde \Pi_{\rm ren} \delta \tilde \phi \,=\frac{3}{\kappa} \int_{\partial \mathcal{M}} {\rm d}^3 x  \sqrt{h}\,\tilde \pi_{\Delta_+}^{\rm ren.} \delta \tilde \phi_- \, ,
\ee
which is the renormalized version of the classical variational problem 
\be\label{eq:non-renormalized-version}
\delta S(\tilde \phi) = \frac{3}{\kappa}  \int_{\partial \mathcal{M}} {\rm d}^3 x  \, \tilde \Pi_{\rm bare} \, \delta \tilde \phi  \, .
\ee
This means that the quantities $\tilde \pi_{\Delta_+}^{\rm ren.}$ and $\tilde \phi_-$ correspond to the VEV and the source of the renormalised (Dirichlet) variational problem (they correspond to the dual symplectic space variables). From the asymptotic expansion of the scalar field, we observe that the source $\tilde \phi_- \sim c_{\rm AdS}+2$. Therefore with Dirichlet BCs, the scalar source is directly controlled by $c_{\rm AdS}$ and vanishes when $c_{\rm AdS} = -2$. It is also possible to consider Neumann BCs, by adding a boundary term\footnote{As discussed in Appendix~\ref{appendix:bnry_scalar}, this boundary term gives no contribution in either the scalar source–free case ($c_{\rm AdS}=-2$) or the non-zero scalar source case ($c_{\rm AdS}\neq -2$).}

\be\label{eq:boundarytermscalar}
S_{\partial \mathcal{M}}^{\rm bare} = - \frac{3}{\kappa} \int_{\partial \mathcal{M}} {\rm d}^3 x \sqrt{h} \tilde \phi  \vec{n} \cdot \partial \tilde \phi   \quad \rightarrow \quad S_{\partial \mathcal{M}}^{\rm ren.} = - \frac{3}{\kappa} \int_{\partial \mathcal{M}} {\rm d}^3 x   \tilde \phi  \tilde{\Pi}_{ren.} \, ,
\ee
and using the renormalised Neumann action $S^{\rm ren.}_N = S_D^{\rm ren.} + S_{\partial \mathcal{M}}^{\rm ren.}$ , where $S_D^{\rm ren.}$ stands for the renormalized Dirichlet action.
In this case/quantisation the source corresponds to $- \tilde \pi_{\Delta_+}^{\rm ren.}$ and the VEV to $\tilde \phi_-$.

We therefore conclude that from the five independent parameters within our ansatz, three are related to the three metric/scalar/E/M sources and the other two determine the set/class of potentials $\tilde{V}(\tilde{\phi})$ we can consider. In particular the potential is completely determined, given its values at $\tilde{\phi}(\tau \rightarrow - \infty),$ and $\tilde{\phi}(\tau = 0)$, fixing $H_{\rm AdS}$ and a combination of $H_{\rm dS}, c_{\rm dS}$ and $A_{\rm dS}$. This means that we can describe within our ansatz a two parameter family of scalar potentials/models that admit wineglass wormhole solutions as long as the conditions in Eq.~\eqref{setofconditions} are satisfied. 

Let us finally mention that in the special case that $c_{\rm AdS} = - 2 \sqrt{A_{\rm AdS}(A_{\rm AdS}-A_{\rm dS})}$ when the metric source gets related to the scalar source (or BCs depending on Dirichlet vs Neumann quantisation of the scalar), all the physical properties of the corresponding wineglass wormhole backgrounds, such as their on-shell action, can be determined analytically. In the rest, when comparing the various on-shell actions, we shall focus in this particular special case.

\subsubsection{The wineglass E/M gauge field background}
In this section, we compute the wineglass electromagnetic (E/M) gauge field with $c_{\rm AdS}=-2 \sqrt{A_{\rm AdS}(A_{\rm AdS}-A_{\rm dS})}$ , for which the on-shell action can be evaluated in a closed analytic form. We recall that, when Dirichlet BCs are assumed, the scalar source is turned off for $c_{\rm AdS}=-2$ , while for $c_{\rm AdS}\neq-2$ the VEV is set to zero. The opposite behavior occurs when Neumann BCs are imposed.

Defining $\bar{\tau}^{\textbf{I}}=2H_{\rm AdS}(\tau-\tau_{\rm min})\,,\,\bar{\tau}^{\textbf{II}}=2H_{\rm dS}\,\tau \,, \, \bar{\tau}^{\textbf{I}'}=2H_{\rm AdS}(\tau+\tau_{\rm min})$, the scale factor takes the form
\begin{equation}
 a_{WG}^2(\tau) = 
    \begin{cases}
   \displaystyle   \frac{1}{4 H_{\rm AdS}^2\cosh(\bar{\tau}^{\textbf{I}})}\left(\sqrt{A_{\rm AdS}}\cosh{(\bar{\tau}^{\textbf{I}})}+\frac{c_{\rm AdS}}{2\sqrt{A_{\rm AdS}}}\right)^2\,, &  (\textbf{I})
\\[0.3cm]
   \displaystyle     \frac{1}{4 H_{\rm dS}^2} \left(A_{\rm dS} \cos{(\bar{\tau}^{\textbf{II}})}+ c_{\rm dS}\right)\,, & (\textbf{II}) 
\\[0.3cm]
  \displaystyle   \frac{1}{4 H_{\rm AdS}^2\cosh{(\bar{\tau}^{\textbf{I}'})}}\left(\sqrt{A_{\rm AdS}}\cosh{(\bar{\tau}^{\textbf{I}'})}+\frac{c_{\rm AdS}}{2\sqrt{A_{\rm AdS}}}\right)^2\,. &  (\textbf{I}')
    \end{cases}
\label{eq:background_ans_simple}
\end{equation}
Since the scale factor is an even function of $\tau$ , the solution for the E/M gauge field which is manifestly $\mathbb{Z}_2$ symmetric reads
\begin{equation}\label{potential in all region}
    H_{WG}(\tau)=H_{0,WG}\,\cosh\left[2\int_0^{\tau}\frac{{\rm d}\tau'}{a_{WG}(\tau')}\right]\, , \qquad \forall\tau \,.
\end{equation}
In Appendix~\ref{computationdetails}, we compute the electromagnetic potential by splitting the integral in the different regions. One finds
\begin{equation}
 \frac{H_{WG}(\tau)}{H_{0,WG}} = 
    \begin{cases}
    \begin{aligned}
   \displaystyle   &\cosh\bigg(\frac{4}{\sqrt{c_{\rm dS}+A_{\rm dS}}}\mathcal{K}\left[\frac{2 A_{\rm dS}}{c_{\rm dS}+A_{\rm dS}}\right]+\frac{4}{\sqrt{A_{\rm AdS}}}\Pi\left[\left.-\frac{c_{\rm AdS}}{2A_{\rm AdS}}\right|-1\right]\\
        &\,\qquad- \frac{4}{\sqrt{A_{\rm AdS}}}\Pi\left[-\frac{c_{\rm AdS}}{2A_{\rm AdS}};\arcsin\left(\cosh^{-1/2}(\bar{\tau}^{\textbf{I}})\right)\Big|-1\right]\bigg)\,,
        \end{aligned}
        \qquad &  (\textbf{I})
\\[1.1cm]
   \displaystyle   \cosh\left(\frac{4}{\sqrt{c_{\rm dS}+A_{\rm dS}}}\mathcal{F}\left[\left.\frac{{\bar{\tau}^{\textbf{II}}}}{2}\right|\frac{2 A_{\rm dS}}{c_{\rm dS}+A_{\rm dS}}\right]\right)\,,\qquad &  (\textbf{II})
\\[0.8cm]
  \begin{aligned}
   \displaystyle   &\cosh\bigg(\frac{4}{\sqrt{c_{\rm dS}+A_{\rm dS}}}\mathcal{K}\left[\frac{2 A_{\rm dS}}{c_{\rm dS}+A_{\rm dS}}\right]+\frac{4}{\sqrt{A_{\rm AdS}}}\Pi\left[ \left.-\frac{c_{\rm AdS}}{2A_{\rm AdS}}\right|-1\right]\\
        & \,\qquad- \frac{4}{\sqrt{A_{\rm AdS}}}\Pi\left[-\frac{c_{\rm AdS}}{2A_{\rm AdS}};\arcsin\left(\cosh^{-1/2}(\bar{\tau}^{\textbf{I}'})\right)\Big|-1\right]\bigg)\,.
        \end{aligned}
        \qquad &  (\textbf{I}') 
    \end{cases}
\label{eq:background_potential}
\end{equation}
$\Pi(n;\psi|m)$ denotes the incomplete elliptic integral of the third kind (see Appendix~\ref{para:incompleteellpticthirdkind}). From the previous formula, one can easily find the relation between the source at the boundary $H_{\rm UV}$ and the value at the center of the wineglass wormhole $H_{0,WG}$ setting $\tau\to\infty$
\begin{equation}
    H_{\rm UV}=H_{0,WG}\,\cosh\left(\frac{4}{\sqrt{c_{\rm dS}+A_{\rm dS}}}\mathcal{K}\left[\frac{2 A_{\rm dS}}{c_{\rm dS}+A_{\rm dS}}\right]+\frac{4}{\sqrt{A_{\rm AdS}}}\Pi\left[ \left.-\frac{c_{\rm AdS}}{2A_{\rm AdS}}\right|-1\right]\right)\,.
\label{eq:Huv_wineglass}
\end{equation}

We can also compute the gauge field momentum $\pi_A $ to find
\begin{equation}
     \pi_A =H_{0,WG}\sinh\left(\frac{4}{\sqrt{c_{\rm dS}+A_{\rm dS}}}\mathcal{K}\left[\frac{2 A_{\rm dS}}{c_{\rm dS}+A_{\rm dS}}\right]+\frac{4}{\sqrt{A_{\rm AdS}}}\Pi\left[ \left.-\frac{c_{\rm AdS}}{2A_{\rm AdS}}\right|-1\right]\right)\,.
\label{eq:VEV_wineglass}
\end{equation}
As already mentioned, we focus on the case $c_{\rm AdS} = -2\sqrt{A_{\rm AdS}(A_{\rm AdS}-A_{\rm dS})} $, for which Eq.~\eqref{eq:Huv_wineglass} can be written as
\begin{equation}
    H_{\rm UV}=H_{0,WG}\,\cosh\left(\frac{4}{\sqrt{c_{\rm dS}+A_{\rm dS}}}\mathcal{K}\left[\frac{2 A_{\rm dS}}{c_{\rm dS}+A_{\rm dS}}\right]+\frac{4}{\sqrt{A_{\rm AdS}}}\Pi\left[ \left. \sqrt{\frac{A_{\rm AdS}-A_{\rm dS}}{A_{\rm AdS}}}\right|-1\right]\right)\,.
\label{eq:Huv_wineglassspecial}
\end{equation}

Indeed, in this case, out of the five parameters, two correspond to the metric and scalar sources (the later being set to zero) and of the three remaining, two correspond to the value of the potential in the UV ($\tilde{V}_{\rm AdS}$) and in the IR ($\tilde{V}_0$). The last one corresponds to the electromagnetic potential at the $\mathbb{Z}_2$ reflection point $H_{0,WG}$ (which is implicitly connected to the boundary source value $H_{\rm UV}$ by \eqref{eq:Huv_wineglass}) and the two values of the potential $\tilde{V}_{{AdS},\,0}$. All other quantities are functions of these parameters. 
% In particular, 
Using the eqs.~\eqref{eq:energydensitywormhole} and~\eqref{eq:rhorad_worm}, together with the condition $A_{\rm dS}<c_{\rm dS}<2$, we also obtain
\begin{align}
\label{eq:H0_wineglass}
 H_{0,WG}^2 =& \frac{(2-A_{\rm dS})}{128\k}  \left(\frac{2A_{\rm dS}}{H_{\rm dS}^2} + \frac{2A_{\rm AdS}-A_{\rm dS}+c_{\rm AdS}}{H_{\rm AdS}^2} \right)
            =  \frac{(2-A_{\rm dS})}{128\k} \frac{(c_{\rm dS} + A_{\rm dS})}{H_{\rm dS}^2}    \,,
\\
\k H_{\rm dS}^2 =& \frac{2(2-A_{\rm dS})A_{\rm dS}\k H_{\rm AdS}^2}{(A_{\rm dS}-2)(2A_{\rm AdS}+c_{\rm AdS}-A_{\rm dS})+128H_{0,WG}^2 \k H_{\rm AdS}^2}\,,
\label{eq:Hds}
\end{align}
and the inequality\footnote{Note that $c_{\rm dS} = A_{\rm dS} + H_{\rm dS}^2 (2A_{\rm AdS}-A_{\rm dS}+c_{\rm AdS})/H_{\rm AdS}^2$ and $A_{\rm dS}<c_{\rm dS}<2$. In the special case $c_{\rm AdS}=-2\sqrt{A_{\rm AdS}(A_{\rm AdS}-A_{\rm dS})}=-2$, $A_{\rm dS}=1$, we obtain $(\k |\tilde{V}_{\rm AdS}|)^{1/2}H_{0,WG}^{\rm min} \simeq {0.074} \Rightarrow (\k |\tilde{V}_{\rm AdS}|)^{1/2}H_{\rm UV}^{\rm min} \simeq  (\k |\tilde{V}_{\rm AdS}|)^{1/2}\pi_A^{\rm min}\simeq { 2.44\times 10^3}$.}
\begin{equation}
\label{eq:H02_ineq}
\k |\tilde{V}_{\rm AdS}| H_{0,WG}^2>\frac{1}{128} \left(2c_{\rm AdS}-A_{\rm dS}^2-2A_{\rm dS}+2A_{\rm dS}A_{\rm AdS}+A_{\rm dS}c_{\rm AdS}+4A_{\rm AdS}\right)\,.
\end{equation}

\begin{figure}[t!]
    \centering
\includegraphics[width=\linewidth]{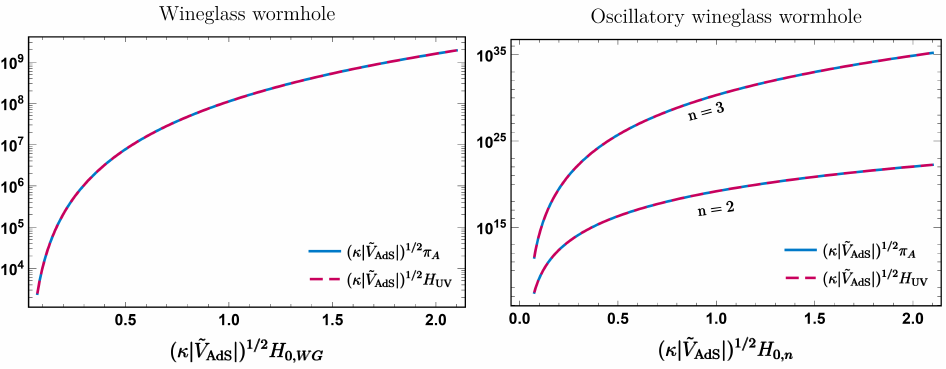}
    \caption{$H_{\rm UV}$ (solid lines), and $\pi_A$ (dashed lines) as a functions of $H_{0,WG}$, and $H_{0,n}$ as given by~\eqref{eq:Huv_wineglass} and~\eqref{eq:Huv_wineglass_OSC} for the wineglass (left) and the oscillatory wineglass (right) wormholes in the case of Dirichlet BCs for the scalar field ($A_{\rm AdS}=\frac{1+\sqrt{5}}{2}, A_{\rm dS}=1$). $(\kappa|\tilde{V}_{\rm AdS}|)^{1/2}|H_{\rm UV}-\pi_A| < 10^{-7}$ is very small and not evident in the scale of this plot. The plot is also restricted to values of $H_{0,WG}$ and $H_{0,n}$ that satisfy the bound in eq.~\eqref{eq:H02_ineq}. We observe that only a single branch of wormhole solutions exists. The relation is a power law for large $H_{0,WG/n}$ (see~\eqref{eq:H_UV/H_0behaviorwineglass}). The plot can be extended to negative $H_{\rm UV},\,H_{0,WG/n}(H_{\rm UV})$ being an odd function.}
    \label{fig:huv_wine}
\end{figure}
It is convenient to rewrite~\eqref{eq:Huv_wineglass} in terms of the potential energy at the UV boundary, $\tilde{V}_{\rm AdS}$, and the parameter $H_{0,WG}$. Using~\eqref{eq:Huv_wineglass},~\eqref{eq:H0_wineglass}, and~\eqref{eq:cond_wine}, in fig.~\ref{fig:huv_wine}, we show the rescaled $H_{\rm UV}$ as a function of $(\kappa |\tilde{V}_{\rm AdS}|)^{1/2} H_{0,WG}$, over the range allowed by the rightmost relation in Eq.~\eqref{eq:H02_ineq}. The plot can be extended to negative values of $H_{\rm UV},\,H_{0,WG}$, being an odd function. 

Evaluating the on-shell Euclidean action for the wineglass wormhole solution imposing Dirichlet quantization for the gauge field and Dirichlet boundary conditions for the scalar field, one finds (see Appendix~\ref{sec:eucl_action in U(1)^3})
\begin{align}
    \mathcal{S}^{\text{WG}}_E =& -\frac{3 \text{Vol}(S^3)}{\kappa}\Bigg[\int_{-\tau_{WG}}^{\tau_{WG}} {\rm d}\tau   a^3(\tau')\tilde{V}(\tau')+\frac{1}{3} \partial_{\tau}a^3\Big|^{\tau_{WG}}_{- \tau_{WG}}\Bigg] \nonumber \\
    &+{24\text{Vol}(S^3)} |H_{\rm UV}| \sqrt{H_{\rm UV}^2 - H_{0,WG}^2}\,,
    \label{eq:on-shell_wineglass}
\end{align}
where we have again remove the cut-off for the radiative part since it remains finite. The relation between $H_{\rm UV}$ and $H_{0 , \, WG}$ is now given by eq.~\eqref{eq:Huv_wineglassspecial}.
Note that compared to the previous case, the scalar field is running leading to a Euclidean time dependence of the potential. This makes the computation of this gravitational part more complicated. In particular, one has to split the integral and evaluate it in the different regions to extract the infrared finite part. Finally, if we adopt the alternative quantization with Neumann BCs, for the gauge field, the boundary term in~\eqref{eq:bt_U1} must be included in~\eqref{eq:on-shell_wineglass}. For the connected wormhole this takes once more a similar form to the simple wormhole, that is
\begin{align}
    S_{\partial \mathcal{M}}^{{\rm U(1)^3},\, {WG}}  = - 48 \text{Vol}(S^3)  |H_{\rm UV}| \sqrt{H_{\rm UV}^2 - H_{0,WG}^2} \, .
\label{eq:neu_wine}
\end{align}

Furthermore, if we wish to understand the wormhole background as providing good initial conditions for phenomenologically realistic inflation as in the scenario proposed in~\cite{Betzios:2024oli,Betzios:2024zhf}, we can constrain the parameters of the model further. In particular assuming that at $\t=t=0$, i.e. at the onset of standard Lorentzian evolution, the field is already located in the plateau region we obtain that,
\begin{equation}
    \tilde{V}_0 \lesssim 3.73\times 10^{-10} \k^{-1} \left( \frac{r_b}{0.036}\right)\,,
    \label{eq:cosmo_bound_Hds}
\end{equation}
where $r$ is the tensor-to-scalar ratio, bounded by $r<r_b=0.036$ at $95\%$ CL when combining the latest Planck, BICEP/Keck, and BAO data~\cite{Planck:2018jri,BICEP:2021xfz}.
Equations~\eqref{eq:rhorad_worm} and~\eqref{eq:Vzero} allow one to relate the value of the scalar potential at $\t=0$ to that of the E/M sector, namely
\begin{equation}
    \tilde{V}_0  = \frac{(2+A_{\rm dS})(2-A_{\rm dS})}{128\k H_{0,WG}^2}\,.
\end{equation}
In the case $A_{\rm dS}=1$ the cosmological bound~\eqref{eq:cosmo_bound_Hds} can be written as
\begin{equation}
    |H_{0,WG}| \gtrsim 7.93\times 10^{3} \left( \frac{0.036}{r_b}\right)^{1/2}\,.
\label{eq:cosmo_bound_H0}
\end{equation}

\subsection{Oscillating wineglass wormhole solutions with running scalar}\label{sec:piecewise}

\subsubsection{Piecewise oscillatory wormhole solutions}

An advantage of the previous piecewise wineglass wormhole solutions is that they can be easily extended to quasi-oscillatory solutions, by keeping further oscillation periods of the scale factor in region $\textbf{II}$, while regions $\textbf{I, I}'$ remain unchanged. 
 
\begin{figure}[t!]
\begin{center}
\includegraphics[width=\textwidth]{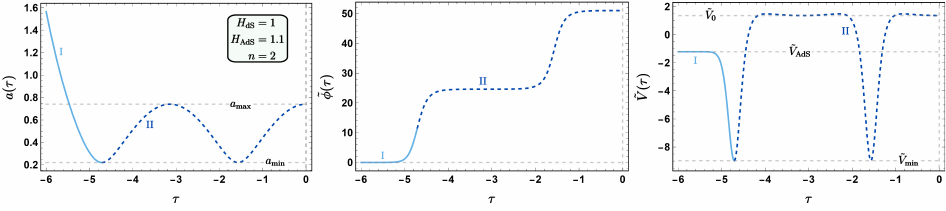}
\end{center}
\caption{Plots of the scale factor $a(\tau)$ (left), the inflaton $\tilde{\phi}(\tau)$ (middle) and the scalar potential $\tilde{V}(\tau)$ (right), for the quasi-oscillatory wormhole with four local minima of the scale factor ($n=2$). The parameters are the same as those used in figs.~\ref{fig:scale_fac} and~\ref{fig:potential} and all are in $\k=1$ units.}
\label{fig:potential_osc}
\end{figure}

For these quasi-oscillatory solutions, the total number of local minima of the scale factor is even ($2n$) and of local maxima is odd $(2n-1)$, with $n \geq 1$ (we did not find any physically consistent backgrounds with an odd number of minima within our ansatz, except the trivial wormhole background with no running scalar). For these cases, the analytic $\mathbb{Z}_2$ symmetric ansatz for the scale factor is now the same as in eq.~\eqref{eq:background_ans}, but region $\textbf{II}$ is extended such that it contains multiple local minima at $\t_{\rm min}^{(m)} = \pm (2m-1) \pi/2H_{\rm dS}$ and multiple local maxima at $\tau^{(m)}_{\rm max} = \pm  (m-1) \pi/H_{\rm dS}$, where $m \leq n$. Such (quasi)-periodic solutions are possible, if the scalar potential also exhibits a similar (quasi)-periodic structure for some field range, see fig.~\ref{fig:potential_osc}. The maximum number of oscillations in the scale factor is therefore fixed in terms of the number of oscillations in the potential (and hence has an upper bound depending on the model).
 
In this (quasi)-periodic case, the analogous matching conditions given in Eq.~\eqref{eq:cond_wine} have to be performed at the two outermost minima i.e. $\pm (2n-1) \pi/2H_{\rm dS}$ of region $(\bf II)$, but remain unchanged. 
The asymptotic values of the potential and the velocity at the Euclidean AdS boundary, as well as their corresponding values at $\tau = 0$, remain unchanged compared to the case with $n = 1$, that corresponds to the wineglass geometry we analyzed in the previous section. Their values at the generic points $\t = \pm k\pi/2H_{\rm dS}$ take the form
\begin{align}
\tilde{\phi}'^2\left(\t = \pm k\pi/2H_{\rm dS}\right) &=  \frac{(-1+(-1)^k)4H_{\rm dS}^2}{3}\frac{c_{\rm dS}-2}{((-1)^k+c_{\rm dS})^2}\,,
\\[0.2cm]
\tilde{V}\left(\t = \pm k\pi/2H_{\rm dS}\right) &= \frac{\left((7+2(-1)^k)c_{\rm dS}+1+8(-1)^k\right)H_{\rm dS}^2}{3((-1)^k+c_{\rm dS})^2}\, ,
\end{align}
showing that the velocity vanishes at the positive maxima of the potential ($k = 2(m-1)$), while being positive at the negative minima of the potential ($k = 2 m -1$). In fig.~\ref{fig:potential_osc}, we display the scale factor, the scalar field, and the potential as functions of $\tau$ for the case $n = 2$ (four local minima). From the figure, it is clear that the emergence of these oscillatory geometries requires the scalar potential itself to take an oscillatory form, ensuring consistency between the potential, the dynamics of the background fields and the corresponding spacetime geometry. In particular each time the motion in Euclidean time traverses a new ``bubble'' of the scale factor $a(\tau)$, the scalar field moves to a new well of the potential $\tilde{V}(\tilde{\phi})$. It is then clear that the maximum allowed number of ``bubbles'' is cut-off by the maximum number of wells present in the scalar potential.

These results for the (quasi)-oscillatory wormholes are very important and contain crucial differences and resolve some paradoxes that appear in analogous (exactly) periodic solutions, such as the axion de-Sitter wormholes of~\cite{Halliwell:1989pu,Aguilar-Gutierrez:2023ril}. The axion de-Sitter wormholes exist for a fixed cosmological constant, have no UV AdS boundaries and give rise to a serious paradox, since their number of periods is not physically bounded by any principle. Moreover, in that case each period lowers the on-shell action, rendering it unbounded from below in the limit of infinite periods\footnote{In a sense this issue is related to the well-known conformal mode problem of Euclidean gravity, exacerbated by the shift symmetry of the constant potential (due to the fixed cosmological constant), see also~\cite{Blommaert:2025bgd}.}. Here the presence of the AdS boundaries and the fact that the potential is a non trivial (quasi)-periodic function whose number of periods is directly correlated to the number of geometric periods in the IR throat part of the geometry, precludes such issues. 

\subsubsection{The oscillatory wormhole E/M gauge field background}

For the oscillatory wormholes with a maximum at $\tau = 0$, the gauge field background is given by Eq.~\eqref{eq:background_ans_simple}, with an extension of the time in region $\textbf{II}$: $- (2n-1) \pi/2H_{\rm dS} \leq \tau \leq (2n-1)\pi/2H_{\rm dS}$ with $n \geq 1$.

Similarly one can easily generalize the solution for the gauge field background Eq.~\eqref{eq:background_potential}, by simply extending the period in region $\textbf{II}$ (the solution can be found analytically once more in the reduced class of models). The relation between the source at the boundary $H_{\rm UV}$ (Dirichlet BCs) and the value of the electromagnetic potential at the center of the wormhole $H_{0,n}$ reads
\begin{equation}
    H_{\rm UV}=H_{0,n}\cosh\left(\frac{4\,(2n-1)}{\sqrt{c_{\rm dS}+A_{\rm dS}}}\mathcal{K}\left[\frac{2A_{\rm dS}}{c_{\rm dS}+A_{\rm dS}}\right]+\frac{4}{\sqrt{A_{\rm AdS}}}\Pi\left[-\frac{c_{\rm AdS}}{2A_{\rm AdS}};-1\right]\right)\, ,
\label{eq:Huv_wineglass_OSC}
\end{equation}
Similarly, for Neumann BCs we have
\begin{equation}
    \pi_A=H_{0,n}\sinh\left(\frac{4\,(2n-1)}{\sqrt{c_{\rm dS}+A_{\rm dS}}}\mathcal{K}\left[\frac{2A_{\rm dS}}{c_{\rm dS}+A_{\rm dS}}\right]+\frac{4}{\sqrt{A_{\rm AdS}}}\Pi\left[-\frac{c_{\rm AdS}}{2A_{\rm AdS}};-1\right]\right)\, ,
\label{eq:piA_OSC}
\end{equation}

In what follows in this subsection, we impose Dirichlet BCs for the gauge field and discuss the two possible BCs for the scalar field. We seek to determine whether the parameter space specified by the bounds in (\ref{wineglass})\footnote{In general, all the bounds described in \ref{wineglass} still hold for the oscillatory solutions.} constrains the possible number of bubbles at a given value of $h_{\rm UV}=(\kappa|\tilde{V}_{\rm AdS}|)^{1/2}H_{\rm UV}$. Indeed, we observe that the wineglass wormhole as well as the oscillatory solutions do not exist for values of $h_{\rm UV}$ smaller than $h_{\rm UV,\,\min}^{\rm (n)}$, where $n$ is the number of oscillations (see fig.~\ref{fig:huv_wine}). To study this effect, we invert the relation \eqref{eq:Huv_wineglass_OSC}, treating the number of bubbles as the variable. The difficulty is that $c_{\rm dS}+A_{\rm dS}$ depends on $H_{0,n}$ (see \eqref{eq:rhorad_worm}). After some algebra, one obtains
\begin{align}
n(h_{0,n},h_{\rm UV},A_{\rm dS},A_{\rm AdS})= &\Bigg[\left(\rm{Argsech}\left[\frac{ h_{0,n}}{h_{\rm UV}} \right]-\frac{4}{\sqrt{A_{\rm AdS}}} \Pi\left[\sqrt{\frac{A_{\rm AdS}-A_{\rm dS}}{A_{\rm AdS}}}\,;\,-1\right]\right)\sqrt{f(h_{0,n})} \nonumber
\\
& +4\mathcal{K}\left[\frac{2A_{\rm dS}}{f(h_{0,n})}\right]\Bigg]\times\Bigg[8\mathcal{K}\left[\frac{2A_{\rm dS}}{f(h_{0,n})}\right]\Bigg]^{-1}\,,
\label{eq:bubblefunction}
\end{align}
with $h_{0,n}=(\kappa|\tilde{V}_{\rm AdS}|)^{1/2}H_{0,n}$ and 
\begin{equation}
f(h_{0,n},A_{\rm AdS},A_{\rm dS})=\frac{256A_{\rm dS}{h^2_{0,n}}}{A_{\rm dS}(2+c_{\rm AdS})-2c_{\rm AdS}-2A_{\rm AdS}(2-A_{\rm dS})-A_{\rm dS}^2+128\,h^2_{0,\,n}}\,.
\end{equation}
The space of parameters is defined by 
\begin{equation}
\begin{aligned}
        &h_{0,n}>0\,,\quad h_{0,n}\leq h_{\rm UV}\,,\quad h_{\rm UV}>0\,,\quad 0<A_{\rm dS}<2\,,\quad A_{\rm AdS}\geq\frac{A_{\rm dS}}{2}+\sqrt{1+\frac{A_{\rm dS}^2}{4}}\,,\\
        &h_{0,n}^2>\frac{1}{128}\left(2c_{\rm AdS}-A_{\rm dS}^2-2A_{\rm dS}+2A_{\rm dS}A_{\rm AdS}+A_{\rm dS}c_{\rm AdS}+4A_{\rm AdS}\right)\,.
\end{aligned}
\end{equation}

\begin{figure}[t!]
    \centering
    \includegraphics[width=0.487\linewidth]{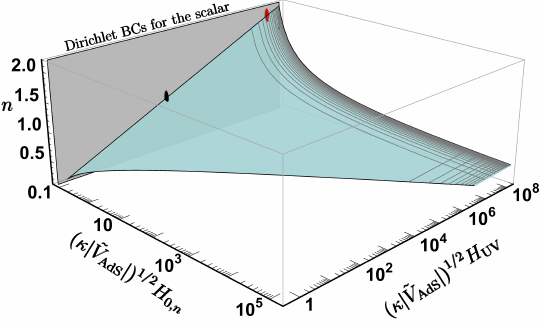}
    \includegraphics[width=0.493\linewidth]{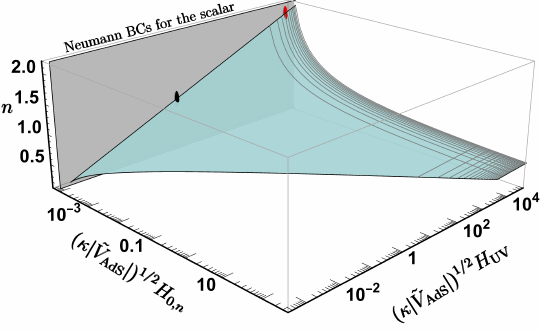}
    \caption{\textit{Left}: the function $n(H_{0,n},A_{\rm dS}=1,A_{\rm AdS}=\frac{1+\sqrt{5}}{2},H_{\rm UV})$ describing the number of bubbles as a function of the (rescaled) electromagnetic potential at the center of the wormhole $H_{0,n}$ and the (rescaled) electromagnetic potential at the boundary $H_{\rm UV}$. This plot describes the situation when we work in the Dirichlet quantization picture. \textit{Right}: same plot but with $A_{\rm AdS}=10^6$. This plot describes the situation when we work in the Neumann quantization picture for the scalar field. In both plots, the black and red dots indicate the minimum values of $h_{\rm UV}$ in the wineglass and $(n=2)$ oscillatory wormhole, respectively.}
    \label{fig:bubbles_function}
\end{figure}
The condition in the second line is precisely the bound \eqref{eq:H02_ineq}. We restrict attention to $A_{\rm dS}=1$. In fig.~\ref{fig:bubbles_function}, we show the case $A_{\rm AdS}=\frac{1+\sqrt{5}}{2}$ ($c_{\rm AdS}=-2$, Dirichlet BCs) and, for comparison, cases with large $A_{\rm AdS}$ (large $c_{\rm AdS}$, Neumann BCs).

As $h_{0,n}$ decreases, the number of bubbles appears to grow rapidly as $h_{0,n}\to 0$. The bound \eqref{eq:H02_ineq}, represented by the grey surface, prevents any divergence. Indeed, at fixed $h_{\rm UV}$, the number of possible bubbles is bounded from above. This means that, for fixed $h_{\rm UV}$, there exists a finite number of oscillatory solutions under consideration. On the other hand, the number of bubbles seems to increase slowly without bound as $h_{\rm UV}$ becomes large. Nevertheless, as mentioned in the previous section, we should keep in mind that there is always an upper cut-off in the number of bubbles, set by the number of wells in the quasi-oscillatory region of the potential $\tilde{V}(\tilde{\phi})$

Finally, our graph can also be understood in the reverse direction. Fix a number of bubbles $n$. The intersection of the surface with the plane $n=\text{const.}$ yields the relation $H_{\rm UV}(H_{0,n})$ plotted in the right panel of fig.~\ref{fig:huv_wine} for $n\in\{2,3\}$. We observe that, contrary to the simple wormhole without a running scalar, there is a one-to-one correspondence between $H_{\rm UV}$ and $H_{0,n}$, resulting in a single branch of wormholes for any fixed $N$. This is also true for wormholes with Neumann BCs for the gauge field (see fig.~\ref{fig:huv_wine} where we also show $\pi_A(H_{0,n})$).

\subsection{``Centaur'' geometries?}

The ``Centaur'' geometries of Anninos and Hofman~\cite{Anninos:2017hhn}, can in a sense be thought of as a finely tuned case of the wineglass wormholes, when the running scalar reaches a local minimum/maximum of the scalar potential with positive value and stops completely there.
On the one hand, it seems that it is always possible to achieve this, for example if the potential $V(\phi)$ becomes exactly flat after a certain value of $\phi$. Unfortunately the presence of the magnetic flux/radiation, which is very important for supporting the throat at its minimum value $a_{\rm min}(\tau_{\rm min})$, prohibits also the smooth closure of the geometry at other Euclidean times.

To make this explicit in two examples, an ansatz for a centaur geometry, is the following: We consider regions $\textbf{I, II}$ of the wineglass geometry, and then we stop region $\textbf{II}$ at $\tau = 0$. Then we glue it to a new region $\textbf{III}$, which is given either by the ansatz
\be
a^2_{{III}}(\tau) = \frac{A_c}{4H_c^2} \cos (2H_c \tau ) \, , \quad \tau_{{III}} \in [0, \pi/4 H_c] \, ,
\ee
or
\be
a_{{III}}(\tau) = \frac{\sqrt{A_c}}{2H_c} \cos (2H_c \tau ) \, , \quad \tau_{III} \in [0, \pi/4 H_c] \, .
\ee
In both ansatzs, the geometry smoothly caps off at $\tau = \pi/ 4H_c$.
Continuity at $\mathcal{O}(a, a' , $ $a'', a''')$ leads to the identifications
\begin{equation}
\frac{A_c}{H_c^2} = \frac{A_{\rm dS}+c_{\rm dS}}{H_{\rm dS}^2} \qquad  \text{and} \qquad A_{\rm dS} = A_c\,,
\end{equation}
for the first ansatz and to
\begin{equation}
\frac{A_c}{H_c^2} = \frac{A_{\rm dS}+c_{\rm dS}}{H_{\rm dS}^2} \qquad  \text{and} \qquad A_{\rm dS} = 2A_c\,,
\end{equation}
for the second.

The velocity and the potential in region (\textbf{III}) for the first ansatz are given by
\begin{align}
\tilde{\phi}_{III}'^2(\t) &= \frac{4 H_{c}^2}{3A_c^2} \left(\frac{A_c^2-16H_{c}^2\tilde{\rho}^E_{\rm rad}-2A_c \cos[2H_{c}\t]}{\cos^2[2H_c\t]}\right)\,,
\\[0.2cm]
\tilde{V}_{III}(\t) &= \frac{H_{c}^2}{6A_c^2}\left(\frac{A_c^2+32H_{c}^2\tilde{\rho}^E_{\rm rad}+16A_c\cos[2H_{c}\t]+3A_c^2\cos[4H_{c}\t]}{\cos^2[2H_{\rm dS}\t]}\right)\, ,
\label{eq:VII_n_A}
\end{align}
while for the second ansatz we find in region \textbf{(III)}
\begin{align}
\tilde{\phi}_{III}'^2(\t) &= -\frac{8H^2_c}{3A_c^2} \left(\frac{8H^2_c\tilde{\rho}^E_{\rm rad} + A_c(1-A_c)\cos^2[2H_c\t]}{\cos^4[2H_c\t]}\right)\,,
\\[0.2cm]
\tilde{V}_{III}(\t) &= \frac{4H^2_c}{3A_c^2}\left(\frac{4H^2_c\tilde{\rho}^E_{\rm rad} +2A_c(1-A_c)\cos^2[2H_c\t] + 3A_c^2\cos^4[2H_c\t]}{\cos^4[2H_c\t]}  \right)\,. 
\label{eq:VIII_A}
\end{align}
We immediately observe a divergence of the scalar velocity (and of the potential) at the point where the geometry smoothly caps-off, rendering such solutions pathological within our model and class of solutions.

Both cases indicate that centaur geometries can only be found if the energy density becomes appropriately time dependent. It is negative at $\tau_{\rm min}$, it diminishes or becomes positive after $\tau=0$ while at the same time the scalar approaches a constant plateau/maximum of the potential. This cannot happen with radiation fields, since their action is scale invariant. The analysis of other types of fields such as axions that can potentially support centaur geometries, goes beyond the scope of the present work.

\section{Comparing the gravitational saddles}\label{sec:comparison}

In this section, we analyze the regimes in which each saddle dominates by comparing them using background subtraction. For the subtraction to be well defined, the boundary manifolds and asymptotic sources must coincide for the pair of geometries that is being compared. For the EFLRW metric~\eqref{EFLWR metric}, the corresponding procedure was described in detail in~\ref{sub:procedure_background}. We classify the results by boundary conditions: we discuss both choices of boundary conditions for the scalar field, while keeping Dirichlet BCs for the gauge field. Switching from Dirichlet to Neumann BCs for the gauge field is found not to qualitatively change the results regarding the dominance of solutions.
We will explicitly give the expression for $\Delta\mathcal{S}_{\rm{Disc-Wormhole(j)}}$, keeping in mind that
\begin{equation}
    \Delta\mathcal{S}_{\rm{Wormhole(i)-Wormhole(j)}}=\Delta\mathcal{S}_{\rm{Disc-Wormhole(j)}}-\Delta\mathcal{S}_{\rm{Disc-Wormhole(i)}}\,.
\end{equation}
At the end of each subsection, we add a summary of the results with a table and additional figures that clarify and summarize our findings.

\subsection{Dirichlet for the scalar field/Dirichlet for the gauge field}
\paragraph{Disconnected geometries \textit{vs} Simple wormholes:}\label{para:disc vs simple wormholes}
We start by expressing the difference between the on-shell actions before sending the cut-off to infinity
\begin{align}
\Delta \mathcal{S}_{D-C}
&= 2 \mathcal{S}^{D}_E - \mathcal{S}^{C}_E \nonumber \\
&= \frac{\mathrm{Vol}(S^3)}{\kappa}\,\mathcal{I}_{C,D}
   + 24\mathrm{Vol}(S^3)\left(H_{\rm UV}^2
   -|H_{\rm UV}|\sqrt{H_{\rm UV}^2-H_{0,C}^2}\right) \, ,
\label{eq:DS_disc_conn}
\end{align}
where we introduce the following quantities to simplify the presentation
\begin{equation}
\label{eq:before-background-subtraction}
        \mathcal{I}_{C,D}
\equiv 3\int_{-\tau_{C}}^{\tau_{C}} \mathrm{d}\tau \,a_C^3\,\tilde{V}_{\mathrm{AdS}}
   + \left.\partial_{\tau}a_C^3\right|_{-\tau_{C}}^{\tau_{C}}
   - 2\left(
        3\int_{0}^{\tau_{D}} \mathrm{d}\tau \,a_D^3\,\tilde{V}_{\mathrm{AdS}}
        + \left.\partial_{\tau}a_D^3\right|^{\tau_D}
     \right)\,.
\end{equation}

\begin{figure}[t!]
    \centering
    \includegraphics[width=0.667\linewidth]{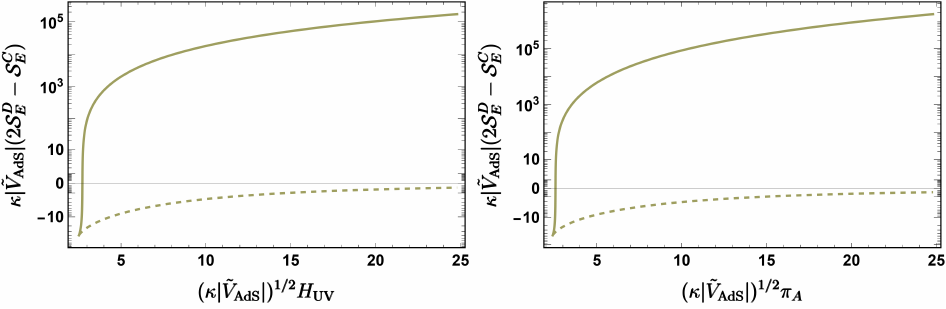}
    \caption{The difference between the on-shell actions of the disconnected and simple connected wormholes defined in eq.~\eqref{eq:DS_disc_conn}, and shown as a function of $H_{\rm UV}$ (left panel, assuming Dirichlet BCs for the gauge field) and $\pi_A$ (right panel, assuming Neumann BCs for the gauge field). The solid line corresponds to the large wormhole branch, while the dashed line to the small wormhole branch.}
    \label{fig:action_diff}
\end{figure}
The second part of the difference of the on-shell actions concerns the radiative part of both geometries and is finite for any value of the regulator, which we can already send to infinity. The first part $\mathcal{I}_{C,D}$ of the last equality contains the gravitational part and while each piece diverges, when $a_C(\tau_C) = a_D(\tau_D)=R\to\infty$, their difference remains finite. This contribution is associated to the geometry of the wormhole (resp. the disconnected geometry) near the throat of the wormhole (resp. near the tips). In this sense, we qualify this residual part of the gravitational piece as an infrared (IR) contribution to the on-shell action. The details of the computation are presented in Appendix~\ref{subtraction1}.  In the end we find the on-shell action difference being a function of the boundary source $H_{\rm UV}$ and the value of the potential $\tilde{V}_{\rm AdS}$
\begin{align}
    \Delta \mathcal{S}_{D-C}(H_{\rm UV} ; \tilde{V}_{\rm AdS})=&-\frac{4\text{Vol}(S^3)\,\zeta^{IR}_{D-C}\left(\sqrt{1-32\kappa H_{0,C}^2(H_{\rm UV})\tilde{V}_{\rm AdS}}\right)}{\kappa\tilde{V}_{\rm AdS}}  \nonumber
\\
    &+24 \text{Vol}(S^3)\left(H_{\rm UV}^2 -|H_{\rm UV}|\sqrt{H_{\rm UV}^2-H_{0,C}^2}\right)\,. 
\label{eq:Delta_S_1_C}
\end{align}
The expression of the IR contribution computed in Appendix~\ref{subtraction1} in terms of the dimensionless parameter $\alpha=\sqrt{1-32\kappa H_{0,C}^2\tilde{V}_{\rm AdS}}$ is
\begin{equation}
\begin{aligned}
    \zeta^{IR}_{D-C}(\alpha) &=1- \frac{3}{4\sqrt{2}}\left[\frac{4\sqrt{\alpha-1}\left[\mathcal{E}(-\frac{\alpha+1}{\alpha-1})-\mathcal{K}(-\frac{\alpha+1}{\alpha-1})\right]}{3}+\frac13\sqrt{\alpha-1}(\alpha+1)\mathcal{K}\left(-\frac{\alpha+1}{\alpha-1}\right)\right]\,.
\end{aligned}
\end{equation}
From this expression, one can study the asymptotic behaviour for large $H_{\rm UV}$ (we recall that for the small wormhole $H_{0,C}\to 0$ while for the large wormhole $H_{0,C}\to \infty$ in this regime (see~\eqref{eq:asymptoticconnected})
\begin{itemize}
    \item Small wormhole
    \begin{subequations}
    \begin{align}
        S_{E\,\,\rm rad}^{D}&= 24\text{Vol}(S^3)H_{\rm UV}^2\underset{H_{0,C}\to 0}{=} S_{E\,\,\rm rad}^{C}+\mathcal{O}\left(H_{0,C}^2\right)\,,
\\
    \zeta^{IR}_{D-C}&\underset{H_{0,C}\to 0}{=}\mathcal{O}\left(H_{0,C}^2\right)\,.
    \end{align}
    \end{subequations}
    \item Large wormhole
    \begin{subequations}
\begin{align}
       S_{E\,\,\rm rad}^{D}&= 24\text{Vol}(S^3)H_{\rm UV}^2\gg S_{E\,\,\rm rad}^{C}\underset{H_{0,C}\to \infty}{\sim}24 \times2^{1/4}\frac{\mathcal{K}(-1)}{(\kappa|\tilde{V}_{\rm AdS}|)^{1/4}}\text{Vol}(S^3)H_{0,C}^{3/2}\,,
\\
    \zeta^{IR}_{D-C}&\underset{H_{0,C}\to \infty}{\sim}-2\times2^{1/4}\left(\kappa|\tilde{V}_{\rm AdS}|\right)^{3/2}\mathcal{K}(-1)H_{\rm UV}^{3/2}\,.
\end{align}
\end{subequations}
\end{itemize}
Thus, we expect for large wormhole $\Delta\mathcal{S}_{D-C}\underset{H_{\rm UV}\to \infty}{\sim} S_{E,\rm rad}^{D}>0$ , \textit{i.e}., the simple large wormhole being more dominant in the gravitational path integral. By contrast, for the small wormhole $\Delta\mathcal{S}_{D-C} \to 0$, since the small wormhole closes off and reproduces the disconnected expression. The complete numerical results are shown fig.~\ref{fig:action_diff} \footnote{In the right panel, we also show the results imposing Neumann BCs for the gauge field. The results are qualitatively similar.}. The large wormhole is always the dominant saddle between the two wormholes and we observe an exchange of dominance between the disconnected geometry and the large wormhole\footnote{This was also observed in two dimensional string theory models that exhibit BH/wormhole transitions (two disconnected cigars vs. a simple Euclidean wormhole)~\cite{Betzios:2023jco}.} revealing a first order phase transition similar to the Hawking-Page transition at $(\k|\tilde{V}_{\rm AdS}|)^{1/2}H_{\rm UV}\simeq  2.73$~\cite{Hawking:1982dh,Marolf:2021kjc}.

\paragraph{Disconnected geometries \textit{vs} wineglass wormholes/oscillatory wormholes:} 

In these cases, we provide the exact result for $\Delta\mathcal{S}_{D-\rm{osc(n)}}$, after performing background subtraction, where $n=1$ corresponds to the wineglass wormhole. Recall from~\ref{piecewisewineglass} (in particular see the discussion after~\eqref{eq:non-renormalized-version}) that for $c_{\rm AdS} \neq -2$, we have a non-trivial scalar source and the wineglass/oscillatory wormholes cannot be compared with the disconnected or the simple wormhole backgrounds since they carry different sources at the boundary of EAdS. When we set $c_{\rm AdS}=-2$ to ensure a vanishing boundary source, no additional counterterm is needed, and performing background subtraction cancels all the divergences in the difference between the on-shell actions (see~\ref{UV-Wineglass} for further details). In this case, we have a three-parameter family of solutions, parameterized by $H_{\rm UV},\,\tilde{V}_{\rm AdS}=-H_{\rm AdS}^2,\,A_{\rm dS}$. Thus, compared to the geometry with a constant scalar, we have an additional parameter, $A_{\rm dS}$, to tune. Formally, the expression before performing background subtraction takes the same form (see~\eqref{eq:before-background-subtraction}). Extracting the remaining constant piece after performing background subtraction is quite more involved and is detailed in Appendix~\ref{IR-Wingelass}. At the end, the difference between the on-shell actions after sending the cut-offs to infinity reads
\begin{align}
\label{eq:Delta_S_1}
    \Delta \mathcal{S}_{D-\rm{osc(n)}}(H_{\rm UV} ; \tilde{V}_{\rm AdS} ; A_{\rm dS})=&-\frac{4\text{Vol}(S^3)\,\zeta^{IR\,\,\rm {Dirichlet}}_{D-\rm{osc(n)}}\left(H_{\rm UV} ;\tilde{V}_{\rm AdS};A_{\rm dS}\right)}{\kappa\tilde{V}_{\rm AdS}} \nonumber
\\
    &+24 \text{Vol}(S^3)\left(H_{\rm UV}^2 -|H_{\rm UV}|\sqrt{H_{\rm UV}^2-H_{0,n}^2}\right)\,. 
\end{align}

For completeness, we provide here the precise expression of the IR contribution, computed in Appendix~\ref{IR-Wingelass}
\begin{align}
    &\zeta^{IR\,\,\rm {Dirichlet}}_{D-\rm{osc(n)}}\left(H_{\rm UV} ;\tilde{V}_{\rm AdS};A_{\rm dS}\right)=1-\Bigg[\frac{1}{4\sqrt{A_{\rm AdS}}}\Bigg(32H_{\rm AdS}^2\kappa H_{0,n}^2\Pi\left[\frac{1}{A_{\rm AdS}};-1\right]\nonumber
\\
    &+2A_{\rm AdS}\mathcal{E}[-1]-(2A_{\rm AdS}-2)\mathcal{K}[-1]\Bigg)+\frac{(2n-1)\tilde{V}_{\rm AdS}\,(c_{\rm dS}+A_{\rm dS})^{1/2}}{16H_{\rm dS}^2}\Bigg(8\mathcal{E}\left[\frac{2A_{\rm dS}}{c_{\rm dS}+A_{\rm dS}}\right] \nonumber
\\
    &-(2-A_{\rm dS})\mathcal{K}\left[\frac{2A_{\rm dS}}{c_{\rm dS}+A_{\rm dS}}\right]\Bigg)\Bigg]\,.
\label{eq:osc_zeta}
\end{align}
We can also analyse this expression for large $H_{\rm UV}$. Contrary to the case discussed earlier, $\Delta \mathcal{S}_{D-\rm{osc(n)}}(H_{\rm UV} ; \tilde{V}_{\rm AdS} ; A_{\rm dS})$ is dominated by the IR contribution $\zeta^{IR\,\,\rm {Dirichlet}}_{D-\rm{osc(n)}}$. In fact, we found (see Appendix~\ref{Asymptoticbehavior})
\begin{align}
    \Delta \mathcal{S}_{D-\rm{osc(n),\, rad}}&\underset{H_{\rm UV}\to \infty}{\sim}12\text{Vol}(S^3)H_{0,n}^2\underset{H_{\rm UV}\to \infty}{\sim}12\text{Vol}(S^3)\left(2e^{-\gamma}H_{\rm UV}\right)^{2/q(n)}\,, \nonumber
\\
    \Delta\mathcal{S}_{D-\rm{osc(n),\, grav}}&\underset{H_{\rm UV}\rightarrow\infty}{\sim}-8\text{Vol}(S^3)\left(2e^{-\gamma}H_{\rm UV}\right)^{2/q(n)}\left(\frac{q(n)-1}{q(n)}\right)\log|2e^{-\gamma}H_{\rm UV}|\,,
\label{eq:asymptoticbehavior}
\end{align}
where $\Delta \mathcal{S}_{D-\rm{osc(n),\, rad}}$ corresponds to the second line of~\eqref{eq:Delta_S_1} and $ \Delta\mathcal{S}_{D-\rm{osc(n),\, grav}}$ corresponds to the first line of~\eqref{eq:Delta_S_1}. The expression of $\gamma$ is given in Appendix~\ref{Asymptoticbehavior} (see in particular~\eqref{eq:a1_expression}) and $q(n)$ is an increasing function of $n$ (see~\eqref{eq:H_UV/H_0behaviorwineglass}). The gravitational contribution is logarithmically dominant with respect to the radiative part at large $H_{\rm UV}$. Based on eq.~\eqref{eq:asymptoticbehavior} we therefore find that, asymptotically for large $H_{\rm UV}$, the disconnected solution dominates the gravitational path integral over the oscillatory solutions. At the same time we also find that wormholes with the largest allowed number of bubbles are preferred in this limit.

\paragraph{Summary of results for Dirichlet BCs:}

\begin{figure}[t!]
    \centering
    \includegraphics[width=1\linewidth]{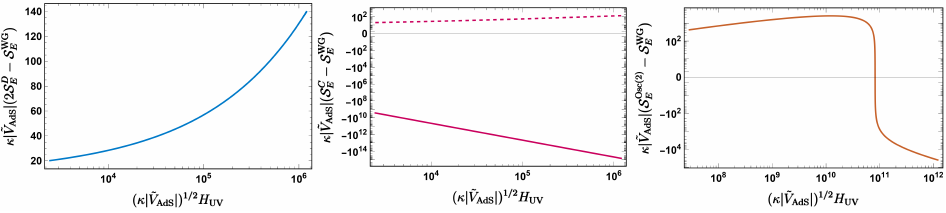}
    \caption{Difference between on-shell actions as a function of $h_{\rm UV}=(\kappa |\tilde{V}_{\rm AdS}|)^{1/2}\,H_{\rm UV}$. \textit{Dirichlet} BCs are assumed for both the scalar and the gauge fields in all panels. The scalar source is zero, i.e. $A_{\rm AdS}=\frac{1+\sqrt{5}}{2} \Rightarrow c_{\rm AdS}=-2$ and we took $A_{\rm dS}=1$. \textit{Left}: Disconnected vs wineglass wormhole. \textit{Center}: Wineglass wormhole vs simple connected wormhole. \textit{Right}: n=2 oscillatory wormhole vs wineglass wormhole.}
    \label{fig:action_diff_2}
\end{figure}

\begin{figure}[t!]
    \centering

\resizebox{\textwidth}{!}{%
\begin{tikzpicture}[x=3.5cm,y=0.95cm]

% equally spaced positions on the x-axis
\def\xA{0.00}
\def\xB{0.55}
\def\xC{1.10}
\def\xD{1.65}
\def\xE{2.20}
\def\xF{2.75}
\def\xG{3.30}

% common vertical positions
\def\yone{-0.55}
\def\ytwo{-1.55}
\def\ythree{-2.55}
\def\yfour{-3.55}
\def\yfive{-4.55}

% sizes
\def\wmain{1.10cm}
\def\wconn{0.92cm}

% curved labels for connected wormholes
% vertical labels for connected wormholes
\newcommand{\connlabel}[3]{%
  \node[
    rotate=90,
    font=\tiny\itshape,
    inner sep=0pt
  ] at ({#1+0.18},{#2}) {#3};
}

% axis with ellipsis before the arrow
\draw[thick] (-0.08,0) -- (\xG-0.03,0);

\node[font=\scriptsize, fill=white, inner sep=1pt]
      at (\xG+0.03,0) {$\cdots$};

\draw[->, thick] (\xG+0.082,0) -- (\xG+0.32,0);

\node[above, font=\scriptsize] at (\xG+0.21,0.02) {$h_{\rm UV}$};

% ticks and labels
\foreach \x/\lab in {
\xA/0,
\xB/2.52,
\xC/2.73,
\xD/2.44\times10^3,
\xE/2.87\times10^7,
\xF/8.19\times10^{10}
}{
  \draw (\x,0.10) -- (\x,-0.10);
  \node[above, font=\scriptsize] at (\x,0.12) {$\lab$};
}

% curly separator lines
\foreach \x in {\xA,\xB,\xC,\xD,\xE,\xF}{
  \draw[
    gray!50,
    decorate,
    decoration={snake, amplitude=1.5pt, segment length=8pt}
  ] (\x,0) -- (\x,-4.85);
}

% -------------------------------------------------
% 0 -- 2.52 : disc
% -------------------------------------------------
\node at ({(\xA+\xB)/2},\yone)
{\includegraphics[width=\wmain]{Figures_paper_1/worm_disc.png}};

% -------------------------------------------------
% 2.52 -- 2.73 : disc --> connected large --> connected small
% -------------------------------------------------
\node at ({(\xB+\xC)/2},\yone)
{\includegraphics[width=\wmain]{Figures_paper_1/worm_disc.png}};

\node at ({(\xB+\xC)/2},\ytwo)
{\includegraphics[width=\wconn]{Figures_paper_1/worm_conn.png}};
\connlabel{(\xB+\xC)/2}{\ytwo}{\textit{large}}

\node at ({(\xB+\xC)/2},\ythree)
{\includegraphics[width=\wconn]{Figures_paper_1/worm_conn.png}};
\connlabel{(\xB+\xC)/2}{\ythree}{\textit{small}}

% -------------------------------------------------
% 2.73 -- 2.44 x 10^3 : connected large --> disc --> connected small
% -------------------------------------------------
\node at ({(\xC+\xD)/2},\yone)
{\includegraphics[width=\wconn]{Figures_paper_1/worm_conn.png}};
\connlabel{(\xC+\xD)/2}{\yone}{\textit{large}}

\node at ({(\xC+\xD)/2},\ytwo)
{\includegraphics[width=\wmain]{Figures_paper_1/worm_disc.png}};

\node at ({(\xC+\xD)/2},\ythree)
{\includegraphics[width=\wconn]{Figures_paper_1/worm_conn.png}};
\connlabel{(\xC+\xD)/2}{\ythree}{\textit{small}}

% -------------------------------------------------
% 2.44 x 10^3 -- 2.87 x 10^7 : connected large --> wine --> disc --> connected small
% -------------------------------------------------
\node at ({(\xD+\xE)/2},\yone)
{\includegraphics[width=\wconn]{Figures_paper_1/worm_conn.png}};
\connlabel{(\xD+\xE)/2}{\yone}{\textit{large}}

\node at ({(\xD+\xE)/2},\ytwo)
{\includegraphics[width=\wmain]{Figures_paper_1/worm_wine.png}};

\node at ({(\xD+\xE)/2},\ythree)
{\includegraphics[width=\wmain]{Figures_paper_1/worm_disc.png}};

\node at ({(\xD+\xE)/2},\yfour)
{\includegraphics[width=\wconn]{Figures_paper_1/worm_conn.png}};
\connlabel{(\xD+\xE)/2}{\yfour}{\textit{small}}

% -------------------------------------------------
% 2.87 x 10^7 -- 8.19 x 10^10 : connected large --> wine --> osc --> disc --> connected small
% -------------------------------------------------
\node at ({(\xE+\xF)/2},\yone)
{\includegraphics[width=\wconn]{Figures_paper_1/worm_conn.png}};
\connlabel{(\xE+\xF)/2}{\yone}{\textit{large}}

\node at ({(\xE+\xF)/2},\ytwo)
{\includegraphics[width=\wmain]{Figures_paper_1/worm_wine.png}};

\node at ({(\xE+\xF)/2},\ythree)
{\includegraphics[width=\wmain]{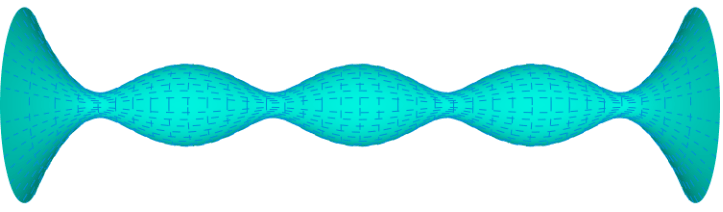}};

\node at ({(\xE+\xF)/2},\yfour)
{\includegraphics[width=\wmain]{Figures_paper_1/worm_disc.png}};

\node at ({(\xE+\xF)/2},\yfive)
{\includegraphics[width=\wconn]{Figures_paper_1/worm_conn.png}};
\connlabel{(\xE+\xF)/2}{\yfive}{\textit{small}}

% -------------------------------------------------
% after 8.19 x 10^10 : connected large --> osc --> wine --> disc --> connected small
% -------------------------------------------------
\node at ({(\xF+\xG)/2},\yone)
{\includegraphics[width=\wconn]{Figures_paper_1/worm_conn.png}};
\connlabel{(\xF+\xG)/2}{\yone}{\textit{large}}

\node at ({(\xF+\xG)/2},\ytwo)
{\includegraphics[width=\wmain]{Figures_paper_1/worm_wine_osc.png}};

\node at ({(\xF+\xG)/2},\ythree)
{\includegraphics[width=\wmain]{Figures_paper_1/worm_wine.png}};

\node at ({(\xF+\xG)/2},\yfour)
{\includegraphics[width=\wmain]{Figures_paper_1/worm_disc.png}};

\node at ({(\xF+\xG)/2},\yfive)
{\includegraphics[width=\wconn]{Figures_paper_1/worm_conn.png}};
\connlabel{(\xF+\xG)/2}{\yfive}{\textit{small}}

% dominant saddle arrow
\draw[thick, gray!50, line width=4pt, ->, opacity=0.8] (\xG+0.20,-5.00) -- (\xG+0.20,-0.35);
\node[rotate=90, font=\footnotesize, fill=white, inner sep=0pt]
      at (\xG+0.12,-2.65) {dominant saddle};

\end{tikzpicture}%
}

\caption{Dominant saddles in the gravitational path integral across a range of
$h_{\rm UV}\equiv(\kappa|\tilde{V}_{\rm AdS}|)^{1/2}H_{\rm UV}$, imposing Dirichlet boundary conditions for the scalar and gauge field. Some saddles cease to exist below certain minimum values $h_{\rm UV}^{\rm min}$. The labels \textit{large} and {\textit{small}} denote the two connected wormhole branches.  The ellipsis on the $h_{\rm UV}$ axis indicates further transitions, where the disconnected saddle becomes dominant first over the wineglass and then over the oscillatory wineglass saddle, at values not accessible numerically.}
\label{fig:saddle_dom}
\end{figure}

In figs.~\ref{fig:action_diff} and \ref{fig:action_diff_2} we present the differences in the on-shell action as one varies the electromagnetic source at the boundary of EAdS, when $c_{\rm AdS} = -2$, so that the scalar source is zero (when $c_{\rm AdS}\neq -2$ and a scalar source is turned on, the dominant solution is always the wineglass one). The table shown in fig.~\ref{fig:saddle_dom} summarizes the results for $A_{\rm dS}=1$. We observe a rich phase diagram with multiple phase transitions as one varies $h_{\rm UV}=(\kappa|\tilde{V}_{\rm AdS}|)^{1/2}H_{\rm UV}$. Here we display only the first two types of oscillatory wormholes ($n=1$ wineglass and $n=2$).

We first notice that the simple connected wormhole (provided it exists and hence $h_{\rm UV}$ is larger than a critical value) is always dominant compared to wormhole solutions with a running scalar, within the subclass of solutions that we analyze. In other words, the scalar prefers not to run. Throughout the range $h_{\rm UV}\gtrsim 2.44\times 10^3$, we observe multiple exchanges of dominance between the disconnected branch and the running scalar wormhole saddles. In particular, when these wormholes first appear, they are dominant relative to the disconnected geometry. However, as $h_{\rm UV}$ increases, the disconnected geometry tends to be preferred. At the same time, the potential depends nontrivially on the radiation and the bubbles start to contain more and more gravitational energy eventually dominating over all other (running scalar) contributions at large $h_{\rm UV}$ (see~\eqref{eq:asymptoticbehavior}). The exchange of dominance between the disconnected and wineglass saddles occurs at values of $h_{\rm UV}$ that are beyond our numerical reach.

Interestingly, we also observe a phase transition between the oscillatory solution and the wineglass solution at $h_{\rm UV}\simeq 8.19\times10^{10}$. This effect is highly nontrivial and stems from the fact that, for a fixed value of $h_{\rm UV}$, the electromagnetic field at the center of the wormhole $H_{0,n}$ decreases as the number of bubbles increases (see fig.~\ref{fig:huv_wine}), while at the same time the gravitational action becomes even more negative. As a consequence, at large $h_{\rm UV}$ the energy contained in the three bubbles of the $n=2$ oscillatory wormhole is smaller than the energy contained in the single bubble of the wineglass solution.

\subsection{Neumann for the scalar field/Dirichlet for the gauge field}

In this section, we analyze the generic case where $c_{\rm AdS}\neq -2$ for the running scalar field. In this case, Neumann BCs are imposed that turn the Dirichlet boundary source into a VEV. Concerning the solutions without a running scalar, this does not affect anything and the expressions given in~\ref{para:disc vs simple wormholes} remain unchanged (the counterterms trivially vanish). For the running solutions, this amounts to adding a counterterm and a boundary term to the scalar field action (see \ref{piecewisewineglass} for further details on the symplectic structure of the solution space at infinity). The difference between the on-shell actions before sending the cut-offs to infinity reads
\begin{align}
\Delta \mathcal{S}_{D-\rm{osc(n)}}
&= 2 \mathcal{S}^{D}_E - \mathcal{S}^{\rm{osc(n)}}_E  \nonumber
\\
&= \frac{\mathrm{Vol}(S^3)}{\kappa}\,\mathcal{I}^{\rm Neumann}_{D,\rm{osc(n)}}
   + 24\mathrm{Vol}(S^3)\left(H_{\rm UV}^2
    - |H_{\rm UV}|\,\sqrt{H_{\rm UV}^2-H_{0,n}^2}\right)\,,
\label{eq:DS_disc_conn_n}
\end{align}
where we introduce the following quantity to simplify the presentation
    \begin{align}
        \mathcal{I}^{\rm Neumann}_{D,\rm{osc(n)}}
\equiv & 3\int_{-\tau_{n}}^{\tau_{n}} \mathrm{d}\tau \,a_{n}^3\,\tilde{V}_{\mathrm{AdS}}
   + \left.\partial_{\tau}a_{n}^3\right|_{-\tau_{n}}^{\tau_{n}}
   - 2\left(
        3\int_{0}^{\tau_{D}} \mathrm{d}\tau \,a_D^3\,\tilde{V}_{\mathrm{AdS}}
        + \left.\partial_{\tau}a_D^3\right|^{\tau_D}
     \right) \nonumber
\\
   &+\frac{3}{2\kappa}\int_{\partial\mathcal{M}} {\rm d}^{3}x\sqrt{h}\,\tilde{\phi}\Delta_-H_{\rm AdS}\tilde{\phi}~-\frac{3}{\kappa}\int_{\partial\mathcal{M}}{\rm d}^3x\sqrt{h}\,\tilde{\phi}(\tilde{\phi}'+H_{\rm AdS}\tilde{\phi})\,.
    \end{align}
    \begin{figure}[t!]
    \centering
    \includegraphics[width=1\linewidth]{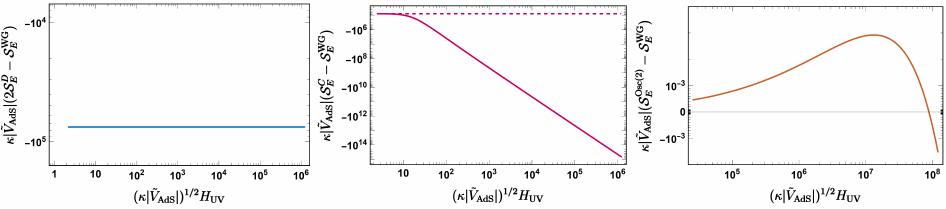}
    \caption{ Difference between on-shell actions as a function of $h_{\rm UV}=(\kappa |\tilde{V}_{\rm AdS}|)^{1/2}\,H_{\rm UV}$. \textit{Neumann} BCs are assumed for the scalar and the Dirichlet BCs for the gauge field in all panels. The scalar source is non-zero, i.e. $A_{\rm AdS}=10^6 \Rightarrow c_{\rm AdS}\neq-2$. \textit{Left}: Disconnected vs wineglass wormhole. \textit{Center}: Wineglass wormhole vs simple connected wormhole. \textit{Right}: n=2 oscillatory wormhole vs wineglass wormhole.}
    \label{fig:action_diff_3}
\end{figure}

\begin{figure}[t!]
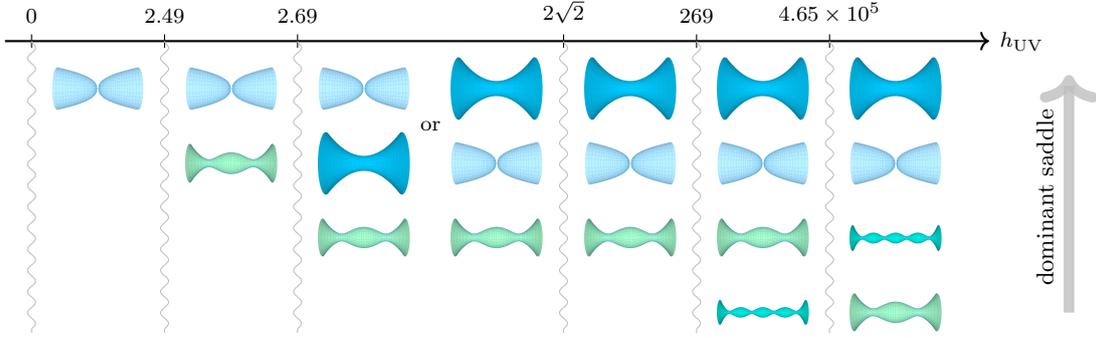

    \centering

\resizebox{\textwidth}{!}{%
\begin{tikzpicture}[x=3.5cm,y=0.95cm]

% equally spaced positions on the x-axis
\def\xA{0.00}
\def\xB{0.55}
\def\xC{1.10}
\def\xD{1.65}
\def\xE{2.20}
\def\xF{2.75}
\def\xG{3.30}

% common vertical positions
\def\yone{-0.55}
\def\ytwo{-1.55}
\def\ythree{-2.55}
\def\yfour{-3.55}
\def\yfive{-4.55}

% sizes
\def\wmain{1.10cm}
\def\wconn{0.92cm}

% vertical labels for connected wormholes
\newcommand{\connlabel}[3]{%
  \node[
    rotate=90,
    font=\scriptsize\itshape,
    inner sep=0pt
  ] at ({#1+0.18},{#2}) {#3};
}

% axis
\draw[->, thick] (-0.08,0) -- (\xG+0.12,0);
\node[above, font=\scriptsize] at (\xG+0.03,0.02) {$h_{\rm UV}$};

% ticks and labels
\foreach \x/\lab in {
\xA/0,
\xB/2.22,
\xC/2.52,
\xD/2.73,
\xE/2.61\times10^4,
\xF/8.8\times10^7
}{
  \draw (\x,0.10) -- (\x,-0.10);
  \node[above, font=\scriptsize] at (\x,0.12) {$\lab$};
}

% curly separator lines
\foreach \x in {\xA,\xB,\xC,\xD,\xE,\xF}{
  \draw[
    gray!50,
    decorate,
    decoration={snake, amplitude=1.5pt, segment length=8pt}
  ] (\x,0) -- (\x,-4.85);
}

% -------------------------------------------------
% 0 -- 2.22 : disc
% -------------------------------------------------
\node at ({(\xA+\xB)/2},\yone)
{\includegraphics[width=\wmain]{Figures_paper_1/worm_disc.png}};

% -------------------------------------------------
% 2.22 -- 2.52 : disc --> wine
% -------------------------------------------------
\node at ({(\xB+\xC)/2},\yone)
{\includegraphics[width=\wmain]{Figures_paper_1/worm_disc.png}};

\node at ({(\xB+\xC)/2},\ytwo)
{\includegraphics[width=\wmain]{Figures_paper_1/worm_wine.png}};

% -------------------------------------------------
% 2.52 -- 2.73 : disc --> conn_L --> conn_S --> wine
% -------------------------------------------------
\node at ({(\xC+\xD)/2},\yone)
{\includegraphics[width=\wmain]{Figures_paper_1/worm_disc.png}};

\node at ({(\xC+\xD)/2},\ytwo)
{\includegraphics[width=\wconn]{Figures_paper_1/worm_conn.png}};
\connlabel{(\xC+\xD)/2}{\ytwo}{large}

\node at ({(\xC+\xD)/2},\ythree)
{\includegraphics[width=\wconn]{Figures_paper_1/worm_conn.png}};
\connlabel{(\xC+\xD)/2}{\ythree}{small}

\node at ({(\xC+\xD)/2},\yfour)
{\includegraphics[width=\wmain]{Figures_paper_1/worm_wine.png}};

% -------------------------------------------------
% 2.73 -- 2.61 x 10^4 : conn_L --> disc --> conn_S --> wine
% -------------------------------------------------
\node at ({(\xD+\xE)/2},\yone)
{\includegraphics[width=\wconn]{Figures_paper_1/worm_conn.png}};
\connlabel{(\xD+\xE)/2}{\yone}{large}

\node at ({(\xD+\xE)/2},\ytwo)
{\includegraphics[width=\wmain]{Figures_paper_1/worm_disc.png}};

\node at ({(\xD+\xE)/2},\ythree)
{\includegraphics[width=\wconn]{Figures_paper_1/worm_conn.png}};
\connlabel{(\xD+\xE)/2}{\ythree}{small}

\node at ({(\xD+\xE)/2},\yfour)
{\includegraphics[width=\wmain]{Figures_paper_1/worm_wine.png}};

% -------------------------------------------------
% 2.61 x 10^4 -- 8.8 x 10^7 : conn_L --> disc --> conn_S --> wine --> osc
% -------------------------------------------------
\node at ({(\xE+\xF)/2},\yone)
{\includegraphics[width=\wconn]{Figures_paper_1/worm_conn.png}};
\connlabel{(\xE+\xF)/2}{\yone}{large}

\node at ({(\xE+\xF)/2},\ytwo)
{\includegraphics[width=\wmain]{Figures_paper_1/worm_disc.png}};

\node at ({(\xE+\xF)/2},\ythree)
{\includegraphics[width=\wconn]{Figures_paper_1/worm_conn.png}};
\connlabel{(\xE+\xF)/2}{\ythree}{small}

\node at ({(\xE+\xF)/2},\yfour)
{\includegraphics[width=\wmain]{Figures_paper_1/worm_wine.png}};

\node at ({(\xE+\xF)/2},\yfive)
{\includegraphics[width=\wmain]{Figures_paper_1/worm_wine_osc.png}};

% -------------------------------------------------
% after 8.8 x 10^7 : conn_L --> disc --> conn_S --> osc --> wine
% -------------------------------------------------
\node at ({(\xF+\xG)/2},\yone)
{\includegraphics[width=\wconn]{Figures_paper_1/worm_conn.png}};
\connlabel{(\xF+\xG)/2}{\yone}{large}

\node at ({(\xF+\xG)/2},\ytwo)
{\includegraphics[width=\wmain]{Figures_paper_1/worm_disc.png}};

\node at ({(\xF+\xG)/2},\ythree)
{\includegraphics[width=\wconn]{Figures_paper_1/worm_conn.png}};
\connlabel{(\xF+\xG)/2}{\ythree}{small}

\node at ({(\xF+\xG)/2},\yfour)
{\includegraphics[width=\wmain]{Figures_paper_1/worm_wine_osc.png}};

\node at ({(\xF+\xG)/2},\yfive)
{\includegraphics[width=\wmain]{Figures_paper_1/worm_wine.png}};

% dominant saddle arrow
\draw[thick, gray!50, line width=4pt, ->, opacity=0.8] (\xG+0.20,-5.00) -- (\xG+0.20,-0.35);
\node[rotate=90, font=\footnotesize, fill=white, inner sep=0pt]
      at (\xG+0.12,-2.65) {dominant saddle};

\end{tikzpicture}%
}

\caption{Dominant saddles in the gravitational path integral across a range of $h_{\rm UV} \equiv (\k |\tilde{V}_{\rm AdS}|)^{1/2}H_{\rm UV}$. Some saddles cease to exist below certain minimum values $h_{\rm UV}^{\rm min}$. The cases shown here match those in figs.~\ref{fig:action_diff} (left) and~\ref{fig:action_diff_3}. he labels \textit{large} and {\textit{small}} denote the two connected wormhole branches. }
\label{fig:saddle_dom_2}
\end{figure}
The second line corresponds precisely to the counterterm and the boundary term. The counterterm is computed in Appendix~\ref{UV-Wineglass}, and the boundary term vanishes on-shell. This is because, when $c_{\rm AdS}\neq -2$, the mode $\tilde{\phi}_+$ (proportional to the source using the Neumann boundary condition) is zero. Note that, in contrast to the preceding section, $c_{\rm AdS}$ is now related to a VEV and thus free to vary, resulting in a four-parameter family which we parametrize in terms of $H_{\rm UV},\,\tilde{V}_{\rm AdS},\,A_{\rm dS},\,A_{\rm AdS}$. After sending the cut-offs to infinity, the difference between the on-shell actions remains finite
\begin{align}
   \Delta \mathcal{S}_{D-\rm{osc(n)}}(H_{\rm UV} ; \tilde{V}_{\rm AdS} ; A_{\rm dS};A_{\rm AdS})=&-\frac{4\text{Vol}(S^3)\,\zeta^{IR\,\,\rm {Neumann}}_{D-\rm{osc(n)}}\left(H_{\rm UV} ;\tilde{V}_{\rm AdS};A_{\rm dS};A_{\rm AdS}\right)}{\kappa\tilde{V}_{\rm AdS}}\nonumber 
\\
    &+24 \text{Vol}(S^3)\left(H_{\rm UV}^2- |H_{\rm UV}|\,\sqrt{H_{\rm UV}^2-H_{0,n}^2}\right)\,.
\label{eq:Delta_S_2}
\end{align}
For completeness, we give the expression of the IR contribution computed in Appendix~\ref{IR-Wingelass} 
\begin{align}
    &\zeta^{IR\,\,\rm {Neumann}}_{D-\rm{osc(n)}}\left(H_{\rm UV} ;\tilde{V}_{\rm AdS};A_{\rm dS};A_{\rm AdS}\right)=1-\Bigg[\frac{1}{4\sqrt{A_{\rm AdS}}} \nonumber
\\
    &\times\Bigg(32H_{\rm AdS}^2\kappa H_{0,n}^2\Pi\left[-\frac{c_{\rm AdS}}{2A_{\rm AdS}};-1\right]+2A_{\rm AdS}\mathcal{E}[-1]-(2A_{\rm AdS}+c_{\rm AdS})\mathcal{K}[-1]\Bigg) \nonumber
\\
    &+\frac{(2n-1)\tilde{V}_{\rm AdS}(c_{\rm dS}+A_{\rm dS})^{1/2}}{16H_{\rm dS}^2}\Bigg(8\mathcal{E}\left[\frac{2A_{\rm dS}}{c_{\rm dS}+A_{\rm dS}}\right]-(2-A_{\rm dS})\mathcal{K}\left[\frac{2A_{\rm dS}}{c_{\rm dS}+A_{\rm dS}}\right]\Bigg)\Bigg]\,.
\end{align}
The asymptotic behavior for large $H_{\rm UV}$ given in the previous section (see \eqref{eq:asymptoticbehavior}) remains unchanged. We emphasize that the difference between the on-shell actions after performing background subtraction remains formally the same as in the case of Dirichlet BCs. In particular, the expression~\eqref{eq:Delta_S_1} can be retrieved by taking the limit $c_{\rm AdS}\to-2$ in~\eqref{eq:Delta_S_2}. 

\paragraph{Summary of results for Neumann BCs:} In figs.~\ref{fig:action_diff} and~\ref{fig:action_diff_3} we present the differences in the on-shell action as one varies the electromagnetic source at the boundary of EAdS. The table shown in fig.~\ref{fig:saddle_dom_2} summarizes the results for $A_{\rm AdS}=10^6,\,A_{\rm dS}=1$. Although the table fig.~\ref{fig:saddle_dom_2} resembles the one for Dirichlet BCs summarized previously in fig.~\ref{fig:saddle_dom}, we shall highlight some striking differences. First, for large values of $A_{\rm AdS}$, we observe that the wineglass wormholes/oscillatory wormholes begin to exist before the simple connected wormholes do. For instance when $A_{\rm AdS}=10^6$, the wineglass wormhole branch starts existing at $h_{\rm UV}\simeq 2.22$, before the simple connected wormhole branch does. In the meantime, the wineglass/oscillatory wormholes are now always subdominant with regard to the disconnected geometry, in contrast to the discussion in the previous section (see \textit{left} panel of fig.\ref{fig:action_diff_3} and fig.~\ref{fig:saddle_dom_2}). The reason is clear: $A_{\rm AdS}$ sets the size of the transverse slices in region (I) for the solutions with running scalar (see~\eqref{eq:background_ans_simple}). In this region the potential is negative and gives a positive contribution to the action of such saddles since $S_{E}^{\rm on-shell}\supset -\int a^3\,V$. We conclude that with Neumann BCs. for the scalar, wineglass wormholes are the dominant (connected) contributions to the gravitational path integral (for our specific model and within the four parameter reduced ansatz), for small values of the E/M source $h_{\rm UV}$ at the AdS boundary, and can give rise to dominant inflationary cosmologies.

\section{Discussion}
\label{sec:Discussions}

The main findings of our paper can be summarized as follows:

\begin{enumerate}
    \item \textit{Analytic control of a rich set of (wormhole) saddles of the gravitational path integral}: Within a general class of Einstein--Scalar--Radiation systems and using an appropriate Euclidean FLRW ansatz, we constructed and classified a broad family of asymptotically Euclidean AdS solutions, including single-boundary geometries, simple wormholes, wineglass wormholes, and quasi-oscillatory configurations, for all choices (Dirichlet and Neumann) of scalar and gauge field boundary conditions.

    \item \textit{Background subtraction in Euclidean FLRW}: We developed a consistent implementation of background subtraction for the asymptotically Euclidean AdS FLRW saddles, allowing for a controlled comparison of the on-shell actions across different boundary conditions. 

    \item \textit{Semiclassical saddle competition in the gravitational path integral and Cosmological ramifications}: By explicitly evaluating the on-shell action differences, we identified the parameter regions in which different saddles dominate the gravitational path integral. As we explain in section~\ref{sec:wavefunction}, the connected two boundary saddles can be used to estimate ratios of probabilities for different cosmological outcomes---either of a cosmology that crunches (usual wormholes), or of a cosmology that is expanding (wineglass and quasi-oscillatory wormholes). In particular, \textbf{wineglass wormholes}~\cite{Betzios:2024oli,Betzios:2024zhf} can become the leading connected contribution for sufficiently small values of the electromagnetic field at the UV boundary, and under analytic continuation they naturally give rise to expanding cosmological spacetimes with an early inflationary phase.
\end{enumerate}

From the class of backgrounds that we constructed, two deserve some further discussion. The wineglass AdS wormholes were first devised~\cite{Betzios:2024oli,Betzios:2024zhf} as a mean to surpass the probabilistic issues inflicting the Hartle-Hawking no-boundary proposal (see~\cite{Lehners:2023yrj,Maldacena:2024uhs} for reviews) and the resulting trivial (one-dimensional) Hilbert space for the closed cosmologies. For these reasons, a clear holographic understanding of closed Universes within the no-boundary proposal has been elusive so far (see though~\cite{Liu:2025cml,Antonini:2025ioh,Harlow:2026hky,Abdalla:2026mxn,Nomura:2026igt,Zhao:2026mpl} for some recent works with a perspective emphasizing the role of observers in this setting). On the other hand, the AdS wormhole proposal of~\cite{Betzios:2024oli,Betzios:2024zhf} can give rise to a non-trivial Hilbert space for (expanding/inflating) cosmologies, through the field theory dual living on the AdS asymptotic boundary/ies, and realised semi-classically via the different competing saddles and asymptotic boundary conditions for the various bulk fields. 

Oscillatory wormholes were first constructed in~\cite{Halliwell:1989pu} and recently discussed in~\cite{Aguilar-Gutierrez:2023ril} using models containing axionic fields. The axion wormhole models seem to suffer from the issue that the flat direction of the (classically) non-existent axion potential (one just considers a fixed positive cosmological constant in these models), leads to the potential presence of an unbound infinite number of oscillations for such solutions and to a badly divergent path integral (an avatar of the notorious conformal mode problem of the Euclidean gravitational path integral). As we found in section~\ref{sec:piecewise}, a breaking of the shift symmetry of the potential precluded this issue, giving rise to an upper bound on the number of oscillations.

 In the context of axionic wormholes, the situation is slightly more complicated, due to the presence of a remnant discrete shift symmetry\footnote{We wish to thank the anonymous referee for emphasizing to us the presence of the discrete shift symmetry in the axionic case.}. In particular it is well known that non-perturbative instanton or gravitational effects (see~\cite{Hebecker:2018ofv} for a review)
lead to an effective potential for the axionic field $\alpha$ of the form
\be
V_{\rm eff}(\alpha) = \sum_n e^{-S_n} \cos \left( \frac{n \alpha}{f} + \delta_n \right) \, ,
\ee
where $S_n$ is the on-shell action of the instanton-like configuration, $f$ is the axion coupling strength ($\mathcal{L}(\alpha) \sim f^2 (\partial \alpha)^2$) and $\delta_n$ are model dependent coefficients. Typically one still finds the presence of a remnant discrete symmetry in the axion potential (what remains of the original shift symmetry after such non-perturbative effects are taken into account)\footnote{When this discrete shift symmetry is absent then the issue of the infinite number of oscillations is completely resolved as in the models studied in this work.}. Nevertheless, the axion dynamics changes dramatically in the presence of such a potential. In particular while in the absence of the potential the axion saddle point equation (in our EFLRW ansatze) is a total derivative with the simple solution ($Q$ is the axion charge)
\be
\alpha'(\tau) = \frac{Q}{a^3(\tau)} \, ,
\ee
and the axion is completely determined from the solution of the metric up to an integration constant,
in the presence of a non-trivial potential it is a fully dynamical
equation of the form
\be
\alpha'' + 3 \frac{a'}{a} \alpha' + \frac{{\rm d} V_{\rm eff}(\alpha)}{ {\rm d} \alpha} = 0  \, ,
\ee
where the derivative of the scale factor acts as an (anti)-friction term. Since this equation is no longer conservative, it is not clear whether the axion can actually form a periodic motion with an infinite number of periods (for this to happen the average axion charge should remain constant per period). In other words, while the discrete shift symmetry can be a symmetry of the Lagrangian, it is not clear whether it is preserved in the (most dominant) saddle point solution of the EOMs, and a more careful analysis is needed to ascertain the fate of the oscillatory axion solutions once such non perturbative corrections are taken into account.

We conclude this manuscript by presenting a list of topics that deserve further analysis. First, it would be interesting to study ours or similar models for more general parameters (even numerically), to explore and carve out the parameter regimes in which wineglass or other oscillatory wormholes that can give rise to an early inflationary cosmological outcome are the dominant saddles of the gravitational path integral.
As a variant of this idea, magnetic types of wormholes can also be found in braneworld scenarios~\cite{Antonini:2024bbm}. It would be interesting to consider more general types of wormholes in such scenarios and verify whether they can lead to inflating braneworld cosmologies.

The family of models that we study in this work constitutes the first instance where a class of models that admits such an intricate variety of wormholes and disconnected saddles, is analysed and discussed in the literature in higher dimensions. In two spacetime dimensions there do exist simple models that contain a similar plethora of different kinds of gravitational saddles (i.e. the ``gravitational Sine-Gordon model''~\cite{Betzios:2025eev}) crucially relying again on a scalar matter potential that can take both positive and negative values. Since these two dimensional models also admit a matrix model description, it is very important to analyse them in great detail, to learn about the non-perturbative and microscopic physics in such rich instances of the gravitational path integral.

At the same time, it is also of utmost importance to perform a careful study of cosmological correlators using these saddles, to understand the similarities and differences with the no-boundary proposal and ascertain their potential role in a more phenomenological setting~\cite{Betzios2025toappear}.

Finally, since the class of models that we consider in this work combines elements that are generic in flux compactifications in string theory (a non-trivial scalar potential containing both AdS/negative  and dS/positive local minima as well as local positive maxima) together with the presence of non-trivial fluxes, we believe that it is worthwhile to extend our study to models that arise in more realistic string compactifications. In particular the recent work~\cite{AbdusSalam:2025twp} discusses such cases of coexisting vacua, which can also realise our proposal within the context of supergravity.

%%%%%%%%%%%%%%%%%%%%%%%%%%%%%%%%%%%%%%%%%%%%%%%%%%%%%%%%%%

%%%%%%%%%%%%%%%%%%%%%%%%%%%%%%%%%%%%%%%%%%%%%%%%%%%%%%%%%%
\section*{Acknowledgments}

We wish to thank Elias Kiritsis for some motivational remarks, Kostas Skenderis for correspondence and the participants of the \href{https://indico.ictp.it/event/10851/overview}{workshops on Approaches to Quantum Gravity in de Sitter space} and \href{https://all-lambda-holography.fzu.cz/}{$\forall \Lambda$ Holography}, for discussions on related topics.
\\
\\
P.B. acknowledges financial support from the European Research Council (grant BHHQG-101040024), funded by the European Union.
\\
The work of IDG was supported by the Estonian Research Council grants PSG1132, TARISTU24-TK10, TARISTU24-TK3, and the CoE program TK202 ``Foundations of the Universe'’. This article is
based upon work from COST Action CA23130, ``Bridging high and low energies in search of Quantum Gravity" (BridgeQG).\\
Views and opinions expressed are those of the authors only and do not necessarily reflect those of the European Union or the European Research Council. Neither the European Union nor the granting authority can be held responsible for them.\\
Finally, IDG wishes to thank the CERN Theoretical Physics Department for hospitality while this research was being carried out.

% \bibliographystyle{JHEP}
% \bibliography{biblio}
% \end{document}

\appendix

%%%%%%%%%%%%%%%%%%%%%%%%%%%%%%%%%%%%%%%%%%%%%%%%%%%%%%%%%%%%%

\section{On-shell Euclidean action for the \texorpdfstring{$U(1)^3$}{U(1)3} model} \label{sec:eucl_action in U(1)^3}
Let's first recall the on-shell Euclidean action 
\begin{equation}
\label{eq:background_app}
    \mathcal{S}_E = \int {\rm d}^4x \sqrt{g_E} \left(-\frac{1}{2\kappa}\mathcal{R} + \frac12 (\partial_\mu{\phi})^2 +V(\phi)\right) +\frac12\int F^{(I)}\wedge \star  F^{(I)} -\frac{1}{\kappa}\int{\rm d}^3x\sqrt{h}K\,.
\end{equation}
First, taking the trace of the Einstein-equation, one obtains
\begin{equation}
G_{\mu\nu}=\kappa(T^{\phi}_{\mu\nu}+T_{\mu\nu}^{\text{\rm rad}}) \Rightarrow G=-\mathcal{R}=\kappa(T^{\phi}+T^{\text{\rm rad}})\,
\end{equation}
The electromagnetic field corresponds to conformal matter in 4d. Hence, the trace of the stress tensor vanishes identically. Concerning the scalar field
\begin{equation}
    T^{\phi}_{\mu\nu}=\partial_\mu{\phi}\partial_\nu{\phi}-g_{\mu\nu}\left(\frac{1}{2}(\partial_\lambda\phi)^2+V(\phi)\right) \Rightarrow T^{\phi}=-(\partial_\mu\phi)^2-4V\,,
\end{equation}
which leads to 
\begin{equation}\label{Einsteinequation}
    -\frac{\mathcal{R}}{2\kappa}=-\frac{1}{2}(\partial_\m\phi)^2-2V(\phi)\,,
\end{equation}
In the FLRW metric, the Gibbons-Hawking-York (GHY) term becomes simply
\begin{equation}
    S_{\text{\rm GHY}}=-\frac{\text{Vol}(S^3)}{\kappa}\partial_{\tau}a^3{\Big|_{-\tau_{0}}^{\tau_{0}}}\,.
\end{equation}
Combining these results, we obtain
\begin{equation}
\label{generalgrounds}
    \mathcal{S}^{\text{on-shell}}_E = -\frac{\text{Vol}(S^3)}{\kappa}\left[\int_{-\tau_{0}}^{\tau_{0}} {\rm d}\tau a^3\kappa V(\phi)+\partial_{\tau}a^3\Big|_{-\tau_{0}}^{\tau_{0}}\right]+6\,\text{Vol}(S^3)\int_{-\tau_{0}}^{\tau_{0}} {\rm d}\tau\left(aH'^2+\frac{4H^2}{a}\right)\,.
\end{equation}

Switching from Dirichlet to Neumann BCs (alternate quantization) gives rise to additional boundary terms for the gauge and scalar fields of the form (for a single boundary component)
\begin{align}
S_{\partial \mathcal{M}}^{\rm U(1)^3} &= - \int_{\partial \mathcal{M}} A_I \wedge \star F_I = - \sum_{I=1}^3 \int_{\partial \mathcal{M}} d^3 x \sqrt{h} n_\mu F^{\m \n}_I A_\n^I = - 12 \text{Vol}(S^3)  a(\tau) H'(\tau) H(\tau)\vert_{\partial \mathcal{M}} \, ,
\\
  S_{\partial \mathcal{M}}^{\rm ren.} &= - \frac{3}{\kappa} \int_{\partial \mathcal{M}} d^3 x \sqrt{h}  \tilde \phi_-  \tilde \pi_{\Delta_+}^{\rm ren.}.
\end{align}

All the fields have to be on-shell in the above formulae. Note that the E/M part remains finite while the first part of~\eqref{generalgrounds}, which we call the ``gravitational part", blows up. In the following we compute the on-shell action for the wormhole geometries introduced in the main text. 

\paragraph{Disconnected solution -}
In this case the inflaton does not run and the potential is constant and plays the role of a negative cosmological constant $\kappa\,V(\phi)=3\tilde{V}_{\rm AdS}$. The solutions to the electromagnetic potential and the scale factor are given in \eqref{eq:aH_sol1}. In this specific case, we can explicitly perform the integral and express the on-shell action as a function of $a_D=a_D(\tau_D)$.
\begin{align}
\label{on-shelldisc}
    \mathcal{S}^{D}_E = & -\frac{\text{Vol}(S^3)}{\kappa}\left[\frac{2 - \sqrt{1- \tilde{V}_{\rm AdS} a_D^2}(2 + \tilde{V}_{\rm AdS} a_D^2)}{\tilde{V}_{\rm AdS}} +3\sqrt{1-\tilde{V}_{\rm AdS}a_D^2}a_D^2\right] \nonumber
\\
   & +12\text{Vol}(S^3)H_{\rm UV}^2x_D^4\,,
\end{align}
where $x_D = \tanh \left(\sqrt{-\tilde{V}_{\rm AdS}}\,\tau_D/2\right)$, that tends to one at infinity. Here we introduced the regulator $\tau_D$ in reference to the disconnected solution.

\paragraph{Wormhole without a running scalar -}
The inflaton is again non running and the potential is still a constant and plays the role of a negative cosmological constant. Introducing the regulator $\tau_C$, one obtains
\begin{equation}\label{eq:actionconnected}
\begin{aligned}
  \mathcal{S}^{\text{C}}_E =& -\frac{\text{Vol}(S^3)}{\kappa}\left[3\int_{-\tau_{C}}^{\tau_{C}} {\rm d}\tau \,a_C^3\,\tilde{V}_{\rm AdS}+\partial_{\tau}a_C^3\Big|_{-\tau_{C}}^{\tau_{C}}\right]\,\\
  &+24\text{Vol}(S^3)|H_C(\tau_C)|\sqrt{H_C(\tau_C)^2-H_{0,C}^2}\,,
\end{aligned}
\end{equation}
where the functions $a_{C}(\tau),~H_{C}(\tau)$ correspond to the solution of the equation of motion in this case, see eqs.~\eqref{eq:scale_connected} and~\eqref{connectedgaugesln}. The gravitational part is quite more involved. It will be treated in the next subsection of this Appendix~\ref{subtraction1}.

\paragraph{Piecewise wineglass wormhole -}

For the wineglass wormhole defined in a piecewise function, it is natural to split the integration range to the regions ($I, II$). Exploiting the $\mathbb{Z}_2$ symmetry, we can write
\bea\label{wineglassconnected}
    \mathcal{S}^{\text{WG} (II)}_E &=& -\frac{3 \text{Vol}(S^3)}{\kappa}\left[\int_0^{-\tau_{\rm min}} {\rm d}\tau \, 2 a^3\,\tilde{V}(\tau) - 4\kappa \int^{-\tau_{\rm min}}_0 {\rm d}\tau\,\left(aH'^2+\frac{4H^2}{a}\right)  \right] \, , \nn \\
    \mathcal{S}^{\text{WG} (I,I')}_E &=& -\frac{\text{3 Vol}(S^3)}{\kappa}\left[\int_{-\tau_{\rm min}}^{\tau_{WG}} {\rm d}\tau \, 2 a^3\,\tilde{V}(\tau)+ \frac{2}{3} \partial_{\tau}a^3\Big|^{\tau_{WG}} - 4\kappa  \int_{-\tau_{\rm min}}^{\tau_{WG}} {\rm d}\tau\,\left(aH'^2+\frac{4H^2}{a}\right) \right] \, , \nn \\
      \mathcal{S}^{\text{WG}}_E &=& \mathcal{S}^{\text{WG} (I,I')}_E + \mathcal{S}^{\text{WG} (II)}_E\, ,
\eea
where one has to substitute the corresponding solutions in regions $(I)$ and $(II)$, see eqs.~\eqref{eq:background_ans_simple}. As in the previous case, the computation of the gravitational part will be further discussed in Appendix~\ref{UV-Wineglass}.

\paragraph{Oscillatory wormholes -} For the oscillatory wormholes the result is similar to the result for the piecewise wineglass wormhole, with the only difference being that one has to extend the integral in the region $(II)$ in order to cover $n$-periods of the oscillatory scale factor and potential.

\subsection{Subtraction of the gravitational pieces}\label{subtraction1}
We now compute the large–cut-off ($R$) asymptotic expansion of the gravitational action for each saddle. For any given geometry there is a near–boundary, ultraviolet (UV) contribution: a universal function of the cut-off radius $R$ that cancels upon performing background subtraction between any two saddles. The remaining constant piece captures the IR physics of the wormhole geometry near the throat (resp. the geometry near the tips for the disconnected saddle). The Gibbons–Hawking–York term, of course, does not contribute to this IR part. In what follows, we verify the universality of the UV divergences and, wherever possible, compute the constant contribution analytically.

\paragraph{Disconnected geometry -}
Let's first study the disconnected geometry. Our starting point is the first part of \eqref{on-shelldisc}. We expand in order of $R=a_D(\tau_D)$ keeping only the part that remains when setting the cut-off radius $R$ to infinity. It yields
\begin{equation}\label{disconnectedgrav}
    2\,\mathcal{S}^{\text{D}}_{\text{Grav}}(R) =-\frac{2\text{Vol}(S^3)\times\sqrt{-\tilde{V}_{\rm AdS}}}{\kappa}\left[2R^3-\frac{3R}{\tilde{V}_{\rm AdS}}-\frac{2}{\left(\sqrt{-\tilde{V}_{\rm AdS}}\right)^3}+\mathcal{O}\left(\frac{1}{R}\right)\right]\,.
\end{equation}
The Gibbons-Hawking term does not yield any finite contribution while the integral exhibits a constant term independent of $a_D(\tau_D)=R$. This constant term arises from the contribution to the integral in the region of vanishing scale factor $a_D$ and hence we understand that it is an IR contribution. The diverging part concerns the universal UV piece: it should be the same for all the geometries for them to be comparable.

\paragraph{Wormhole without a running scalar (UV part) -}
Now we wish to do the same for the simple wormhole without any running scalar. We first focus on the diverging part of the on-shell action. For that, we shall expand the terms in powers of the cut-off radius $R$. To simplify notations, we introduce $\tilde{\tau}=\sqrt{-\tilde{V}_{\rm AdS}}\tau$. From \eqref{eq:actionconnected}, we can isolate the gravitational contribution to the on-shell action
\begin{subequations}
    \begin{align}
        a_C(\tau)&=\sqrt{\frac{1}{2\tilde{V}_{\rm AdS}}-\frac{(1+4\tilde{\rho}^E_{\rm rad}\tilde{V}_{\rm AdS})^{1/2}}{4\tilde{V}_{\rm AdS}}\left[e^{2\tilde{\tau}}+e^{-2\tilde{\tau}}\right]}\,
\\
        \mathcal{S}^{\text{C}}_{\text{Grav}} &= -\frac{\text{Vol}(S^3)}{\kappa}\left[3\int_{-\tau_{C}}^{\tau_{C}} {\rm d}\tau a_C^3(\t)\tilde{V}_{\rm AdS}\,+\partial_{\tau}a_C^3\Big|_{-\tau_C}^{\tau_C}\right]\,.
    \end{align}
\end{subequations}

To extract the asymptotic expansion of $a_C^3$, it is first useful to obtain the asymptotic expansion of $a_C$ up to $\exp\left(-\sqrt{\tilde{V}_{\rm AdS}}\tau\right)$ as $\tau\to\infty$. One easily finds
\begin{equation}
    a_C(\tau)=\frac{(1+4\tilde{\rho}^E_{\rm rad}\tilde{V}_{\rm AdS})^{1/4}}{2\sqrt{-\tilde{V}_{\rm AdS}}}\left[e^{\tilde{\tau}}-\frac{e^{-\tilde{\tau}}}{(1+4\tilde{\rho}^E_{\rm rad}\tilde{V}_{\rm AdS})^{1/2}}+\mathcal{O}\left(e^{-3\tilde{\tau}}\right)\right]\,.
\end{equation}
With this expression we can obtain the UV pieces
\begin{align}
    a_C^3(\tau)&=\frac{(1+4\tilde{\rho}^E_{\rm rad}\tilde{V}_{\rm AdS})^{3/4}}{8\left(\sqrt{-\tilde{V}_{\rm AdS}}\right)^3}\left[e^{3\tilde{\tau}}-\frac{3e^{\tilde{\tau}}}{(1+4\tilde{\rho}^E_{\rm rad}\tilde{V}_{\rm AdS})^{1/2}}+\mathcal{O}\left(e^{-\tilde{\tau}}\right)\right]\,, \nonumber
\\
    \partial_\tau a_C^3(\tau)&=\frac{3\,(1+4\tilde{\rho}^E_{\rm rad}\tilde{V}_{\rm AdS})^{3/4}}{8\left(\sqrt{-\tilde{V}_{\rm AdS}}\right)^2}\left[e^{3\tilde{\tau}}-\frac{e^{\tilde{\tau}}}{(1+4\tilde{\rho}^E_{\rm rad}\tilde{V}_{\rm AdS})^{1/2}}+\mathcal{O}\left(e^{-\tilde{\tau}}\right)\right]\,, \nonumber
\\
    \int_{0}^{\tau_C}{\rm d}\tau\,a_C^3(\tau)&=\frac{(1+4\tilde{\rho}^E_{\rm rad}\tilde{V}_{\rm AdS})^{3/4}}{8\left(\sqrt{-\tilde{V}_{\rm AdS}}\right)^4}\left[\frac{e^{3\tilde{\tau}}}{3}-\frac{3 e^{\tilde{\tau}}}{(1+4\tilde{\rho}^E_{\rm rad}\tilde{V}_{\rm AdS})^{1/2}}\right] + \text{const.}\,.
\end{align}
Note that in the last line there is a $\mathcal{O}(1)$ term corresponding to the IR part of the geometry. To obtain it, one has to keep higher order terms in the expansions, or to perform the integral exactly. We postpone this issue for the next paragraph and focus on the UV part. Expressing the Gibbons-Hawking-York term and the integral as expansions of the scale factor itself, we obtain
\begin{subequations}
\begin{align}
\label{expansionina}
 \partial_\tau a_C^3 &= 3a_C^3\sqrt{-\tilde{V}_{\rm AdS}}+\frac{3a_C}{2\sqrt{-\tilde{V}_{\rm AdS}}}+\mathcal{O}\left(1/a_C \right)\,,
 \\
 \int_{0}^{\tau_C} {\rm d}\tau\,a_C^3(\tau) &= \frac{1}{\sqrt{-\tilde{V}_{\rm AdS}}}\left[\frac{a_C^3}{3}+\frac{a_C}{2\tilde{V}_{\rm AdS}}\right] + \text{const.}\,.
\end{align}
 \end{subequations}

After substituting $a_C(\tau_C)=R$ and summing the two contributions computed just above, one obtains the gravitational on-shell action as a function of the cut-off radius
\begin{equation}\label{connectedgrav}
    \mathcal{S}^{\text{C}}_{\text{Grav}}(R) =-\frac{\text{Vol}(S^3)\times2\sqrt{-\tilde{V}_{\rm AdS}}}{\kappa}\left[2R^3-\frac{3R}{\tilde{V}_{\rm AdS}}+\mathcal{O}\left(1\right)\right]\,.
\end{equation}
If one compares~\eqref{disconnectedgrav} and~\eqref{connectedgrav}, we can see that the UV pieces are exactly the same and will exactly cancel when computing the differences of the on-shell actions.

\paragraph{Wormhole without a running scalar (IR part) -}

The simplest way to obtain it, is to compute exactly the following integral
\begin{align}
\label{FC}
    F_C(\tau_C)&\equiv\int_{0}^{\tau_C} {\rm d}\tau\,a_C^3(\tau) \nonumber
\\
    &\,\,=\frac{1}{2^{3/2}\tilde{V}_{\rm AdS}^2}\int_{0}^{\tau_C}{\rm d}\tau\sqrt{-\tilde{V}_{\rm AdS}}\left[\left(1+4\tilde{\rho}^E_{\rm rad}\tilde{V}_{\rm AdS}\right)^{1/2}\cosh\left(2\sqrt{-\tilde{V}_{\rm AdS}}\tau\right)-1\right]^{3/2} \nonumber
\\
    &\,\,=\frac{1}{2^{3/2}\tilde{V}_{\rm AdS}^2}\int_{0}^{\tilde{\tau}_C}{\rm d}\tilde{\tau}\left[\alpha\cosh\left(2\tilde{\tau}\right)-1\right]^{3/2}\,,
\end{align}
where we define the dimensionless parameter $\alpha=(1+4\tilde{\rho}^E_{\rm rad}\tilde{V}_{\rm AdS})^{1/2}>1$. Let's forget about the prefactor and focus on the dimensionless quantity function of $\alpha$ which we denote by
\begin{equation}\label{GC}
    G(\tilde{\tau_C},\alpha)=\int_{0}^{\tilde{\tau}_C}{\rm d}\tilde{\tau}\left[\alpha\cosh\left(2\tilde{\tau}\right)-1\right]^{3/2}\,.
\end{equation}
Importantly, the $\alpha=1$ limit should correspond to the disconnected geometry. We can directly perform the integral yielding $\frac{4\sqrt{2}}{3}$ as a constant independent of $\tilde{\tau}_C$. In general, after integration, we have the exact formula
\begin{align}
\label{completeGC}
G(\tilde{\tau}_C,\alpha)
=& \frac{1}{3\sqrt{\alpha\cosh(2\tilde{\tau}_C)-1}}
\Biggl[
  4\,i\,\sqrt{(\alpha-1)\bigl(\alpha\cosh(2\tilde{\tau}_C)-1\bigr)}\,
  \mathcal{E}\!\left(i\tilde{\tau}_C,\frac{2\alpha}{\alpha-1}\right)
\\
&-i\,(\alpha^2-1)\sqrt{\frac{\alpha\cosh(2\tilde{\tau}_C)-1}{\alpha-1}}\mathcal{F}\!\left(i\tilde{\tau}_C,\frac{2\alpha}{\alpha-1}\right)+\alpha(\alpha\cosh(2\tilde{\tau}_C)-1)\sinh(2\tilde{\tau}_C)
\Biggr]\,, \nonumber
\end{align}
where $\mathcal{F}(\psi|m),~\mathcal{E}(\psi|m)$ denote the incomplete elliptic integral of the first and the second kind (see Appendix~\ref{incompleterecipe}). We are only interested in the constant part of the asymptotic expansion, since we already know that the diverging parts will cancel. This contribution comes from the constant parts of the elliptic functions $\mathcal{E,F}$. The incomplete elliptic function of the first kind converges as $\tilde{\tau}_C$ goes to infinity while that of the second kind diverges ( see Appendix~\ref{asymptoticellipticrecipe}). Using \ref{asymptoticellipticrecipe} and focusing on the constant part of $\lim_{\tilde{\tau}_C\to\infty}G(\tilde{\tau}_C,\alpha)$, one finds
\begin{equation}   \lim_{\tilde{\tau}_C\to\infty}G(\tilde{\tau}_C,\alpha)\Big|_{\text{const}}=\frac{4\sqrt{\alpha-1}\left[\mathcal{E}(-\frac{\alpha+1}{\alpha-1})-\mathcal{K}(-\frac{\alpha+1}{\alpha-1})\right]}{3}+\frac13\sqrt{\alpha-1}(\alpha+1)\mathcal{K}\left(-\frac{\alpha+1}{\alpha-1}\right)\,.
\end{equation}
We can check using the basic asymptotic formulae given in Appendix~\ref{asymptoticellipticrecipe}, that when $\alpha\to1$ (which is equivalent to $\tilde{\rho}_{\rm rad}^{E} = 0$), it goes to the correct value $4\sqrt{2}/3$. Hence, we exactly recover the correct term in the limit where the wormhole throat closes off to give back the product of disconnected geometries. Once more the constant piece of the gravitational part of the on-shell action is associated with the IR part of the geometry.

\paragraph{Piecewise wineglass wormhole (region \textbf{I,I'}/UV part) -}\label{UV-Wineglass}

We start by showing the cancellation of the UV part. For that, we need the asymptotic expansion of $a,~a^3,~\partial_\tau a^3,~\tilde{V}$, $~a^3\tilde{V}$. We focus on the region (I') for which $\t^{\textbf{I}'}>0$ since the ansatz is $\mathbb{Z}_2$ symmetric. One finds (using the complete expression of the scale factor~\eqref{eq:background_ans_simple} and the potential~\eqref{eq:full_V_I})
    \begin{align}       a_{WG}(\tau)&=\frac{\sqrt{A_{\rm AdS}}\,e^{\t^{\textbf{I}'}/2}}{2\sqrt{2}H_{\rm AdS}}+\mathcal{O}\left(e^{-\t^{\textbf{I}'}/{2}}\right)\,, \nonumber
\\
        \partial_{\tau}a_{WG}^3(\t)&=\frac{3}{16}\frac{A_{\rm AdS}^{3/2}e^{3\t^{\textbf{I}'}/{2}}}{\sqrt{2}H_{\rm AdS}^2}+\frac{3c_{\rm AdS}\sqrt{A_{\rm AdS}}}{16\sqrt{2}H_{\rm AdS}^2}e^{\t^{\textbf{I}'}/{2}}+\mathcal{O}\left(e^{-\t^{\textbf{I}'}/{2}}\right)\, ,\nonumber
\\
        \int_{-\tau_{\rm min}}^{\tau_{WG}}{\rm d}\tau a_{WG}^3(\t)\tilde{V}_{I'}(\t)&=-\frac{A_{\rm AdS}^{3/2}e^{3\t^{\textbf{I}'}/{2}}}{48\sqrt{2}H_{\rm AdS}^2}+\frac{(16-c_{\rm AdS})\sqrt{A_{\rm AdS}}}{48\sqrt{2}H_{\rm AdS}^2}e^{\t^{\textbf{I}'}/{2}}+\rm{const.}\,.
    \label{eq:limits_worm}
    \end{align}
We can now express everything directly in terms of the asymptotic form of the scale factor itself,
\begin{subequations}
    \begin{align}
        \partial_\tau a_{WG}^3(\t)&=3\sqrt{-\tilde{V}_{\rm AdS}}a_{WG}^3-\frac{3c_{\rm AdS}}{4\sqrt{-\tilde{V}_{\rm AdS}}} a_{WG}  + \mathcal{O}(1/a_{WG})\,,
\\
        \int_{-\tau_{\rm min}}^{\tau_{WG}} {\rm d}\tau a_{WG}^3(\t)\tilde{V}_{I'}(\t)&=-\frac{\sqrt{-\tilde{V}_{\rm AdS}}}{3} a_{WG}^3 + \frac{8+c_{\rm AdS}}{12\sqrt{-\tilde{V}_{\rm AdS}}}a_{WG} + {\rm const.}\,.
    \end{align}
\end{subequations}
Combining the two expressions above, one obtains the UV contribution to the gravitational action,
\begin{equation}
    S_E^{WG,\,UV}=-\frac{2\text{Vol}(S^3)\sqrt{-\tilde{V}_{\rm AdS}}}{\kappa}\left(2R^3-\frac{3R}{\tilde{V}_{\rm AdS}}\left(\frac{4-c_{\rm AdS}}{6}\right)\right)\,,
\label{eq:S_wg_reg}
\end{equation}
where $R= a(\t_{WG} \rightarrow \infty)$.
In the absence of a scalar source with Dirichlet BCs, that is for $c_{\rm AdS}=-2$, the coefficient of the term linear in $R$ coincides with the corresponding coefficient in eqs.~\eqref{disconnectedgrav} and~\eqref{connectedgrav}. If a scalar source is present, the linear divergence is modified and must be cancelled by the appropriate counterterm. The required counterterm, given by eq.~\eqref{eq:scalar_ct}, reads
\begin{align}
S_{ct.}& =  \frac{3}{2\k} \int_{\partial \mathcal{M}} {\rm d}^3 x \sqrt{h} \tilde{\phi} \Delta_- H_{\rm AdS} \tilde{\phi}=
 \frac{3}{2\k} 2\text{Vol}(S^3) a^3(\t_{WG}) H_{\rm AdS}\tilde{\phi}^2(\t_{WG}) \nonumber
 \\
 & = - \frac{2\text{Vol}(S^3)\sqrt{-\tilde{V}_{\rm AdS}}}{\k} \left( - \frac{3R}{\tilde{V}_{\rm AdS} } \left( \frac{2+c_{\rm AdS}}{6} \right) \right)\,,
 \label{eq:ct_action_scalar}
\end{align}
which precisely removes the extra $R$-dependent contribution generated by the scalar profile.

After adding this counterterm, the sum of eqs.~\eqref{eq:S_wg_reg},~\eqref{eq:ct_action_scalar} and boundary term for the scalar (see eq.~\eqref{eq:boundarytermscalar}) which vanishes (see Appendix~\ref{appendix:bnry_scalar}) yields the properly renormalized UV action for the wineglass geometry in the alternative quantization scheme (Neumann BCs). The result can then be compared directly with the actions of the disconnected and simple wormholes (constant scalar) in eqs.~\eqref{disconnectedgrav} and~\eqref{connectedgrav}.

\paragraph{Piecewise wineglass wormhole (IR part) -}
\label{IR-Wingelass}
In region (II), the integral $\int {\rm d} \t a^3(\t) \tilde{V}(\t)$ can be evaluated analytically using eqs.~\eqref{eq:background_ans_simple} and~\eqref{eq:VII_r3}, yielding
\begin{align}
\label{eq:grav_II_wg}
\int^{-\t_{\rm min}}_0 {\rm d} \t a(\t)^3 \tilde{V}(\t) = &  -\frac{2H_{\rm dS}\t_{\rm min}}{\pi} \frac{\left(c_{\rm dS}+A_{\rm dS}\right)^{1/2}}{24H_{\rm dS}^2}\bigg(8\mathcal{E}\left[\frac{2A_{\rm dS}}{c_{\rm dS}+A_{\rm dS}}\right] \nonumber
\\
& -(2-A_{\rm dS})\mathcal{K}\left[\frac{2A_{\rm dS}}{c_{\rm dS}+A_{\rm dS}}\right] \bigg)\,.
\end{align}
In regions (I) and (I$'$), the computation is considerably more involved, and we therefore present a more detailed analysis in the following.
We start from the integral $\int_{-\t_{\rm min}}^{\t_{WG}} {\rm d} \t a^3(\t) \tilde{V}(\t)$ where we substitute the expressions~\eqref{eq:background_ans_simple} and~\eqref{eq:VII_r}. The first step is to decompose the potential in the basis $\cosh^{\alpha}(\bar{\tau}^{\textbf{I}'})$. Afterwards, this
requires the evaluation of 
\begin{equation}
    \begin{aligned}
\mathcal{I}_{\alpha}=\int_0^{\infty}\frac{{\rm d}\bar{\tau}^{\textbf{I}'}\cosh^{\alpha}(\bar{\tau}^{\textbf{I}'})}{\sqrt{A_{\rm AdS}}\cosh(\bar{\tau}^{\textbf{I}'})-\sqrt{A_{\rm AdS}-A_{\rm dS}}}\Bigg|_{\text{const.}}\,, \qquad \alpha\in\{5/2,3/2, ... , -7/2\}\,.
    \end{aligned}
\end{equation}
The "const." at the end means that we shall discard the divergent part of the integrals. By performing a partial fraction decomposition, all these integrals can be related, so that it suffices to compute only one of them. Indeed,
\begin{align}
\label{magic}
    \mathcal{I}_{\alpha}=&\sqrt{\frac{A_{\rm AdS}-A_{\rm dS}}{A_{\rm AdS}}} \int_0^{\infty}\frac{{\rm d}\bar{\tau}^{\textbf{I}'}\cosh^{\alpha-1}(\bar{\tau}^{\textbf{I}'})}{\sqrt{A_{\rm AdS}}\cosh(\bar{\tau}^{\textbf{I}'})-\sqrt{A_{\rm AdS}-A_{\rm dS}}}\Bigg|_{\text{const.}}  \nonumber
\\
   &+\frac{1}{\sqrt{A_{\rm AdS}}}\int_0^{\infty}{\rm d}\bar{\tau}^{\textbf{I}'}\cosh^{\alpha-1}(\bar{\tau}^{\textbf{I}'})\Bigg|_{\text{const.}}\,, \nonumber
\\
    =& -\frac{c_{\rm AdS}}{2A_{\rm AdS}} \mathcal{I}_{\alpha-1}+\frac{1}{\sqrt{A_{\rm AdS}}}\int_0^{\infty}{\rm d}\bar{\tau}^{\textbf{I}'}\cosh^{\alpha-1}(\bar{\tau}^{\textbf{I}'})\Bigg|_{\text{const.}}\,.
\end{align}
We therefore see that once the integrals in the second line and a single $\alpha$–integral are known, all others follow. The simplest case corresponds to $\alpha=1/2$, since it reduces to the same type of integral that appears in the electromagnetic radiation term ${\rm d}\tau/a$. One finds
\begin{equation}
    \mathcal{I}_{1/2}=\frac{2}{\sqrt{A_{\rm AdS}}}\Pi\left[-\frac{c_{\rm AdS}}{2A_{\rm AdS}};-1\right]\,.
\end{equation}
Using the relations in Eq.~\eqref{magic}, we obtain
\begin{equation}
\begin{aligned}
    \mathcal{I}_{3/2}&=-\frac{c_{\rm AdS}}{2A_{\rm AdS}} \mathcal{I}_{1/2}+\frac{2}{\sqrt{A_{\rm AdS}}}\left(\mathcal{K}(-1)-\mathcal{E}(-1)\right)\,,
\\
    \mathcal{I}_{5/2}&=-\frac{c_{\rm AdS}}{2A_{\rm AdS}} \mathcal{I}_{3/2}+\frac{2}{3\sqrt{A_{\rm AdS}}}\mathcal{K}(-1)\,,
\\
    \mathcal{I}_{-1/2}&=-\frac{2A_{\rm AdS}}{c_{\rm AdS}} \mathcal{I}_{1/2}+\frac{4\sqrt{A_{\rm AdS}}}{c_{\rm AdS}}\mathcal{K}(-1)\,,
\\
    \mathcal{I}_{-3/2}&=-\frac{2A_{\rm AdS}}{c_{\rm AdS}} \mathcal{I}_{-1/2}-\frac{4\sqrt{A_{\rm AdS}}}{c_{\rm AdS}}\left(\mathcal{K}(-1)-\mathcal{E}(-1)\right)\,,
\\
    \mathcal{I}_{-5/2}&=-\frac{2A_{\rm AdS}}{c_{\rm AdS}} \mathcal{I}_{-3/2}+\frac{4\sqrt{A_{\rm AdS}}}{3c_{\rm AdS}}\mathcal{K}(-1)\,,
\\
    \mathcal{I}_{-7/2}&=-\frac{2A_{\rm AdS}}{c_{\rm AdS}} \mathcal{I}_{-5/2}-\frac{12\sqrt{A_{\rm AdS}}}{5c_{\rm AdS}}\left(\mathcal{K}(-1)-\mathcal{E}(-1)\right)\,.
\end{aligned}
\end{equation}
It then remains to express $a^3\tilde{V}_I$ in the basis $\cosh^i(\tau)$. We write 
\begin{equation}
    a^3\tilde{V}_I\equiv\sum_{i\in\mathbb{Z}+\frac{1}{2}}\frac{2H_{\rm AdS}\,a_i\cosh^{i}(\bar{\tau}^{\textbf{I}'})}{\sqrt{A_{\rm AdS}}\cosh(\bar{\tau}^{\textbf{I}'})-\sqrt{A_{\rm AdS}-A_{\rm dS}}}\,,
\end{equation}
where we recall that
\begin{equation}
    \bar{\tau}^{\textbf{I}'}=2H_{\rm AdS}(\tau+\tau_{\rm min})\,,\quad a(\tau)=\frac{\sqrt{A_{\rm AdS}}\cosh( \bar{\tau}^{\textbf{I}'})-\sqrt{A_{\rm AdS}-A_{\rm dS}}}{2H_{\rm AdS}\sqrt{\cosh( \bar{\tau}^{\textbf{I}'})}}\,.
\end{equation}
Hence,
\begin{equation}
    \int_{-\tau_{\rm min}}^{\infty} {\rm d} \tau a^3\tilde{V}_I\Bigg|_{\rm const.} 
    =  \frac{1}{2H_{\rm AdS}}\int_{0}^{\infty} {\rm d} \bar{\tau}^{\textbf{I}'} a^3\tilde{V}_I\Bigg|_{\rm const.}
    =  \sum_{i\in\mathbb{Z}+\frac{1}{2}}\mathcal{I}_i\times a_i\,.
\end{equation}
It only remains to extract the coefficients $a_i$ (see the potential~\eqref{eq:otherbasis})
\begin{align}
    a_{-7/2}&=\frac{5(A_{\rm AdS}-A_{\rm dS})^2}{48H_{\rm AdS}^2}\,, \quad a_{-5/2}=\frac{(A_{\rm AdS}-A_{\rm dS})c_{\rm AdS}}{12H_{\rm AdS}^2}\,, \nonumber
\\
    a_{-3/2}&=\frac{-A_{\rm AdS}^2+4A_{\rm AdS}A_{\rm dS}-3A_{\rm dS}^2}{48H_{\rm AdS}^2}\,, \quad a_{-1/2}=\frac{8A_{\rm AdS}-8A_{\rm dS}-2c_{\rm AdS}(A_{\rm AdS}-A_{\rm dS})}{48H_{\rm AdS}^2}\,, \nonumber
\\
    a_{1/2}&=\frac{1}{48H_{\rm AdS}^2}\left(-A_{\rm AdS}^2+2A_{\rm AdS}A_{\rm dS}+8c_{\rm AdS}+16H_{\rm AdS}^2\tilde{\rho}^{E}_{\rm rad}\right)\,, \nonumber
\\
    a_{3/2}&=\frac{8A_{\rm AdS}-2A_{\rm AdS}c_{\rm AdS}}{48H_{\rm AdS}^2}\,, \quad a_{5/2}=-\frac{3A_{\rm AdS}^2}{48H_{\rm AdS}^2}\,.
\end{align}
Combining all contributions and using the relation $c_{\rm AdS}=-2\sqrt{A_{\rm AdS}(A_{\rm AdS}-A_{\rm dS})}$, we obtain
\begin{align}
 \int_{-\t_{\rm min}}^{\t_{WG}} {\rm d} \t a(\t)^3 \tilde{V}(\t) =& \frac{1}{6\sqrt{A_{\rm AdS}}H_{\rm AdS}^2}\bigg(4H_{\rm AdS}^2 \tilde{\rho}^E_{\rm rad} \Pi\left[ \left.-\frac{c_{\rm AdS}}{2A_{\rm AdS}}\right|-1\right]-2A_{\rm AdS}\mathcal{E}[-1] \nonumber
\\
 & +(2A_{\rm AdS}+c_{\rm AdS})\mathcal{K}[-1]\bigg)\,,
\label{eq:grav_I_wg}
\end{align}
where
\begin{equation}
 \tilde{\rho}^E_{\rm rad} = \frac{(A_{\rm dS}-2)(A_{\rm dS}+c_{\rm dS})}{16H_{\rm dS}^2}\,, \qquad c_{\rm dS} = A_{\rm dS} + \left(2A_{\rm AdS}-A_{\rm dS}+c_{\rm AdS}\right) \frac{H_{\rm dS}^2}{H_{\rm AdS}^2}\,,
\end{equation}
and $c_{\rm AdS} = -2\sqrt{A_{\rm AdS}(A_{\rm AdS}-A_{\rm dS})}$.
Note that in~\eqref{eq:grav_I_wg} we have taken the limit $\t_{WG}\rightarrow\infty$, and we have removed the divergent terms given by~\eqref{eq:S_wg_reg}. The expressions~\eqref{eq:grav_II_wg} and~\eqref{eq:grav_I_wg} apply to both the single wineglass wormhole and the oscillatory case; in the former $\tau_{\rm min}=-\pi/(2H_{\rm dS})$, while in the latter $\tau_{\rm min}=-(2n-1)\pi/(2H_{\rm dS})$.

\section{Useful formulas}
\label{appendix:Useful formulas}
In this Appendix, we provide analytic expressions for the scalar field velocity and the scalar potential in regions $I$ and $II$ for the general wineglass ansatz~\eqref{eq:background_ans} and for the restricted version given in~\eqref{eq:background_ans_simple}.

\subsection{General wineglass ansatz~\eqref{eq:background_ans}}
Assuming that the background~\eqref{eq:background_ans} solves the eqs.~\eqref{eq:equAp} and~\eqref{eq:diffA}, we can extract the velocity of the field and the potential as functions of the Euclidean time $\tau$. In region (\textbf{I})
\begin{align}
\tilde{\phi}_I'^2(\t)= &\frac{
H_{\rm AdS}^2}{{
3 \Big(
B_{\rm AdS} + \cosh\big[2 H_{\rm AdS} (\tau - \tau_{\min}) \big] (c_{\rm AdS} + A_{\rm AdS} \cosh[2 H_{\rm AdS} (\tau - \tau_{\min}) ])
\Big)^2
}}  \nonumber
\\ &\times \Big[
-2 \big(A_{\rm AdS}^2 - 4 A_{\rm AdS} B_{\rm AdS} + 2 c_{\rm AdS} + 16 H_{\rm AdS}^2 \tilde{\rho}^E_{\rm rad} \big) 
- (3 A_{\rm AdS} + 4 B_{\rm AdS})   \nonumber
\\ & \times (2 + c_{\rm AdS}) \cosh\big[2 H_{\rm AdS} (\tau - \tau_{\min}) \big]- 2 \big(A_{\rm AdS}^2 + 4 A_{\rm AdS} B_{\rm AdS} + 2 c_{\rm AdS}   \nonumber
\\ & + 16 H_{\rm AdS}^2 \tilde{\rho}^E_{\rm rad} \big) 
\cosh\big[ 4 H_{\rm AdS} (\tau - \tau_{\min}) \big] - A_{\rm AdS} (2 + c_{\rm AdS}) \cosh\big[ 6 H_{\rm AdS} (\tau - \tau_{\min}) \big]  \nonumber
\\ & + 8 B_{\rm AdS} c_{\rm AdS} {\rm sech} \big[2 H_{\rm AdS} (\tau - \tau_{\min}) \big] 
+ 4 B_{\rm AdS}^2 {\rm sech}^2\big[2 H_{\rm AdS} (\tau - \tau_{\min}) \big] 
\Big]\,,
\\[0.2cm]
\tilde{V}_{I}(\t) =& \frac{
H_{\rm AdS}^2}{
24 \Big(
B_{\rm AdS} + \cosh\big[2 H_{\rm AdS} (\tau - \tau_{\min}) \big] (c_{\rm AdS} + A_{\rm AdS} \cosh[2 H_{\rm AdS} (\tau - \tau_{\min}) ])
\Big)^2} \nonumber
\\ &  \times\Big[
-5 A_{\rm AdS}^2 + 8 A_{\rm AdS} B_{\rm AdS} - 24 B_{\rm AdS}^2 + 32 c_{\rm AdS} + 64 H_{\rm AdS}^2 \tilde{\rho}^E_{\rm rad} \nonumber
\\ &
- 4 (3 A_{\rm AdS} + 4 B_{\rm AdS}) (-4 + c_{\rm AdS}) \cosh\big[2 H_{\rm AdS} (\tau - \tau_{\min}) \big] \nonumber
\\ &
- 8 (A_{\rm AdS}^2 + A_{\rm AdS} B_{\rm AdS} - 4 c_{\rm AdS} - 8 H_{\rm AdS}^2 \tilde{\rho}^E_{\rm rad}) \cosh\big[ 4 H_{\rm AdS} (\tau - \tau_{\min}) \big] \nonumber
\\ &
- 4 A_{\rm AdS} (-4 + c_{\rm AdS}) \cosh\big[ 6 H_{\rm AdS} (\tau - \tau_{\min}) \big] - 3 A_{\rm AdS}^2 \cosh\big[ 8 H_{\rm AdS} (\tau - \tau_{\min}) \big] \nonumber
\\ & 
+ 32 B_{\rm AdS} c_{\rm AdS} {\rm sech}\big[2 H_{\rm AdS} (\tau - \tau_{\min}) \big] 
+ 40 B_{\rm AdS}^2 {\rm sech}^2\big[2 H_{\rm AdS} (\tau - \tau_{\min}) \big]
\Big]\,,
\label{eq:full_V_I}
\end{align}
and in region (\textbf{II})
\begin{align}
\tilde{\phi}_{II}'^2(\t) &= \frac{4H_{\rm dS}^2}{3} \left(\frac{1-2c_{\rm dS}-16H_{\rm dS}^2\tilde{\rho}^E_{\rm rad}+(c_{\rm dS}-2)\cos[2H_{\rm dS}\t]}{\left(c_{\rm dS}+\cos[2H_{\rm dS}\t]\right)^2}\right)\,,
\\[0.2cm]
\tilde{V}_{II}(\t) &= \frac{H_{\rm dS}^2}{6}\left(\frac{1+16c_{\rm dS}+32H_{\rm dS}^2\tilde{\rho}^E_{\rm rad}+4(c_{\rm dS}+4)\cos[2H_{\rm dS}\t]+3\cos[4H_{\rm dS}\t]}{\left(c_{\rm dS}+\cos[2H_{\rm dS}\t]\right)^2}\right)\,. 
\label{eq:VII}
\end{align}

\subsection{Restricted wineglass ansatz~\eqref{eq:background_ans_simple}}
Assuming the restricted 3-parameter ansatz~\eqref{eq:background_ans_simple} we obtain
\begin{align}
\tilde{V}_{I}(\t) =& \frac{H_{\rm AdS}^2}{3(\sqrt{A_{\rm AdS}}\cosh(\bar{\tau}^{\textbf{I}'})-\sqrt{A_{\rm AdS}-A_{\rm dS}})^4} \Big[ 8 (A_{\rm AdS}-A_{\rm dS})\cosh(\bar{\tau}^{\textbf{I}'})
\\ 
&  + \cosh^2(\bar{\tau}^{\textbf{I}'})  (8(c_{\rm AdS}+2H_{\rm AdS}^2\tilde{\rho}^E_{\rm rad})-2A_{\rm AdS}(A_{\rm AdS}-A_{\rm dS})-A_{\rm AdS}^2 \sinh^2(\bar{\tau}^{\textbf{I}'})) \nonumber
\\ 
&  -2A_{\rm AdS} (c_{\rm AdS}-4)\cosh^3(\bar{\tau}^{\textbf{I}'}) -2A_{\rm AdS}^2\cosh^4(\bar{\tau}^{\textbf{I}'}) -\dfrac12(A_{\rm AdS}-A_{\rm  dS}){\rm sech}^2(\bar{\tau}^{\textbf{I}'})  \nonumber
\\ 
& \times \left((A_{\rm AdS}-3A_{\rm dS})\cosh(2\bar{\tau}^{\textbf{I}'}) -9 A_{\rm AdS} +7A_{\rm dS} - 5c_{\rm AdS}\cosh(\bar{\tau}^{\textbf{I}'}) +c_{\rm AdS}\cosh(3\bar{\tau}^{\textbf{I}'})   \right)\Big]\,, \nonumber
\\[0.2cm]
\tilde{V}_{II}(\t) =& \frac{H_{\rm dS}^2}{6}\left(\frac{1+16c_{\rm dS}+32H_{\rm dS}^2\tilde{\rho}^E_{\rm rad}+4(c_{\rm dS}+4)\cos(\btt)+3\cos(2\btt)}{\left(c_{\rm dS}+\cos(\btt)\right)^2}\right)\,,
\label{eq:VII_r3}
\\[0.2cm]
\tilde{V}_{I'}(\t) =& \tilde{V}_{I}(\t) \quad \text{with} \quad \bto \rightarrow \bar{\tau}^{\textbf{I}'}\,.
\label{eq:VII_r}
\end{align}
It is also useful to re-express~\eqref{eq:VII_r} in the other basis,
\begin{align}
\tilde{V}_{I'}(\t)
=&
\frac{H_{\rm AdS}^{2}}{3\left(\sqrt{A_{\mathrm{AdS}}}\cosh\tau-\sqrt{A_{\mathrm{AdS}}-A_{\mathrm{dS}}}\right)^{4}}
\Big[
\bigl(8A_{\mathrm{AdS}}-2A_{\mathrm{AdS}}c_{\mathrm{AdS}}\bigr)\cosh^{3}(\bar{\tau}^{\textbf{I}'})\nonumber
\\
&
-3A_{\mathrm{AdS}}^{2}\cosh^{4}(\bar{\tau}^{\textbf{I}'}) +\left(2A_{\mathrm{AdS}}A_{\mathrm{dS}}-A_{\mathrm{AdS}}^{2}+8c_{\mathrm{AdS}}+16H_{\rm AdS}^{2}\tilde{\rho}^E_{\rm rad}\right)\cosh^{2}(\bar{\tau}^{\textbf{I}'}) \nonumber
\\
&
+\bigl[8(A_{\mathrm{AdS}}-A_{\mathrm{dS}})
-2c_{\mathrm{AdS}}(A_{\mathrm{AdS}}-A_{\mathrm{dS}})\bigr]\cosh(\bar{\tau}^{\textbf{I}'})+\bigl(4A_{\mathrm{AdS}}A_{\mathrm{dS}}-A_{\mathrm{AdS}}^{2}-3A_{\mathrm{dS}}^{2}\bigr) \nonumber
\\
&
+4c_{\mathrm{AdS}}(A_{\mathrm{AdS}}-A_{\mathrm{dS}})\cosh^{-1}(\bar{\tau}^{\textbf{I}'}) +5(A_{\mathrm{AdS}}-A_{\mathrm{dS}})^{2}\cosh^{-2}(\bar{\tau}^{\textbf{I}'})
\Big].
\label{eq:otherbasis}
\end{align}

\section{Elliptic integrals}
\label{incompleterecipe}
In this Appendix, we review the definitions and collect useful identities and asymptotic expansions for elliptic integrals used in the main text.

\subsection{Incomplete elliptic integrals}

\paragraph{First kind -}The incomplete elliptic integral of the first kind denoted by $\mathcal{F}(\psi|m)$ is defined by 
\begin{equation}
\label{definitionElliptic}
        \mathcal{F}(\psi|m)\equiv\int_0^{\psi}\frac{{\rm d}\theta}{\sqrt{1-m\sin^2(\theta)}}=\int_0^{\sin(\psi)}\frac{{\rm d}x}{\sqrt{1-m\,x^2}\,\sqrt{1-x^2}}\,.
\end{equation}
There is also a useful formula in terms of the complementary modulus parameter $m'=1-m$
\begin{equation}
    \mathcal{F}(\psi|m)=\int_0^{\tan(\psi)}\frac{{\rm d}\tilde{x}}{\sqrt{1+m'\,\tilde{x}^2}\,\sqrt{1+\tilde{x}^2}}\,.
\end{equation}
A useful relation can be obtained via an imaginary modulus transformation
\begin{equation}\label{imaginarytransfoF}
    \mathcal{F}(i\,\psi|m)=i\,\mathcal{F}(\beta|m')\,\qquad \sinh(\psi)=\tan(\beta)\,.
\end{equation}
\paragraph{Second kind -} The incomplete elliptic function of the second kind$\mathcal{E}(\psi,m)$ is defined by
\begin{equation}
    \mathcal{E}(\psi,m)\equiv\int_0^{\psi}\sqrt{1-m\sin^2(\theta)}{\rm d}\theta~=\int_0^{\sin(\psi)}\sqrt{\frac{1-m\,x^2}{1-x^2}}{\rm d}x\,.
\end{equation}
A useful relation can be obtained via an imaginary modulus transformation
\begin{equation}\label{imaginarytransfoE}
    \mathcal{E}(i\,\psi|m)=i\,\left[\mathcal{F}(\beta|m')-\mathcal{E}(\beta,m')+\tanh(\psi)\sqrt{1+m\sinh^2(\psi)}\right]\,\qquad \sinh(\psi)=\tan(\beta)\,.
\end{equation}

\paragraph{Third kind -}\label{para:incompleteellpticthirdkind} The incomplete elliptic function of the third kind $\Pi(n;\psi|m)$ is defined by
\begin{equation}\label{eq:incompleteellipticthirdkind}
    \Pi(n;\psi|m)\equiv\int_0^{\psi}\frac{{\rm d}\theta}{(1-n\sin(\theta)^2)\sqrt{1-m\sin^2(\theta)}}~=\int_0^{\sin(\psi)}\frac{{\rm d}x}{(1-nx^2)\sqrt{1-m\,x^2}\sqrt{1-x^2}}\,.
\end{equation}

\subsection{Complete elliptic integrals}\label{completerecipe}

\paragraph{First kind -} The complete elliptic integral of the first kind is defined through the incomplete elliptic integral of the first kind via 
\begin{equation}
    \mathcal{K}(m)\equiv\mathcal{F}\left(\frac{\pi}{2}\Big|m\right)\,.
\end{equation}
\paragraph{Second kind -}
The complete elliptic integral of the second kind is defined through the incomplete elliptic integral of the second kind via 
\begin{equation}
    \mathcal{E}(m)\equiv\mathcal{E}\left(\frac{\pi}{2}\Big|m\right)=\int_0^{\pi/2}\sqrt{1-m\sin^2(\theta)}{\rm d}\theta\,.
\end{equation}

\paragraph{Third kind -} The complete elliptic integral of the third kind $\Pi(n|m)$ is defined by
\begin{equation}
    \Pi(n|m)\equiv\int_0^{\pi/2}\frac{{\rm d}\theta}{(1-n\sin(\theta)^2)\sqrt{1-m\sin^2(\theta)}}~=\int_0^{1}\frac{{\rm d}x}{(1-nx^2)\sqrt{1-m\,x^2}\sqrt{1-x^2}}\,.
\end{equation}

\subsection{Asymptotic formulas}\label{asymptoticellipticrecipe}
In this paragraph, we give useful asymptotics for the elliptic integrals that we use in the core of the text.

\begin{equation}
\begin{aligned}
    \lim_{\psi\to+\infty}\mathcal{F}(i\,\psi|m)&=i\,\mathcal{K}(m')=i\,\mathcal{K}(1-m)\,,
\\
    \mathcal{E}(i\,\psi|m)&=i\frac{\sqrt{m}}{2}e^{\psi}+i\,\left[\mathcal{K}(m')-\mathcal{E}(m')+\mathcal{O}\left(e^{-\psi}\right)\right]\,,
\\
    & \hspace{-1.26cm}\mathcal{K}(m)\underset{m\to -\infty}{\sim}\frac{1}{\sqrt{-m}} \log(4\sqrt{-m})\,,
\\
   & \hspace{-1.18cm} \mathcal{K}(m)\underset{m\to 1^{-}}{\sim}\frac{1}{2}\log\left(\frac{16}{1-m}\right)\,,
\\
    & \hspace{-1.18cm}\mathcal{E}(m)\underset{m\to -\infty}{\sim}\sqrt{-m}\,.
\end{aligned}
\end{equation}

\section{Computation of the electromagnetic potential for wormhole solutions}\label{computationdetails}

The computation directly stem from the computation of the conformal time as a function of $\tau$, $u(\tau)=\int_0^{\tau}\frac{d\tilde{\tau}}{a(\tilde{\tau})}$.

\paragraph{Simple wormhole without running scalar -} Recall the formula of the scale factor in this region \eqref{eq:scale_connected}. Let $\alpha=\sqrt{1+4\tilde{\rho}^{E}_{\rm rad}\tilde{V}_{\rm AdS}}$. We find
\begin{equation}
\begin{aligned}
    \int_0^{\tau}\frac{{\rm d}\tau'}{a_C(\tau')}&=\left(\frac{2}{\sqrt{1+4\tilde{\rho}^{E}_{\rm rad}\tilde{V}_{\rm AdS}}-1}\right)^{1/2}\mathcal{F}\Bigg[\arcsin{\left(\tanh{\tilde{\tau}}\right)}\Bigg|m\Bigg]\,,\\
    m&=-\frac{\alpha+1}{\alpha-1}=-\frac{\sqrt{1+4\tilde{\rho}^{E}_{\rm rad}\tilde{V}_{\rm AdS}}+1}{\sqrt{1+4\tilde{\rho}^{E}_{\rm rad}\tilde{V}_{\rm AdS}}-1}<0\,,
\end{aligned}
\end{equation}
where we defined $\tilde{\tau}=H_{\rm AdS}\,\tau=\sqrt{-\tilde{V}_{\rm AdS}}\,\tau$. $\mathcal{F}(\psi|m)$ denotes the incomplete elliptic integral of the first kind (see Appendix~\ref{definitionElliptic} for further details). Using~\eqref{eq:equationH} and choosing $c_1,c_2$ such that we obtain a $\mathbb{Z}_2$ symmetric solution whose derivative vanishes at $\tilde{\tau}=0$, we obtain
\begin{equation}
H_C(\tau)=H_{0,C} \cosh\left[2\left(\frac{2}{\sqrt{1+4\tilde{\rho}^{E}_{\rm rad}\tilde{V}_{\rm AdS}}-1}\right)^{1/2}\mathcal{F}\left[\arcsin{\left(\tanh{\tilde{\tau}}\right)}\big|m\right]\right]\,.
\end{equation}

\paragraph{Wormhole solutions with running scalar -}
Recall the form of the scale factor \eqref{eq:background_ans_simple}. We start by region \textbf{II}. The integral in the central region simply reads
\begin{equation}
 u(\tau) =  \int_{0}^{\tau}\frac{{\rm d}\tau'}{a_{\textbf{II}}(\tau')}=\frac{2}{\sqrt{c_{\rm dS}+A_{\rm dS}}}\mathcal{F}\left[\left.\frac{{\bar{\tau}^{\textbf{II}}}}{2}\right|\frac{2 A_{\rm dS}}{c_{\rm dS}+A_{\rm dS}}\right]\,\quad\bar{\tau}^{\textbf{II}}=2 H_{\rm dS}\,\tau\,. 
\end{equation}
If we want to find the solution in region $\textbf{I}'$, $\tau>-\tau_{\rm min}$, we can extend the integral in the following manner
    \begin{align}
   u(\tau) = \int_{0}^{\tau}\frac{{\rm d}\tau'}{a(\tau')} = & \int_{0}^{-\tau_{\rm min}}\frac{{\rm d}\tau'}{a_{\textbf{II}}(\tau')}+\int_{-\tau_{\rm min}}^{\tau}\frac{{\rm d}\tau'}{a_{\textbf{I}'}(\tau')} \nonumber
\\
        =&\frac{2}{\sqrt{c_{\rm dS}+A_{\rm dS}}}\mathcal{K}\left[\frac{2 A_{\rm dS}}{c_{\rm dS}+A_{\rm dS}}\right]+\frac{2}{\sqrt{A_{\rm AdS}}}\Pi\left[ \left.-\frac{c_{\rm AdS}}{2A_{\rm AdS}}\right|-1\right] \nonumber
\\
        &- \frac{2}{\sqrt{A_{\rm AdS}}}\Pi\left[-\frac{c_{\rm AdS}}{2A_{\rm AdS}};\arcsin\left(\cosh^{-1/2}(\bar{\tau}^{\textbf{I}'})\right)\Big|-1\right]\,,
    \end{align}
where we define $\bar{\tau}^{\textbf{I}'}= 2 H_{\rm AdS}(\tau+\tau_{\rm min})$ and $\Pi\left(n;\psi|m\right)$ is the incomplete elliptic integral of the third kind and $\Pi\left(n|m\right)$ is the corresponding complete elliptic integral (see Appendix~\ref{definitionElliptic} for further details). In region \textbf{I}, $\tau<\tau_{\rm min}$, thus we split the integral in the following manner
\begin{align}
    u(\tau) = \int_0^{\tau}\frac{{\rm d}\tau'}{a(\tau')}=&-\left(\int_{\tau}^{\tau_{\rm min}}\frac{{\rm d}\tau'}{a(\tau')}+\int_{\tau_{\rm min}}^{0}\frac{{\rm d}\tau'}{a(\tau')}\right) \nonumber
\\
        =&-\frac{2}{\sqrt{c_{\rm dS}+A_{\rm dS}}}\mathcal{K}\left[\frac{2 A_{\rm dS}}{c_{\rm dS}+A_{\rm dS}}\right]-\frac{2}{\sqrt{A_{\rm AdS}}}\Pi\left[ \left.-\frac{c_{\rm AdS}}{2A_{\rm AdS}}\right|-1\right] \nonumber
\\
        &+ \frac{2}{\sqrt{A_{\rm AdS}}}\Pi\left[-\frac{c_{\rm AdS}}{2A_{\rm AdS}};\arcsin\left(\cosh^{-1/2}(\bar{\tau}^{\textbf{I}})\right)\Big|-1\right]\,,
\end{align}
where we define $\bar{\tau}^{\textbf{I}}=2H_{\rm AdS}(\tau-\tau_{\rm min})$. The previous results allow to write the electromagnetic potential in a piecewise fashion given  by~\eqref{eq:background_potential}.

\section{Boundary term for the scalar}
\label{appendix:bnry_scalar}
    
As discussed in Sec.~\ref{piecewisewineglass}, imposing Neumann BCs for the scalar field requires the addition of the boundary term
\begin{equation}
    S_{\partial \mathcal{M}}^{\rm ren.} = - \frac{3}{\kappa} \int_{\partial \mathcal{M}} {\rm d}^3 x   \tilde \phi  \tilde{\Pi}_{ren.}\,.
\end{equation}
Integrating Eq.~\eqref{eq:vel_inf}, we obtain
\begin{equation}
\tilde{\phi}(\t\rightarrow\infty)\simeq \begin{cases}
   \displaystyle     \frac{\mathcal{C}_1}{H_{\rm AdS}} e^{-H_{\rm AdS}\t} + \frac{\mathcal{C}_3}{3H_{\rm AdS}} e^{-3H_{\rm AdS}\t} +\cdots\,, & \text{non-zero Dirichlet scalar source}\\[0.4cm]
   \displaystyle     \frac{\mathcal{C}_2}{2H_{\rm AdS}} e^{-2H_{\rm AdS}\t} + \frac{\mathcal{C}_4}{4H_{\rm AdS}} e^{-4H_{\rm AdS}\t} +\cdots\,, & \text{zero Dirichlet scalar source}\,,
    \end{cases}
\end{equation}
where the parameters $\mathcal{C}_i$ denote combinations of the parameters appearing in the metric ansatz~\eqref{eq:background_ans}, introduced here
for notational simplicity. Using this asymptotic behaviour, the boundary term becomes
\begin{equation}
\frac{ S_{\partial \mathcal{M}}^{\rm ren.}}{3\text{Vol}(S^3)/\k} = -  a^3\tilde{\phi} ( \tilde{\phi}' +H_{\rm AdS}\tilde{\phi}) \simeq \begin{cases}
   \displaystyle      \frac{A_{\rm AdS}^{3/2}\mathcal{C}_1\mathcal{C}_3}{8\sqrt{2}H_{\rm AdS}^4}  e^{-H_{\rm AdS}\t}\,,  & \text{non-zero Dirichlet scalar source}\\[0.4cm]
   \displaystyle  \frac{A_{\rm AdS}^{3/2}\mathcal{C}_2^2}{64\sqrt{2}H_{\rm AdS}^4}  e^{-H_{\rm AdS}\t}\,,  & \text{zero Dirichlet scalar source}\,.
    \end{cases}
\end{equation}
We therefore conclude that, in both cases, namely for a non-zero scalar source ($c_{\rm AdS}\neq -2$) and for a vanishing scalar source
($c_{\rm AdS}=-2$), the boundary term vanishes.

\section{Asymptotic behaviour of the on-shell action for wineglass and oscillatory solutions}\label{Asymptoticbehavior}

In this Appendix we analyse the asymptotic behaviour of $\Delta\mathcal{S}_{D-\rm{osc(n)}}$ as a function of $H_{\rm UV}$ for both Dirichlet or Neumann BCs for the scalar field. We recall that the wineglass wormhole corresponds to $n=1$, while solutions with more oscillations correspond to $n>1$. For notational simplicity we denote the value of the electromagnetic field at the center by $H_{0,WG}$ and the one of the more general oscillatory solutions as $H_{0,n}$, we commonly refer to them by $H_{0,n}$ where again $n=1$ refers to the wineglass solution and $n>1$ to the more general oscillatory solutions.

\paragraph{Relation between $H_{\rm UV}$ and $H_{0,n}$ -}
We first determine the asymptotic relation between $H_{\rm UV}$ and $H_{0,n}$. Recall that for the simple wormhole $H_{\rm UV}\sim H_{0,C}$, while for the disconnected solution $H_{\rm UV}=H_0$. The exact relation reads
\begin{equation}
    H_{\rm UV}=H_{0,n}\cosh\left(\frac{4\,(2n-1)}{\sqrt{c_{\rm dS}+A_{\rm dS}}}\mathcal{K}\left[\frac{2A_{\rm dS}}{c_{\rm dS}+A_{\rm dS}}\right]+\frac{4}{\sqrt{A_{\rm AdS}}}\Pi\left[-\frac{c_{\rm AdS}}{2A_{\rm AdS}};-1\right]\right)\, ,
\label{eq:Huv_wineglass_OSC2}
\end{equation}
To extract the large-$H_{0,n}$ behaviour we expand $c_{\rm dS}+A_{\rm dS}$ as
\begin{equation}
    c_{\rm dS}+A_{\rm dS}=2A_{\rm dS}-2A_{\rm dS}\frac{A_{\rm dS}(2+c_{\rm AdS})-2c_{\rm AdS}-2A_{\rm AdS}(2-A_{\rm dS})-A_{\rm dS}^2}{128\kappa H_{\rm AdS}^2H_{0,n}^2}+\mathcal{O}\left(H_{0,n}^{-4}\right)\,,
\end{equation}
which implies
\begin{equation}
   \frac{2A_{\rm dS}}{c_{\rm dS}+A_{\rm dS}}=1+\frac{\beta}{H_{0,n}^{2}}+\mathcal{O}\left(H_{0,n}^{-4}\right)\,, \quad \beta =- \frac{(2-A_{\rm dS})(2A_{\rm AdS}-A_{\rm dS}+c_{\rm AdS})}{128\k |\tilde{V}_{\rm AdS}|}<0  \,,
\end{equation}
since $c_{\rm dS}+A_{\rm dS}>2A_{\rm dS}$.
Since the complete elliptic integral of the first kind diverges logarithmically as its argument approaches unity (see Appendix~\ref{asymptoticellipticrecipe}), the leading behaviour is
\begin{equation}\label{eq:ellipticblow}
    \mathcal{K}\left[\frac{2A_{\rm dS}}{c_{\rm dS}+A_{\rm dS}}\right]\simeq\frac{1}{2}\log(H_{0,n}^2)+\frac{1}{2}\log\left(\frac{16}{-\beta}\right)\,.
\end{equation}
The second term inside the $\cosh$ in eq.~\eqref{eq:Huv_wineglass_OSC2} is constant, while the first diverges as
\begin{equation}
     \frac{4(2n-1)}{\sqrt{c_{\rm dS}+A_{\rm dS}}}\mathcal{K}\left[\frac{2A_{\rm dS}}{c_{\rm dS}+A_{\rm dS}}\right]\underset{H_{\rm UV}\to \infty}{\simeq}\log\left(|H_{0,n}|^{\frac{4(2n-1)}{\sqrt{2A_{\rm dS}}}}\right)+\frac{2(2n-1)}{\sqrt{2A_{\rm dS}}}\log\left(\frac{16}{-\beta}\right)\,.
\end{equation}
We therefore obtain
\begin{equation}\label{eq:H_UV/H_0behaviorwineglass}
    H_{\rm UV}\underset{H_{\rm UV}\to \infty}{\sim}e^{\gamma}\frac{H_{0,n}^{q(n)}}{2}\gg H_{0,n}\,, \qquad \text{with} \qquad q(n) = 1+\frac{4(2n-1)}{\sqrt{2A_{\rm dS}}} \,,
\end{equation}
and
\begin{equation}\label{eq:a1_expression}
        \gamma(n) =  \frac{1}{\sqrt{A_{\rm AdS}A_{\rm dS}}}\left(4\sqrt{A_{\rm dS}} \Pi\left[-\frac{c_{\rm AdS}}{2A_{\rm AdS}};-1\right]+\sqrt{2A_{\rm AdS}}(2n-1)\log\left[\frac{-16}{\b}\right] \right)\,.
\end{equation}

\paragraph{Radiative contribution -}
Using the above relation we can find the large $H_{\rm UV}$ expansion of the wineglass (and oscillatory) action. First we focus on the radiative part,
\begin{equation}
    |H_{\rm UV}|\sqrt{H_{\rm UV}^2-H_{0,n}^2}=H_{\rm UV}^2-\frac{\left(2e^{-\gamma}H_{\rm UV}\right)^{2/q(n)}}{2}+\mathcal{O}\left(H_{\rm UV}^{4/q(n)-2}\right)\,.
\end{equation}
The leading term cancels against the disconnected contribution.  So when taking the difference, disconnect-oscillatory(wineglass), one obtains
\begin{equation}
    \Delta \mathcal{S}_{\rm rad.}\underset{H_{\rm UV}\to \infty}{\sim}12\text{Vol}(S^3)H_{0,n}^2\underset{H_{\rm UV}\to \infty}{\sim}12\text{Vol}(S^3)\left(2e^{-\gamma}H_{\rm UV}\right)^{2/q(n)}\,.
\end{equation}

\paragraph{Gravitational contribution -}
Since $H_{0,n}\to\infty$ as $H_{\rm UV}\to\infty$, it suffices to extract the asymptotic dependence on $H_{0,n}$. From eq.~\eqref{eq:osc_zeta} the first term scales as $\propto H_{0,n}^2$. The piece which concerns the integral in the region of the throat is quite more involved since $H_{\rm dS}$ is itself a function of the electromagnetic field, see~\eqref{eq:Hds}. In particular,
\begin{equation}
    \kappa H_{\rm dS}^2\underset{H_{\rm UV}\rightarrow\infty}{=}\frac{(2-A_{\rm dS})A_{\rm dS}}{64H_{0,n}^2}+\mathcal{O}\left(H_{0,n}^{-4}\right)\,.
\end{equation}
which tends to zero, implying $\tau_{\rm min}\to -\infty$. The prefactor behaves as
\begin{equation}
    \frac{\tilde{V}_{\rm AdS}(c_{\rm dS}+A_{\rm dS})^{1/2}}{16H_{\rm dS}^2}\underset{H_{\rm UV}\rightarrow\infty}{=}-\frac{4\kappa |\tilde{V}_{\rm AdS}|\sqrt{2}}{(2-A_{\rm dS})\sqrt{A_{\rm dS}} }H_{0,n}^2 +{\rm const.} +\mathcal{O}\left(H_{0,n}^{-2}\right)\,.
\end{equation}
Finally, recall that the complete elliptic integral of the first kind diverges logarithmically when the argument approaches one, see~\eqref{eq:ellipticblow} (this is not the case of the complete elliptic integral of the second kind). Hence, the dominant term in the region of large $H_{\rm UV}$ is dominated by the term $\propto H_{0,n}^2\log|H_{0,n}|$. The coefficient and the sign in front of this term is crucial. We find
\begin{equation}
     \zeta^{IR}_{D,\rm osc(n)}\underset{H_{\rm UV}\rightarrow\infty}{\sim}-\frac{8\kappa |\tilde{V}_{\rm AdS}| (2n-1)}{\sqrt{2A_{\rm dS}}}H_{0,n}^2\log|H_{0,n}|\,.
\end{equation}
Thus
\begin{align}
    \Delta\mathcal{S}_{\rm grav.}&\underset{H_{\rm UV}\rightarrow\infty}{\sim}-\frac{\text{Vol}(S^3) 32(2n-1)}{\sqrt{2A_{\rm dS}}}H_{0,n}^2\log|H_{0,n}| \nonumber
\\
    &\underset{H_{\rm UV}\rightarrow\infty}{\sim}-8\text{Vol}(S^3)\left(2e^{-\gamma}H_{\rm UV}\right)^{2/q(n)}\left(\frac{q(n)-1}{q(n)}\right)\log|2H_{\rm UV}e^{-\gamma}|\,.
\end{align}
At this stage, there are crucial details to observe. First, $\Delta\mathcal{S}_{\rm grav.}$ is the leading contribution and is negative meaning that the disconnected geometry is always dominant asymptotically. This conclusion holds for all $n$. Moreover, as $n$ increases, the logarithmic divergence of $\Delta\mathcal{S}{D\text{-}\rm osc(n)}$ becomes progressively weaker. In particular,
\begin{equation}\label{eq:bubbledifference}
    \Delta\mathcal{S}_{\rm osc(n) - osc(n+1)}=\Delta\mathcal{S}_{D-\rm osc(n+1)}-\Delta\mathcal{S}_{D-\rm osc(n)}\underset{H_{\rm UV}\rightarrow\infty}{\sim}-\Delta\mathcal{S}_{D-\rm osc(n)}>0\,.
\end{equation}

%%%%%%%%%%%%%%%%%%%%%%%%%%%%%%%%%%%%%%%%%%%%%%%%%%%%%%%%%%%%%
\bibliographystyle{JHEP}
\bibliography{biblio}

\providecommand{\noopsort}[1]{}\providecommand{\singleletter}[1]{#1}%
\providecommand{\href}[2]{#2}\begingroup\raggedright\begin{thebibliography}{10}

\bibitem{Hawking:1982dh}
S.~W. Hawking and D.~N. Page, \emph{{Thermodynamics of Black Holes in anti-De Sitter Space}}, \href{http://dx.doi.org/10.1007/BF01208266}{\emph{Commun. Math. Phys.} {\bf 87} (1983) 577}.

\bibitem{Maldacena:2004rf}
J.~M. Maldacena and L.~Maoz, \emph{{Wormholes in AdS}}, \href{http://dx.doi.org/10.1088/1126-6708/2004/02/053}{\emph{JHEP} {\bf 02} (2004) 053}, [\href{http://arxiv.org/abs/hep-th/0401024}{{\tt hep-th/0401024}}].

\bibitem{Betzios:2019rds}
P.~Betzios, E.~Kiritsis and O.~Papadoulaki, \emph{{Euclidean Wormholes and Holography}}, \href{http://dx.doi.org/10.1007/JHEP06(2019)042}{\emph{JHEP} {\bf 06} (2019) 042}, [\href{http://arxiv.org/abs/1903.05658}{{\tt arXiv:1903.05658}}].

\bibitem{Betzios:2021fnm}
P.~Betzios, E.~Kiritsis and O.~Papadoulaki, \emph{{Interacting systems and wormholes}}, \href{http://dx.doi.org/10.1007/JHEP02(2022)126}{\emph{JHEP} {\bf 02} (2022) 126}, [\href{http://arxiv.org/abs/2110.14655}{{\tt arXiv:2110.14655}}].

\bibitem{VanRaamsdonk:2021qgv}
M.~Van~Raamsdonk, \emph{{Cosmology from confinement?}}, \href{http://dx.doi.org/10.1007/JHEP03(2022)039}{\emph{JHEP} {\bf 03} (2022) 039}, [\href{http://arxiv.org/abs/2102.05057}{{\tt arXiv:2102.05057}}].

\bibitem{Antonini:2022blk}
S.~Antonini, P.~Simidzija, B.~Swingle and M.~Van~Raamsdonk, \emph{{Cosmology from the vacuum}}, \href{http://dx.doi.org/10.1088/1361-6382/ad1d46}{\emph{Class. Quant. Grav.} {\bf 41} (2024) 045008}, [\href{http://arxiv.org/abs/2203.11220}{{\tt arXiv:2203.11220}}].

\bibitem{Antonini:2022ptt}
S.~Antonini, P.~Simidzija, B.~Swingle and M.~Van~Raamsdonk, \emph{{Accelerating Cosmology from a Holographic Wormhole}}, \href{http://dx.doi.org/10.1103/PhysRevLett.130.221601}{\emph{Phys. Rev. Lett.} {\bf 130} (2023) 221601}, [\href{http://arxiv.org/abs/2206.14821}{{\tt arXiv:2206.14821}}].

\bibitem{Betzios:2023obs}
P.~Betzios and O.~Papadoulaki, \emph{{Wilson loops and wormholes}}, \href{http://dx.doi.org/10.1007/JHEP03(2024)066}{\emph{JHEP} {\bf 03} (2024) 066}, [\href{http://arxiv.org/abs/2311.09289}{{\tt arXiv:2311.09289}}].

\bibitem{Maloney:2025tnn}
A.~Maloney, V.~Meruliya and M.~Van~Raamsdonk, \emph{{Ordinary wormholes}},  \href{http://arxiv.org/abs/2503.12227}{{\tt arXiv:2503.12227}}.

\bibitem{VanRaamsdonk:2026tnv}
M.~Van~Raamsdonk and A.~Vilar~L{\'o}pez, \emph{{An AS${}^2$ Menagerie}},  \href{http://arxiv.org/abs/2601.10906}{{\tt arXiv:2601.10906}}.

\bibitem{Belin:2025ako}
A.~Belin and J.~de~Boer, \emph{{Baby Universes in AdS$_3$}},  \href{http://arxiv.org/abs/2512.02098}{{\tt arXiv:2512.02098}}.

\bibitem{Betzios:2024oli}
P.~Betzios and O.~Papadoulaki, \emph{{Inflationary Cosmology from Anti-de Sitter Wormholes}}, \href{http://dx.doi.org/10.1103/PhysRevLett.133.021501}{\emph{Phys. Rev. Lett.} {\bf 133} (2024) 021501}, [\href{http://arxiv.org/abs/2403.17046}{{\tt arXiv:2403.17046}}].

\bibitem{Betzios:2024zhf}
P.~Betzios, I.~D. Gialamas and O.~Papadoulaki, \emph{{Magnetic anti{\textendash}de Sitter wormholes as seeds for Higgs inflation}}, \href{http://dx.doi.org/10.1103/9w85-fyhs}{\emph{Phys. Rev. D} {\bf 111} (2025) 123542}, [\href{http://arxiv.org/abs/2412.03639}{{\tt arXiv:2412.03639}}].

\bibitem{DeAlwis:2019rxg}
S.~P. De~Alwis, F.~Muia, V.~Pasquarella and F.~Quevedo, \emph{{Quantum Transitions Between Minkowski and de Sitter Spacetimes}}, \href{http://dx.doi.org/10.1002/prop.202000069}{\emph{Fortsch. Phys.} {\bf 68} (2020) 2000069}, [\href{http://arxiv.org/abs/1909.01975}{{\tt arXiv:1909.01975}}].

\bibitem{Fu:2019oyc}
Z.~Fu and D.~Marolf, \emph{{Bag-of-gold spacetimes, Euclidean wormholes, and inflation from domain walls in AdS/CFT}}, \href{http://dx.doi.org/10.1007/JHEP11(2019)040}{\emph{JHEP} {\bf 11} (2019) 040}, [\href{http://arxiv.org/abs/1909.02505}{{\tt arXiv:1909.02505}}].

\bibitem{Freivogel:2005qh}
B.~Freivogel, V.~E. Hubeny, A.~Maloney, R.~C. Myers, M.~Rangamani and S.~Shenker, \emph{{Inflation in AdS/CFT}}, \href{http://dx.doi.org/10.1088/1126-6708/2006/03/007}{\emph{JHEP} {\bf 03} (2006) 007}, [\href{http://arxiv.org/abs/hep-th/0510046}{{\tt hep-th/0510046}}].

\bibitem{Lavrelashvili:1988un}
G.~V. Lavrelashvili, V.~A. Rubakov and P.~G. Tinyakov, \emph{{Loss of Quantum Coherence Due to Topological Changes: A Toy Model}}, \href{http://dx.doi.org/10.1142/S0217732388001483}{\emph{Mod. Phys. Lett. A} {\bf 3} (1988) 1231--1242}.

\bibitem{Hartle:1983ai}
J.~B. Hartle and S.~W. Hawking, \emph{{Wave Function of the Universe}}, \href{http://dx.doi.org/10.1103/PhysRevD.28.2960}{\emph{Phys. Rev. D} {\bf 28} (1983) 2960--2975}.

\bibitem{Lehners:2023yrj}
J.-L. Lehners, \emph{{Review of the no-boundary wave function}}, \href{http://dx.doi.org/10.1016/j.physrep.2023.06.002}{\emph{Phys. Rept.} {\bf 1022} (2023) 1--82}, [\href{http://arxiv.org/abs/2303.08802}{{\tt arXiv:2303.08802}}].

\bibitem{Maldacena:2024uhs}
J.~Maldacena, \emph{{Comments on the no boundary wavefunction and slow roll inflation}},  \href{http://arxiv.org/abs/2403.10510}{{\tt arXiv:2403.10510}}.

\bibitem{Betzios2025toappear}
P.~Betzios, P.~Ghiringhelli, I.~D. Gialamas and O.~Papadoulaki, To appear.

\bibitem{Betzios:2017krj}
P.~Betzios, N.~Gaddam and O.~Papadoulaki, \emph{{Antipodal correlation on the meron wormhole and a bang-crunch universe}}, \href{http://dx.doi.org/10.1103/PhysRevD.97.126006}{\emph{Phys. Rev. D} {\bf 97} (2018) 126006}, [\href{http://arxiv.org/abs/1711.03469}{{\tt arXiv:1711.03469}}].

\bibitem{Quantum:Cosmology}
S.~Coleman, J.~B. Hartle, T.~Piran and S.~Weinberg, \emph{Quantum Cosmology and Baby Universes}.
\newblock WORLD SCIENTIFIC, 1991, \href{http://dx.doi.org/10.1142/1190}{10.1142/1190}.

\bibitem{Abdalla:2026mxn}
A.~I. Abdalla, S.~Antonini, R.~Bousso, L.~V. Iliesiu, A.~Levine and A.~Shahbazi-Moghaddam, \emph{{Consistent Evaluation of the No-Boundary Proposal}},  \href{http://arxiv.org/abs/2602.02682}{{\tt arXiv:2602.02682}}.

\bibitem{Banks:2025nfe}
T.~Banks, \emph{{Old Ideas for New Physicists III: String Theory Parameters are NOT Vacuum Expectation Values}},  \href{http://arxiv.org/abs/2501.17697}{{\tt arXiv:2501.17697}}.

\bibitem{Sen:2025bmj}
A.~Sen, \emph{{Are Moduli Vacuum Expectation Values or Parameters?}},  \href{http://arxiv.org/abs/2502.07883}{{\tt arXiv:2502.07883}}.

\bibitem{Betzios:2022oef}
P.~Betzios, N.~Gaddam and O.~Papadoulaki, \emph{{Baby universes born from the void}}, \href{http://dx.doi.org/10.1142/S0218271822420214}{\emph{Int. J. Mod. Phys. D} {\bf 31} (2022) 2242021}, [\href{http://arxiv.org/abs/2204.01764}{{\tt arXiv:2204.01764}}].

\bibitem{Bezrukov:2007ep}
F.~L. Bezrukov and M.~Shaposhnikov, \emph{{The Standard Model Higgs boson as the inflaton}}, \href{http://dx.doi.org/10.1016/j.physletb.2007.11.072}{\emph{Phys. Lett. B} {\bf 659} (2008) 703--706}, [\href{http://arxiv.org/abs/0710.3755}{{\tt arXiv:0710.3755}}].

\bibitem{Barvinsky:2008ia}
A.~O. Barvinsky, A.~Y. Kamenshchik and A.~A. Starobinsky, \emph{{Inflation scenario via the Standard Model Higgs boson and LHC}}, \href{http://dx.doi.org/10.1088/1475-7516/2008/11/021}{\emph{JCAP} {\bf 11} (2008) 021}, [\href{http://arxiv.org/abs/0809.2104}{{\tt arXiv:0809.2104}}].

\bibitem{DeSimone:2008ei}
A.~De~Simone, M.~P. Hertzberg and F.~Wilczek, \emph{{Running Inflation in the Standard Model}}, \href{http://dx.doi.org/10.1016/j.physletb.2009.05.054}{\emph{Phys. Lett. B} {\bf 678} (2009) 1--8}, [\href{http://arxiv.org/abs/0812.4946}{{\tt arXiv:0812.4946}}].

\bibitem{Barbon:2009ya}
J.~L.~F. Barbon and J.~R. Espinosa, \emph{{On the Naturalness of Higgs Inflation}}, \href{http://dx.doi.org/10.1103/PhysRevD.79.081302}{\emph{Phys. Rev. D} {\bf 79} (2009) 081302}, [\href{http://arxiv.org/abs/0903.0355}{{\tt arXiv:0903.0355}}].

\bibitem{Barvinsky:2009fy}
A.~O. Barvinsky, A.~Y. Kamenshchik, C.~Kiefer, A.~A. Starobinsky and C.~Steinwachs, \emph{{Asymptotic freedom in inflationary cosmology with a non-minimally coupled Higgs field}}, \href{http://dx.doi.org/10.1088/1475-7516/2009/12/003}{\emph{JCAP} {\bf 12} (2009) 003}, [\href{http://arxiv.org/abs/0904.1698}{{\tt arXiv:0904.1698}}].

\bibitem{Barvinsky:2009ii}
A.~O. Barvinsky, A.~Y. Kamenshchik, C.~Kiefer, A.~A. Starobinsky and C.~F. Steinwachs, \emph{{Higgs boson, renormalization group, and naturalness in cosmology}}, \href{http://dx.doi.org/10.1140/epjc/s10052-012-2219-3}{\emph{Eur. Phys. J. C} {\bf 72} (2012) 2219}, [\href{http://arxiv.org/abs/0910.1041}{{\tt arXiv:0910.1041}}].

\bibitem{Bezrukov:2010jz}
F.~Bezrukov, A.~Magnin, M.~Shaposhnikov and S.~Sibiryakov, \emph{{Higgs inflation: consistency and generalisations}}, \href{http://dx.doi.org/10.1007/JHEP01(2011)016}{\emph{JHEP} {\bf 01} (2011) 016}, [\href{http://arxiv.org/abs/1008.5157}{{\tt arXiv:1008.5157}}].

\bibitem{Bezrukov:2014ipa}
F.~Bezrukov, J.~Rubio and M.~Shaposhnikov, \emph{{Living beyond the edge: Higgs inflation and vacuum metastability}}, \href{http://dx.doi.org/10.1103/PhysRevD.92.083512}{\emph{Phys. Rev. D} {\bf 92} (2015) 083512}, [\href{http://arxiv.org/abs/1412.3811}{{\tt arXiv:1412.3811}}].

\bibitem{Gialamas:2025kef}
I.~D. Gialamas, A.~Karam, A.~Racioppi and M.~Raidal, \emph{{Has ACT measured radiative corrections to the tree-level Higgs-like inflation?}}, \href{http://dx.doi.org/10.1103/6fpc-67s1}{\emph{Phys. Rev. D} {\bf 112} (2025) 103544}, [\href{http://arxiv.org/abs/2504.06002}{{\tt arXiv:2504.06002}}].

\bibitem{Coleman:1985rnk}
S.~Coleman, \emph{{Aspects of Symmetry}: {Selected Erice Lectures}}.
\newblock Cambridge University Press, Cambridge, U.K., 1985, \href{http://dx.doi.org/10.1017/CBO9780511565045}{10.1017/CBO9780511565045}.

\bibitem{Antonini:2024bbm}
S.~Antonini and L.~G.~C. Bariuan, \emph{{Magnetic braneworlds: cosmology and wormholes}}, \href{http://dx.doi.org/10.1007/JHEP09(2024)070}{\emph{JHEP} {\bf 09} (2024) 070}, [\href{http://arxiv.org/abs/2405.18465}{{\tt arXiv:2405.18465}}].

\bibitem{Marolf:2021kjc}
D.~Marolf and J.~E. Santos, \emph{{AdS Euclidean wormholes}}, \href{http://dx.doi.org/10.1088/1361-6382/ac2cb7}{\emph{Class. Quant. Grav.} {\bf 38} (2021) 224002}, [\href{http://arxiv.org/abs/2101.08875}{{\tt arXiv:2101.08875}}].

\bibitem{Lan:2024gnv}
Q.-Y. Lan and Y.-S. Piao, \emph{{Prepare inflationary universe via the Euclidean charged wormhole}},  \href{http://arxiv.org/abs/2411.13844}{{\tt arXiv:2411.13844}}.

\bibitem{Rey:1989th}
S.-J. Rey, \emph{{Space-time Wormholes With {Yang-Mills} Fields}}, \href{http://dx.doi.org/10.1016/0550-3213(90)90346-F}{\emph{Nucl. Phys. B} {\bf 336} (1990) 146--156}.

\bibitem{Hawking:1995ap}
S.~W. Hawking and S.~F. Ross, \emph{{Duality between electric and magnetic black holes}}, \href{http://dx.doi.org/10.1103/PhysRevD.52.5865}{\emph{Phys. Rev. D} {\bf 52} (1995) 5865--5876}, [\href{http://arxiv.org/abs/hep-th/9504019}{{\tt hep-th/9504019}}].

\bibitem{Witten:2003ya}
E.~Witten, \emph{{SL(2,Z) action on three-dimensional conformal field theories with Abelian symmetry}},  in \emph{{From Fields to Strings: Circumnavigating Theoretical Physics: A Conference in Tribute to Ian Kogan}}, pp.~1173--1200, 7, 2003.
\newblock \href{http://arxiv.org/abs/hep-th/0307041}{{\tt hep-th/0307041}}.

\bibitem{Skenderis:2002wp}
K.~Skenderis, \emph{{Lecture notes on holographic renormalization}}, \href{http://dx.doi.org/10.1088/0264-9381/19/22/306}{\emph{Class. Quant. Grav.} {\bf 19} (2002) 5849--5876}, [\href{http://arxiv.org/abs/hep-th/0209067}{{\tt hep-th/0209067}}].

\bibitem{Jonas:2023ipa}
C.~Jonas, G.~Lavrelashvili and J.-L. Lehners, \emph{{Zoo of axionic wormholes}}, \href{http://dx.doi.org/10.1103/PhysRevD.108.066012}{\emph{Phys. Rev. D} {\bf 108} (2023) 066012}, [\href{http://arxiv.org/abs/2306.11129}{{\tt arXiv:2306.11129}}].

\bibitem{Halliwell:1989pu}
J.~J. Halliwell and R.~C. Myers, \emph{{Multiple Sphere Configurations in the Path Integral Representation of the Wave Function of the Universe}}, \href{http://dx.doi.org/10.1103/PhysRevD.40.4011}{\emph{Phys. Rev. D} {\bf 40} (1989) 4011}.

\bibitem{Aguilar-Gutierrez:2023ril}
S.~E. Aguilar-Gutierrez, T.~Hertog, R.~Tielemans, J.~P. van~der Schaar and T.~Van~Riet, \emph{{Axion-de Sitter wormholes}}, \href{http://dx.doi.org/10.1007/JHEP11(2023)225}{\emph{JHEP} {\bf 11} (2023) 225}, [\href{http://arxiv.org/abs/2306.13951}{{\tt arXiv:2306.13951}}].

\bibitem{Anninos:2017hhn}
D.~Anninos and D.~M. Hofman, \emph{{Infrared Realization of dS$_2$ in AdS$_2$}}, \href{http://dx.doi.org/10.1088/1361-6382/aab143}{\emph{Class. Quant. Grav.} {\bf 35} (2018) 085003}, [\href{http://arxiv.org/abs/1703.04622}{{\tt arXiv:1703.04622}}].

\bibitem{Breitenlohner:1982bm}
P.~Breitenlohner and D.~Z. Freedman, \emph{{Positive Energy in anti-De Sitter Backgrounds and Gauged Extended Supergravity}}, \href{http://dx.doi.org/10.1016/0370-2693(82)90643-8}{\emph{Phys. Lett. B} {\bf 115} (1982) 197--201}.

\bibitem{Breitenlohner:1982jf}
P.~Breitenlohner and D.~Z. Freedman, \emph{{Stability in Gauged Extended Supergravity}}, \href{http://dx.doi.org/10.1016/0003-4916(82)90116-6}{\emph{Annals Phys.} {\bf 144} (1982) 249}.

\bibitem{Petkou_1999}
A.~Petkou and K.~Skenderis, \emph{A non-renormalization theorem for conformal anomalies}, \href{http://dx.doi.org/10.1016/s0550-3213(99)00514-3}{\emph{Nuclear Physics B} {\bf 561} (Nov., 1999) 100–116}.

\bibitem{Papadimitriou:2007sj}
I.~Papadimitriou, \emph{{Multi-Trace Deformations in AdS/CFT: Exploring the Vacuum Structure of the Deformed CFT}}, \href{http://dx.doi.org/10.1088/1126-6708/2007/05/075}{\emph{JHEP} {\bf 05} (2007) 075}, [\href{http://arxiv.org/abs/hep-th/0703152}{{\tt hep-th/0703152}}].

\bibitem{Planck:2018jri}
{\scshape Planck} collaboration, Y.~Akrami et~al., \emph{{Planck 2018 results. X. Constraints on inflation}}, \href{http://dx.doi.org/10.1051/0004-6361/201833887}{\emph{Astron. Astrophys.} {\bf 641} (2020) A10}, [\href{http://arxiv.org/abs/1807.06211}{{\tt arXiv:1807.06211}}].

\bibitem{BICEP:2021xfz}
{\scshape BICEP, Keck} collaboration, P.~A.~R. Ade et~al., \emph{{Improved Constraints on Primordial Gravitational Waves using Planck, WMAP, and BICEP/Keck Observations through the 2018 Observing Season}}, \href{http://dx.doi.org/10.1103/PhysRevLett.127.151301}{\emph{Phys. Rev. Lett.} {\bf 127} (2021) 151301}, [\href{http://arxiv.org/abs/2110.00483}{{\tt arXiv:2110.00483}}].

\bibitem{Blommaert:2025bgd}
A.~Blommaert, J.~Kudler-Flam and E.~Y. Urbach, \emph{{Absolute entropy and the observer{\textquoteright}s no-boundary state}}, \href{http://dx.doi.org/10.1007/JHEP11(2025)113}{\emph{JHEP} {\bf 11} (2025) 113}, [\href{http://arxiv.org/abs/2505.14771}{{\tt arXiv:2505.14771}}].

\bibitem{Betzios:2023jco}
P.~Betzios, N.~Gaddam and O.~Papadoulaki, \emph{{Black hole {\textemdash} wormhole transitions in two dimensional string theory}}, \href{http://dx.doi.org/10.1007/JHEP05(2024)132}{\emph{JHEP} {\bf 05} (2024) 132}, [\href{http://arxiv.org/abs/2312.02257}{{\tt arXiv:2312.02257}}].

\bibitem{Liu:2025cml}
H.~Liu, \emph{{Towards a holographic description of closed universes}},  \href{http://arxiv.org/abs/2509.14327}{{\tt arXiv:2509.14327}}.

\bibitem{Antonini:2025ioh}
S.~Antonini, P.~Rath, M.~Sasieta, B.~Swingle and A.~Vilar~L{\'o}pez, \emph{{The baby universe is fine and the CFT knows it: on holography for closed universes}}, \href{http://dx.doi.org/10.1007/JHEP12(2025)159}{\emph{JHEP} {\bf 12} (2025) 159}, [\href{http://arxiv.org/abs/2507.10649}{{\tt arXiv:2507.10649}}].

\bibitem{Harlow:2026hky}
D.~Harlow, \emph{{Observers, $\alpha$-parameters, and the Hartle-Hawking state}},  \href{http://arxiv.org/abs/2602.03835}{{\tt arXiv:2602.03835}}.

\bibitem{Nomura:2026igt}
Y.~Nomura and T.~Ugajin, \emph{{Physical Predictions in Closed Quantum Gravity}},  \href{http://arxiv.org/abs/2602.13387}{{\tt arXiv:2602.13387}}.

\bibitem{Zhao:2026mpl}
Y.~Zhao, \emph{{''It from Bit'': The Hartle-Hawking state and quantum mechanics for de Sitter observers}},  \href{http://arxiv.org/abs/2602.05939}{{\tt arXiv:2602.05939}}.

\bibitem{Hebecker:2018ofv}
A.~Hebecker, T.~Mikhail and P.~Soler, \emph{{Euclidean wormholes, baby universes, and their impact on particle physics and cosmology}}, \href{http://dx.doi.org/10.3389/fspas.2018.00035}{\emph{Front. Astron. Space Sci.} {\bf 5} (2018) 35}, [\href{http://arxiv.org/abs/1807.00824}{{\tt arXiv:1807.00824}}].

\bibitem{Betzios:2025eev}
P.~Betzios, \emph{{A microscopic normal matrix model for (A)dS$_{2}$}}, \href{http://dx.doi.org/10.1007/JHEP01(2026)008}{\emph{JHEP} {\bf 01} (2026) 008}, [\href{http://arxiv.org/abs/2505.23891}{{\tt arXiv:2505.23891}}].

\bibitem{AbdusSalam:2025twp}
S.~AbdusSalam, C.~Hughes, F.~Quevedo and A.~Schachner, \emph{{Coexisting flux string vacua from numerical K{\"a}hler moduli stabilisation}}, \href{http://dx.doi.org/10.1007/JHEP01(2026)056}{\emph{JHEP} {\bf 01} (2026) 056}, [\href{http://arxiv.org/abs/2507.00615}{{\tt arXiv:2507.00615}}].

\end{thebibliography}\endgroup

\end{document}